\documentclass[journal]{IEEEtran}

\usepackage{graphicx}
\usepackage{amsmath}
\usepackage{amssymb}
\usepackage{tabu}
\usepackage{cite}
\usepackage{indentfirst}
\usepackage[utf8]{inputenc}
\usepackage[english]{babel}
\usepackage{bbm} 
\usepackage{epstopdf}
\usepackage{amsxtra}
\usepackage{amstext}
\usepackage{amsthm}
\usepackage{latexsym}
\usepackage{dsfont} 
\usepackage{xcolor}
\usepackage{units}
\usepackage{multirow}
\usepackage{lipsum}
\usepackage{bm}
\usepackage{multicol}
\usepackage{setspace}
\usepackage{caption}
\usepackage{subcaption}
\definecolor{darkgreen}{rgb}{0,0.75,0}
\usepackage[colorlinks=true]{hyperref}
\hypersetup{
	pdfauthor = {Asmaa W. Abdallah},
	pdfcreator = {Asmaa W. Abdallah},
	pdfnewwindow,
	pdffitwindow=true,
	colorlinks=true,
	linkcolor=blue,       
	citecolor=darkgreen,      
	filecolor=red,        
	urlcolor=cyan
}
\definecolor{maroon}{RGB}{186,0,0}
\definecolor{purple}{RGB}{96,26,149}
\definecolor{mavi}{RGB}{46,76,255}
\definecolor{haki}{RGB}{38,99,33}

\makeatletter
\let\NAT@parse\undefined
\makeatother

\theoremstyle{remark}

\theoremstyle{definition}

\usepackage{algorithm}
\usepackage{algpseudocode}
\algnewcommand{\LineComment}[1]{\State \(\triangleright\) #1}

\newcommand{\nth}[1]{{#1}^{\text{th}}}
\newcommand{\mbf}[1]{\mathbf{#1}}

\newcommand{\abs}[1]{\left|{#1}\right|}
\newcommand{\norm}[1]{\left\|{#1}\right\|}

\newcommand{\expec}[1]{{\mathbb{E}\!\left[{#1}\right]}}

\begin{document}
    \title{Deep Learning Based Frequency-Selective Channel Estimation for Hybrid mmWave MIMO Systems}
	\author
	{
Asmaa~Abdallah,~\IEEEmembership{Member,~IEEE},
Abdulkadir~Celik,~\IEEEmembership{Senior~Member,~IEEE},\\
Mohammad~M.~Mansour,~\IEEEmembership{Senior Member,~IEEE},
and Ahmed~M.~Eltawil,~\IEEEmembership{Senior Member,~IEEE}.
	}
\maketitle

\begin{abstract}	
Millimeter wave (mmWave) massive multiple-input multiple-output (MIMO) systems typically employ hybrid mixed signal processing to avoid expensive hardware and high training overheads. {However, the lack of fully digital beamforming at mmWave bands imposes additional challenges in channel estimation.
Prior art on hybrid architectures has mainly focused on greedy optimization algorithms to estimate frequency-flat narrowband mmWave channels, despite the fact that in practice, the large bandwidth associated with mmWave channels results in frequency-selective channels. In this paper, we consider a frequency-selective wideband mmWave system and propose two deep learning (DL) compressive sensing (CS) based algorithms for channel estimation.} The proposed algorithms learn critical apriori information from training data to provide highly accurate channel estimates with low training overhead. In the first approach, a DL-CS based algorithm simultaneously estimates the channel supports in the frequency domain, which are then used for channel reconstruction. The second approach exploits the estimated supports to apply a low-complexity multi-resolution fine-tuning method to further enhance the estimation performance. Simulation results demonstrate that the proposed DL-based schemes significantly outperform conventional orthogonal matching pursuit (OMP) techniques in terms of the normalized mean-squared error (NMSE), computational complexity, and spectral efficiency, particularly in the low signal-to-noise ratio regime. {When compared to OMP approaches that achieve an NMSE gap of $\unit[\{4-10\}]{dB}$ with respect to the Cramer Rao Lower Bound (CRLB), the proposed algorithms reduce the CRLB gap to only  $\unit[\{1-1.5\}]{dB}$, while significantly reducing complexity by two orders of magnitude.}




\end{abstract}
	

\section{Introduction}
\IEEEPARstart Millimeter wave (mmWave) communication has emerged as a key technology to fulfill beyond fifth-generation (B5G) network requirements, such as enhanced mobile broadband, massive connectivity, and ultra-reliable low-latency communications. The mmWave band offers an abundant frequency spectrum (30-300 GHz) at the cost of low penetration depth and high propagation losses. Fortunately, its short-wavelength mitigates these drawbacks by allowing the deployment of large antenna arrays into small form factor transceivers, paving the way for multiple-input multiple-output (MIMO) systems with high directivity gains \cite{Pi2011InroMmwave,Heath2016OverviewMmwave,Khateeb2014covcap,Andrews20145g}. 

Hybrid MIMO structures have been introduced to operate at mmWave frequencies because an all-digital architecture, { with a dedicated radio frequency (RF) chain for each antenna element}, results in expensive system architecture and high power consumption at these frequencies \cite{Heath2016OverviewMmwave}. {In these hybrid architectures, phase-only analog beamformers are employed to steer the beams using steering vectors of quantized angles. The down-converted signal is then processed by low-dimensional baseband beamformers, each of which is dedicated to a single RF chain \cite{Khateeb2014mmwave,Heath2016shiftOrSwitches}. The number of RF chains is significantly reduced with this combination of high-dimensional phase-only analog and low-dimensional baseband digital beamformers \cite{Heath2016shiftOrSwitches}.} Moreover, optimal configuration of the digital/analog precoders and combiners requires instantaneous channel state information (CSI) to achieve spatial diversity and multiplexing gain \cite{Khateeb2014Chanest}.  However, acquiring mmWave CSI is challenging with a hybrid architecture due to the following reasons  \cite{Khateeb2014mmwave}: 1) {There is no direct access to the different antenna elements in the array since the channel is seen through the analog combining network, which forms a compression stage for the received signal when the number of RF chains is much smaller than the  number of antennas}, 2)  the large channel bandwidth yields high noise power and low received signal-to-noise-ratio (SNR) before beamforming, and 3) the large size of channel matrices increases the complexity and overheads associated with traditional precoding and channel estimation algorithms.  {Therefore, low complexity channel estimation for mmWave MIMO systems with hybrid architecture is necessary.}


   
\subsection{Related Work}
Channel estimation techniques typically leverage the sparse nature of mmWave MIMO channels by formulating the estimation as a sparse recovery problem and apply compressive sensing (CS) methods to solve it. Compressive sensing is a general framework for estimation of sparse vectors from linear measurements \cite{eldar2011CS}. The estimated \emph{supports} of the sparse vectors using CS help identify the indices of Angle-of-Arrival (AoA) and Angle-of-Departure (AoD) pairs for each path in the mmWave channel, while the \emph{amplitudes} of the non-zero coefficients in the sparse vectors represent the channel gains for each path. Therefore, these \emph{supports} and \emph{amplitudes} are key components to be estimated to obtain accurate CSI. Moreover, it has been shown that pilot training overhead can be reduced with compressive estimation, unlike the conventional approaches such as those based on least squares (LS) estimation \cite{Heath2016shiftOrSwitches}. 

Several channel estimation methods based on CS tools that explore the mmWave channel sparsity have been investigated in the literature \cite{Heath2016shiftOrSwitches,Gao2016CEfreqSelec,Heath2017TDCE,Heath2018CEMain,Ma2018CE}.  
A  distributed grid matching pursuit (DGMP) channel estimation scheme is presented in \cite{Gao2016CEfreqSelec}, where the dominant entries of the line-of-sight (LoS) channel path are detected and updated iteratively. In \cite{Heath2017TDCE}, an orthogonal matching pursuit (OMP) channel estimation scheme to detect multiple channel paths support entries is also considered.{ Likewise, a simultaneous weighted orthogonal matching pursuit (SW-OMP) channel estimation scheme based on a weighted OMP method is developed in \cite{Heath2018CEMain} for frequency-selective mmWave systems. A sparse reconstruction problem was formulated in \cite{Heath2018CEMain} to estimate the channel independently for every subcarrier by exploiting common sparsity in the frequency domain.} However, such optimization and CS-based channel estimation schemes detect the support indices of the mmWave channel sequentially and greedily, and hence are not globally optimal~\cite{Ma2018CE}.
	
Alternatively, deep learning (DL) approaches and data-driven algorithms have recently received much attention as key enablers for beyond 5G networks. Traditionally, signal processing and numerical optimization techniques have been heavily used to address channel estimation at mmWave bands \cite{Gao2016CEfreqSelec,Heath2017TDCE,Heath2018CEMain,Ma2018CE}. However, optimization algorithms often demand considerable computational complexity overhead, which creates a barrier between theoretical design/analysis and real-time processing requirements.  {Hence, the prior data-set observations and deep neural network (DNN) models can be leveraged to learn the non-trivial mapping from compressed received pilots to channels.} DNNs can be used to approximate the optimization problems by selecting the suitable set of parameters that minimize the approximation error. The use of DNNs is expected to substantially reduce computational complexity and processing overhead since it only requires several layers of simple operations such as matrix-vector multiplications.  {Moreover, several successful DL applications have been demonstrated in wireless communications problems such as channel estimation \cite{Ye2018PwrDLCE,DOng2019DLDNNCE,He2018DLCE,M2019DLCE,Ma2020SparseDLCE,wei2019knowledge,Chun2019DLCEmassiveMIMO,Jin2019CellFreeDLCE,andrew2020MIMOCEDL,Bjornson2020CEbayes},  analog beam selection \cite{long2018DD}, \cite{Hodge2019RFmeta}, and hybrid beamforming \cite{long2018DD,Khateeb2018DLbeam,Huang2019DLHP,Elbir2019CNNHP,Elbir2020JASHP,elbir2019online}. Besides, DL-based techniques, when compared with other conventional optimization methods, have been shown  \cite{DOng2019DLDNNCE,Elbir2019CNNHP,Elbir2020JASHP,dorner2018DLair} to be more computationally efficient in searching for beamformers and more tolerant to imperfect channel inputs.  }
In \cite{He2018DLCE}, a learned denoising-based approximate message passing (LDAMP) network is presented to estimate the mmWave communication system with lens antenna array, where the noise term is detected and removed to estimate the channel. However, channel estimation for mmWave massive MIMO systems with hybrid architecture is not considered in \cite{He2018DLCE}. 

{
Prior work on channel estimation for hybrid mmWave MIMO architecture \cite{Khateeb2018DLbeam,Huang2019DLHP,Elbir2019CNNHP,M2019DLCE, Xu2019DLCEMultiuser, He2018DLCE, Kang2018DLCEEnergy,Ma2020SparseDLCE,wei2019knowledge,Chun2019DLCEmassiveMIMO,Jin2019CellFreeDLCE,andrew2020MIMOCEDL,Bjornson2020CEbayes, Asmaa2020CF,Asmaa2019CF} consider the narrow-band flat fading channel model for tractability, while the practical mmWave channels exhibit the wideband frequency-selective fading due to the very large bandwidth, short coherence time and different delays of multipath\cite{Heath2018CEMain, Heath2016FSF,emil2019sub6mmWave}.	
MmWave environments such as indoor and vehicular communications are highly variable with short coherence time  \cite{emil2019sub6mmWave} which requires channel estimation algorithms that are robust to the rapidly changing channel characteristics \footnote{The coherence time is within few milliseconds such as $\unit[5]{ms}$ when operating at $\unit[60]{GHz}$ with $\unit[1 ]{GHz}$ bandwidth \cite{emil2019sub6mmWave}.}.} 
Accordingly, this paper presents combination of DL and CS methods to identify AoA/AoD pairs' indices and estimate the channel amplitudes for frequency-selective channel estimation of hybrid MIMO systems.

	
	
\subsection{Contributions of the Paper}
In this paper, we propose a frequency-selective channel estimation framework for mmWave MIMO systems with hybrid architecture. By considering the mmWave channel sparsity, the developed method aims at reaping the full advantages of both CS and DL methods. We consider the received pilot signal as an image, and then employ a denoising convolutional neural network (DnCNN) from \cite{Zhang2017Dncnn} for channel amplitude estimation. Thereby, we treat image denoising as a plain discriminative learning problem, i.e., separating the noise from a noisy image by feed-forward convolutional neural networks (CNNs). The main motivations behind using CNNs are twofold: First, deep CNNs have been recognized to effectively extract image features \cite{Zhang2017Dncnn}.  Second, considerable advances have been achieved on regularization and learning methods for training CNNs, including Rectifier Linear Unit (ReLU), batch normalization, and residual learning \cite{he2016deep}. These methods can be adopted in CNNs to speed up the training process and improve the denoising performance. The main contributions of the paper can be summarized as follows:
	\begin{enumerate}
	
	\item 
	{We propose a deep learning compressed sensing channel estimation (DL-CS-CE) scheme for wideband mmWave massive MIMO systems. The proposed DL-CS-based channel estimation (DL-CS-CE) algorithm aims at exploiting the information on the support coming from every subcarrier in the MIMO-OFDM system. It is executed in two steps: channel amplitude estimation through deep learning and channel reconstruction.  We train a DnCNN using \emph{real} mmWave channel realizations obtained from Raymobtime \footnote{ Available at https://www.lasse.ufpa.br/raymobtime/}. The correlation between the received signal vectors and the measurement matrix is fed into the trained DnCNN to predict the  channel amplitudes. Using the obtained channel amplitudes, the indices of dominant entries of the channel are obtained, based on which the channel can be reconstructed. Unlike the existing work of~\cite{Gao2016CEfreqSelec,Heath2017TDCE,Heath2018CEMain} that estimates the dominant channel entries sequentially, we estimate dominant entries simultaneously, which is able to save in computational complexity and improve estimation performance.}
		
	\item 
	{Using the DL-CS-CE for support detection, we propose a refined DL-CS-CE algorithm that  exploits the spatially common sparsity within the system bandwidth.   A channel reconstruction with a low complexity multi-resolution fine-tuning approach is developed that further improves NMSE performance by enhancing the accuracy of the estimated AoAs/AoDs. The channel reconstruction is performed by consuming a very small amount of pilot training frames, which significantly reduces the training overhead and computational complexity.}
		
	\item 
	{Simulation results in the low SNR regime show that both proposed algorithms significantly outperform the frequency domain approach developed in \cite{Heath2018CEMain}. Numerical results also show that using a reasonably small pilot training frames, approximately in the range of 60-100 frames, leads to substantially low channel estimation errors. The proposed algorithms are also compared with existing solutions by analyzing the trade-off between delivered performance and incurred computational complexity. Our analysis reveals that both proposed channel estimation methods achieve the desired performance at significant lower complexity. The developed approaches are shown to attain an NMSE gap of $\unit[1-1.5]{dB}$ with the Cramer Rao Lower Bound (CRLB) compared to the $\unit[4-10]{dB}$ gap attained by the SW-OMP technique, while reducing the computational complexity by two orders of magnitude.}
	\end{enumerate}

\begin{figure*}[t]
		\centering		\includegraphics[width=\textwidth]{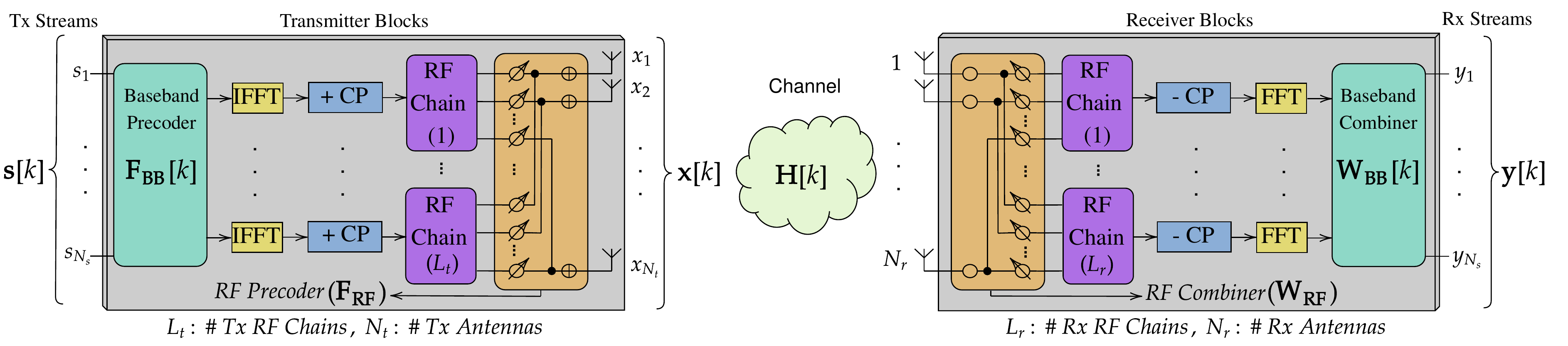}
		\caption{Hybrid architecture system model of a mmWave MIMO system, which includes analog/digital precoders and combiners.}\label{fig:model}
	\end{figure*}
	
\subsection{Notation and Paper Organization}
Bold upper case, bold lower case, and lower case letters correspond to matrices, vectors, and scalars, respectively. Scalar norms, vector $\text{L}_2$ norms, and Frobenius norms, are denoted by $\abs{\cdot}$, $\norm{\cdot}_2$, and $\norm{\cdot}_{\text{F}}$, respectively. We use $\mathcal{X}$ to denote a set. $\mbf{I}_X$ denotes a $X\times X$ identity matrix. $\expec{\cdot}$, $(\cdot)^{\mathsf{T}}$, $\bar {(\cdot)}$, and $(\cdot)^*$ 
	stand for expected value, transpose, complex conjugate, and Hermitian. 
	{$\bf{X}^\dagger $ stands for the Moore-Penrose pseudo-inverse of $\bf{X}$.} 
	$[\mbf{x}]_i$ represents $\nth{i}$ element of a vector $\mbf{x}$. The $\nth{(i, j)}$ entry of a matrix  $\mbf{X}$ is denoted by $[\mbf{X}]_{i,j}$. {In addition, $[\mbf{X}]_{:,j}$ and $[\mbf{X}]_{:,\Omega}$ denote the $\nth{j}$ column vector of matrix $\mbf{X}$ and the sub-matrix consisting of the columns of matrix $\mbf{X}$ with indices in set $\Omega$. $\{a\}\bmod\: b$ means $a$ modulo $b$. }$\mathcal{CN}(\boldsymbol{\mu},\mbf{C})$ refers to a circularly-symmetric complex Gaussian distribution with mean $\boldsymbol{\mu}$ and covariance matrix $\mbf{C}$. The operations $\mathrm{vec}{(\bf X)}$, $\mathrm{vec2mat}({\bf x}, sz)$, $\mathrm{sub2ind}(sz, [{ r},{ c}])$, and $\mathrm{ind2sub}(sz, { i})$ correspond to transforming a matrix into a vector, transforming a vector into a matrix for a defined size ($sz$), transforming the row $r$ and column $c$ subscripts of a matrix into their corresponding linear index, and transforming the linear index $i$ into its corresponding row and column subscripts  for a matrix of a defined size ($sz$), respectively. $\mbf{X} \otimes \mbf{Y}$ is the Kronecker product of $\mbf{X}$ and $\mbf{Y}$. Key model-related notation is listed in Table \ref{table:model}. 
	
The rest of the paper is organized as follows. The system model for the frequency selective mmwave  MIMO system  is described in Section~\ref{s:System_Model}. In Section~\ref{sec:proposed_schem}, the proposed two deep learning-based compressive sensing channel estimation schemes in the frequency domain are introduced.  Moreover, complexity analysis  in terms of convergence and computational analysis is presented in Section~\ref{sec:Comp_analysis}. Case studies with numerical results are simulated and analyzed based on the proposed schemes in Section~\ref{sec:simulation}. Section~\ref{sec:conclusion} concludes the paper.

\begin{table}[t]
	\centering
\captionsetup{justification=centering}
	\caption{\textsc{Notation}}
		\footnotesize
	\begin{tabular}{|l|c|}
		\hline
		\textbf{Notation} & \textbf{ Definition} \\\hline
		 ${\mathbf {F}}_{\mathrm {RF}}\in \mathbb{C}^{N
		 _{\mathrm{t}}\times L_{\mathrm{t}}} $  & RF analog precoder  (time domain (TD))\\ \hline
		 ${\mathbf {W}}_{\mathrm {RF}} \in \mathbb{C}^{N
		 _{\mathrm{r}}\times L_{\mathrm{r}}}$ & RF analog combiner  (TD)\\ \hline
		  $\boldsymbol {\mathsf {F}} _{{{\text{BB}}}}[k] \in \mathbb{C}^{L_{\mathrm{t}}\times N_{\mathrm{s}}} $ & Baseband digital precoder (frequency domain(FD)) \\\hline
		 $ \boldsymbol {\mathsf {W}} _{{{\text{BB}}}}[k] \in \mathbb{C}^{L_{\mathrm{r}}\times N_{\mathrm{s}}} $ & Baseband digital combiner (FD) \\\hline
	 	 $ \boldsymbol {\mathsf {s}} [k] \in \mathbb{C}^{ N_{\mathrm{s}}\times 1} $ & Data symbol vector (FD) \\\hline
	 	 $\mathbf{H}_{d} \in \mathbb{C}^{N_{\mathrm{r}}\times N_{\mathrm{t}}}$ & $\nth{d}$ delay tap of the channel (TD)\\\hline
	 	 $\mathbf{\Delta}_d \in \mathbb{C}^{L\times L}$ &  Complex diagonal matrix (time domain)\\\hline 
	 	 $\mathbf{A}_{\mathrm{R}} \in \mathbb{C}^{N_{\mathrm{r}}\times L}$ & Receive array steering matrix\\\hline
	 	 $\mathbf{A}_{\mathrm{T}} \in \mathbb{C}^{N_{\mathrm{t}}\times L}$ & Transmit array steering matrix \\\hline
	 	  $\boldsymbol {\mathsf {H}}[k]\in \mathbb{C}^{N_{\mathrm{r}}\times N_{\mathrm{t}}}$ & Channel at $\nth{k}$ subcarrier (FD)\\\hline
	 	  $\mathbf{\Delta}[k] \in \mathbb{C}^{L\times L}$ &  Complex diagonal matrix (FD)\\\hline 
	 	  $\tilde { {\mathbf {A}}}_{\mathrm{R}} \in \mathbb{C}^{N_{\mathrm{r}}\times G_{\mathrm {r}}}$ & Dictionary matrix for receive array response\\\hline 
 	  	  $\tilde { {\mathbf {A}}}_{\mathrm{T}}\in \mathbb{C}^{N_{\mathrm{t}}\times G_{\mathrm {t}}}$ & Dict. matrix for transmit array response\\\hline
 	  	  $\tilde { {\mathbf {A}}}_{\mathrm{R}}^{\mathrm{r}} \in \mathbb{C}^{N_{\mathrm{r}}\times G_{\mathrm {r}}^{\mathrm{r}}}$ & Refining dict. matrix for receive array response\\\hline 
 	  	  $\tilde { {\mathbf {A}}}_{\mathrm{T}}^{\mathrm{r}}\in \mathbb{C}^{N_{\mathrm{t}}\times G_{\mathrm {t}}^{\mathrm{r}}}$ & Refining dict. matrix for transmit array response\\\hline 
 	  	  $\boldsymbol{\Delta }_{d}^{v}\in \mathbb {C}^{ G_{\mathrm {r}}\times G_{\mathrm {t}}}$ & Path gains sparse matrix of the \emph{virtual} channel (TD)\\\hline
 	  	  $\boldsymbol{\Delta }^{\mathrm{v}}[k] \in \mathbb {C}^{ G_{\mathrm {r}}\times G_{\mathrm {t}}}$ & Path gains sparse matrix of the \emph{virtual} channel (FD)\\\hline
 	  	   $\boldsymbol{\Phi } \in \mathbb {C}^{M L_{{{\text{r}}}} \times N_{{{\text{t}}}} N_{{{\text{r}}}}}$ & Measurement matrix \\\hline
 	  	   $\boldsymbol{\Psi } \in \mathbb {C}^{N_{{{\text{t}}}} N_{{{\text{r}}}} \times G_{\mathrm{t}} G_{\mathrm{r}}}$ & Dictionary matrix\\\hline 
 	  	   ${\boldsymbol {\mathsf {h}}}^{\mathrm{v}}[k]\in \mathbb {C}^{G_{\mathrm{r}}G_{\mathrm{t}}\times 1}$ & Sparse vector containing complex channel gains (FD)\\\hline
 	  	   $\boldsymbol \Upsilon \in \mathbb {C}^{M L_{{{\text{r}}}} \times G_{\mathrm{t}} G_{\mathrm{r}}}$ & Equivalent measurement matrix\\\hline 
 	  	   ${\boldsymbol {\mathsf {y}}}[k] \in \mathbb {C}^{M L_{{{\text{r}}}} \times1}$& Received signal (FD)\\\hline
 	  	    ${\boldsymbol {\mathsf {c}}}[k] \in \mathbb {C}^{G_{\mathrm{r}} G_{\mathrm{t}}}$ & Correlation vector (FD)\\\hline
 	  	   ${\mathbf {C}_{\text{w}}}\in \mathbb {C}^{M L_{{{\text{r}}}} \times M L_{{{\text{r}}}}}$ &  Noise covariance matrix of ${\boldsymbol {\mathsf {y}}}[k]$\\\hline
 	  	   ${\mathbf {D}}_{{{\text{w}}}} \in \mathbb {C}^{M L_{{{\text{r}}}} \times M L_{{{\text{r}}}}}$ & Whitening matrix (upper triangular matrix)\\\hline
 	  	   ${\boldsymbol {\mathsf {y}}}_{{{\text{w}}}}[k] \in \mathbb {C}^{M L_{{{\text{r}}}} \times1}$& Whitened received signal (FD)\\\hline
 	  	   $\boldsymbol \Upsilon _{{{\text{w}}}} \in \mathbb {C}^{M L_{\mathrm {r}}\times G_{\mathrm {t}} G_{\mathrm {r}} }$ & Whitened measurement matrix\\\hline
 	  	   $\boldsymbol \Upsilon _{{{\text{w}}}}^{\mathrm{d}} \in \mathbb {C}^{M L_{\mathrm {r}}\times G_{\mathrm {t}} G_{\mathrm {r}}^\mathrm{r} }$ & White. meas. matrix to remove detection uncertainty \\\hline
 	  	   $\boldsymbol \Upsilon _{{{\text{w}}}}^{\mathrm{r}} \in \mathbb {C}^{M L_{\mathrm {r}}\times G_{\mathrm {t}} G_{\mathrm {r}} }$ & White. meas. matrix for refining \\\hline
 	  	   $ {\boldsymbol {\mathsf {C}}}_{\bm{\alpha}}[k]\in \mathbb{R}^{ G_{\mathrm {r}}\times G_{\mathrm {t}} }$ & Input matrix to the DnCNN (FD)\\\hline
 	  	   ${\bf  G}[k]\in \mathbb {R}^{G_{\mathrm{r}}\times G_{\mathrm{t}} }$ & Output matrix of the DnCNN (FD)\\\hline
 	  	   $\boldsymbol{g}[k] \in \mathbb {R}^{G_{\mathrm{r}} G_{\mathrm{t}} \times 1} $& Vectorized form of  ${\bf  G}[k]$ (FD) \\\hline
 	  	   $\boldsymbol \xi [k] \in \mathbb{C}^{L\times 1}$& Vector of actual channel gains (FD) \\\hline
 	  	   ${\mathbf {P}}\in \mathbb {C}^{M L_{\mathrm {r}}\times M L_{\mathrm {r}}}$& Projection matrix\\\hline
 	  	   ${\boldsymbol {\mathsf {r}}}[k] \in \mathbb {C}^{M L_{{{\text{r}}}} \times1}$& Residual vector (FD)\\\hline
 	  	   $\mathcal{T}$& Sparse channel support set\\\hline
 	  	   $\mathcal{K}$& Subset from total $K$ subcarriers\\\hline
 	  	   
	\end{tabular}
	\label{table:model}
\end{table}

\section{System Model and Problem Formulation}
\label{s:System_Model}

This section first provides the system and channel models of frequeny-selective hybrid mmWave transceivers. Then, it {formulates a sparse recovery problem} to estimate the sparse channel in the frequency domain.

\subsection {System Model}
As shown in Fig.~\ref{fig:model}, we consider an OFDM-based mmWave MIMO link employing a total of $K$ subcarriers to send $N_{\mathrm{s}}$ data streams from a transmitter with $N_{\mathrm{t}}$ antennas to a receiver with $N_{\mathrm{r}}$ antennas. The system is based on a hybrid MIMO architecture, with $L_{\mathrm{t}}<N_{\mathrm{t}}$ and $L_{\mathrm{r}}< N_{\mathrm{r}}$ radio frequency (RF) chains at the transmitter and receiver sides.  Following the notation of~\cite{Heath2018CEMain}, we define  a frequency-selective hybrid precoder $\boldsymbol {\mathsf {F}}[k]= {\mathbf {F}}_{\mathrm {RF}} \boldsymbol {\mathsf {F}} _{{{\text{BB}}}}[k] \in {\mathbb {C}}^{ N_{\mathrm {t}}\times N_{\mathrm {s}}}$, $k = 0,\dots,K - 1$, where ${\mathbf {F}}_{\mathrm {RF}} $ and $\boldsymbol {\mathsf {F}} _{{{\text{BB}}}}[k] $ are the analog and digital precoders, respectively.  Although, the analog precoder is considered to be frequency-flat, the digital precoder is different for every subcarrier. The RF precoder and combiner are deployed using a fully connected network of quantized phase shifters, as described in \cite{Heath2016shiftOrSwitches}. During transmission, the transmitter (TX) first precodes data symbols ${\boldsymbol {\mathsf {s}}}[k] \in \mathbb{C}^{N_{\mathrm{s}}\times 1}$  at each subcarrier by applying the subcarrier-dependent baseband precoder $\boldsymbol {\mathsf {F}} _{{{\text{BB}}}}[k] $. The symbol blocks are then transformed into the time domain using $L_{\mathrm{t}}$ parallel $K$-point inverse Fast Fourier transform (IFFT). After adding the cyclic prefix (CP), the transmitter employs the subcarrier-independent RF precoder ${\mathbf {F}}_{\mathrm {RF}}$ to form the transmitted signal. The complex baseband signal at the $\nth{k}$ subcarrier can be expressed as
\begin{equation} {\boldsymbol {\mathsf {x}}}[k] = {\mathbf {F}}_{\mathrm {RF}} \boldsymbol {\mathsf {F}} _{{{\text{BB}}}}[k] {\boldsymbol {\mathsf {s}}}[k], \end{equation}
where ${\boldsymbol {\mathsf {s}}}[k]$ denotes the transmitted symbol sequence at the $\nth{k}$ subcarrier of size $N_{\mathrm{s}}\times 1$.

\subsubsection{Channel Model}	
We consider a frequency-selective MIMO channel between the transmitter and the receiver, with a delay tap length of $N_{\mathrm{c}}$ in the time domain. The $\nth{d}$ delay tap of the channel is denoted by an $N_{\mathrm{r}}\times N_{\mathrm{t}}$ matrix $\mathbf{H}_{d}$, $d=0,1, \dots ,N_{\mathrm{c}}-1$. Assuming a geometric channel model~\cite{Heath2018CEMain}, $\mathbf{H}_{d}$ can be written as
\begin{equation} 
	\mathbf {H}_{d} = \sqrt {\tfrac {N_{{{\text{t}}}} N_{{{\text{r}}}}}{L\rho _{{{\text{L}}}}}}\sum _{\ell = 1}^{L}\alpha _{\ell } p_{\mathrm {rc}}(dT_{{{\text{s}}}}-\tau _{\ell }) {\mathbf {a}}_{\mathrm {R}}(\phi _{\ell }) {\mathbf {a}}_{\mathrm {T}}^{*}(\theta _{\ell }), 
\end{equation}
where $\rho _{{{\text{L}}}}$ represents the path loss between the transmitter and the receiver; $L$ corresponds to the number of paths; $T_s$ denotes the sampling period; $p_{\mathrm {rc}}(\tau)$ is a filter that includes the effects of pulse-shaping and other lowpass filtering evaluated at $\tau$; $\alpha _{\ell } \in {\mathbb {C}}$ is the complex gain of the $\nth{\ell}$ path; $\tau _{\ell } \in {\mathbb {R}}$ is the delay of the $\nth{\ell}$ path; $\phi _{\ell } \in [0, 2\pi]$ and $\theta _{\ell } \in [0, 2\pi]$ are the AoA and AoD of the $\nth{\ell}$ path, respectively; and ${\mathbf {a}}_{\mathrm {R}}(\phi _{\ell }) \in {\mathbb {C}}^{ N_{\mathrm {r}}\times 1}$ and ${\mathbf {a}}_{\mathrm {T}}(\theta _{\ell }) \in {\mathbb {C}}^{ N_{\mathrm {t}}\times 1}$ are the array steering vectors for the receive and transmit antennas, respectively. Both the transmitter and the receiver are assumed to use Uniform Linear Arrays (ULAs) with half-wavelength separation. Such an ULA has steering vectors obeying the expressions 
		 $$\left[ {\mathbf {a}}_{\mathrm{T}}(\theta _\ell)\right]_{n} = \sqrt {\tfrac {1}{N_{{{\text{t}}}}}} e^{ {\mathrm {j}} n \pi \cos {(\theta _\ell)}}, \quad n = 0,\ldots,N_{{{\text{t}}}}-1,$$ 
		 $$\left[ {\mathbf {a}}_{\mathrm{R}}(\phi _\ell)\right]_{m} = \sqrt {\tfrac {1}{N_{{{\text{r}}}}}} e^{ {\mathrm {j}} m \pi \cos {(\phi _\ell)}}, \quad m= 0,\ldots,N_{{{\text{r}}}}-1.$$
The channel can be expressed more compactly in the following form:
\begin{align}
	\mathbf{H}_d=\mathbf{A}_R\mathbf{\Delta}_d\mathbf{A}^*_T
\end{align}
where $\mathbf{\Delta}_d \in \mathbb{C}^{L\times L}$  is diagonal with non-zero complex diagonal entries, and $\mathbf{A}_R \in \mathbb{C}^{N_{\mathrm{r}}\times L}$ and $\mathbf{A}_T \in \mathbb{C}^{N_{\mathrm{t}}\times L}$ contain the receive and transmit array steering vectors $\mathbf{a}_R(\phi_l)$ and $\mathbf{a}_T(\theta_l)$, respectively. 	The channel at subcarrier $k$ can be written in terms of the different delay taps as
\begin{equation} 
\label{equ:FreqChan}
	\boldsymbol {\mathsf {H}}[k]=\sum _{d=0}^{ N_{\mathrm {c}}-1} \mathbf {H}_{d} e^{- {\mathrm {j}}\frac {2\pi k}{K}d} = {\mathbf {A}}_{\mathrm{R}} \boldsymbol \Delta [k] {\mathbf {A}}_{\mathrm{T}}^{*}. 
\end{equation}
where $\boldsymbol{\Delta }[k]\in \mathbb{C}^{L\times L}$  is diagonal with non-zero complex diagonal entries such that  $\boldsymbol{\Delta }[k] = \sum _{d=0}^{ N_{\mathrm {c}}-1} \boldsymbol{\Delta }_{d} e^{-j\frac {2\pi k}{N}d}$, $k = 0,\ldots,K-1$.

\subsubsection{Extended Virtual Channel Model} 
According to \cite{Heath2016OverviewMmwave}, we can further approximate the channel $\mathbf {H}_{d} $ using the extended virtual channel model   as	
\begin{equation} 
	\mathbf {H}_{d} \approx \tilde { {\mathbf {A}}}_{\mathrm{R}} \boldsymbol{\Delta }_{d}^{v} \tilde { {\mathbf {A}}}_{\mathrm{T}}^{*}, 
	\end{equation}
where $\boldsymbol{\Delta }_{d}^{v} \in \mathbb {C}^{ G_{\mathrm {r}}\times G_{\mathrm {t}}}$ corresponds to a sparse matrix that contains the path gains in the non-zero elements. Moreover, the dictionary matrices $\tilde { {\mathbf {A}}}_{\mathrm{T}}$ and $\tilde { {\mathbf {A}}}_{\mathrm{R}} $ contain the transmitter and receiver array response vectors evaluated on a grid of size $G_{\mathrm {r}}\gg L$ for the AoA and a grid of size $G_{\mathrm {t}}\gg L$ for the AoD, i.e., $\tilde \theta_{\ell} \in \{0,\frac{2\pi}{G_{\mathrm {r}}},\ldots,\frac{2\pi(G_{\mathrm {r}}-1)}{G_{\mathrm {r}}}\}$ and $\tilde\phi_{\ell} \in \{0,\frac{2\pi}{G_{\mathrm {t}}},\ldots,\frac{2\pi(G_{\mathrm {t}}-1)}{G_{\mathrm {t}}}\}$, respectively: 
\begin{align}
    \tilde{\mathbf {A}}_{\mathrm{T}}&=[{\bf a}_{\mathrm{T}}(\tilde\theta_1) \dots {\bf a}_{\mathrm{T}}(\tilde\theta_{G_{\mathrm{t}}})], \\
    \tilde{\mathbf {A}}_{\mathrm{R}}&=[{\bf a}_{\mathrm{R}}(\tilde\phi_1) \dots {\bf a}_{\mathrm{R}}(\tilde\phi_{G_{\mathrm{r}}})].
\end{align}
Since we have few scattering clusters in mmWave channels, the sparse assumption for $\boldsymbol{\Delta }_{d}^{v} \in \mathbb {C}^{ G_{\mathrm {r}}\times G_{\mathrm {t}}}$ is commonly accepted. To help expose the sparse structure, we can express the channel at subcarrier $k$ in terms of the sparse matrices $\boldsymbol{\Delta }_{d}^{v}$ and the dictionaries as follows
\begin{equation} \label{equ:channel_model}
	\boldsymbol {\mathsf {H}}[k] \approx \tilde { {\mathbf {A}}}_{\mathrm{R}} {\bigg (}\sum _{d=0}^{ N_{\mathrm {c}}-1} \boldsymbol{\Delta }_{d}^{v} e^{- {\mathrm {j}}\frac {2\pi k}{K}d} {\bigg) } \tilde { {\mathbf {A}}}_{\mathrm{T}}^{*} \approx \tilde { {\mathbf {A}}}_{\mathrm{R}} \boldsymbol{\Delta }^{\mathrm{v}}[k] \tilde { {\mathbf {A}}}_{\mathrm{T}}^{*} .
\end{equation}
where $\boldsymbol{\Delta }[k] = \sum _{d=0}^{ N_{\mathrm {c}}-1} \boldsymbol{\Delta }_{d}^{v} e^{-j\frac {2\pi k}{N}d}$, $k = 0,\ldots,K-1$, is a $G_{\mathrm{r}}\times G_{\mathrm{t}}$ complex sparse matrix containing the channel gains of the virtual channel.

\subsubsection{Signal Reception} 	

Considering that the receiver (RX) applies a hybrid combiner ${ \boldsymbol {\mathsf {W}}}[k]={ {\mathbf {W}}_{\mathrm {RF}} \boldsymbol {\mathsf {W}} _{{{\text{BB}}}}[k]} \in {\mathbb {C}}^{ N_{\mathrm {r}}\times N_{{{\text{s}}}}}$, the received signal at subcarrier $k$ can be expressed as
\begin{align} \label{equ:rec_w}
{\boldsymbol {\mathsf {y}}}[k] = \boldsymbol {\mathsf {W}}_{{{\text{BB}}}}^{*}[k] {\mathbf {W}}_{\mathrm {RF}}^{*} {\mathbf {H}}[k] {\mathbf {F}}_{\mathrm {RF}} \boldsymbol {\mathsf {F}} _{{{\text{BB}}}}[k] {\boldsymbol {\mathsf {s}}}[k]+ \boldsymbol {\mathsf {W}}_{{{\text{BB}}}}^{*}[k] {\mathbf {W}}_{\mathrm {RF}}^{*} {\boldsymbol {\mathsf {n}}}[k], \!\!\!\!\!\!\notag \\ {}
\end{align}
where ${\boldsymbol {\mathsf {n}}}[k] \sim \mathcal {CN}\left ({0,\sigma ^{2} \mathbf {I}}\right)$ corresponds to the circularly symmetric complex Gaussian distributed additive noise vector. The received signal model in ~(\ref{equ:rec_w}) corresponds to the data transmission phase. As explained in Section \ref{sec:proposed_schem}, during the channel acquisition phase, frequency-flat training precoders and combiners will be considered to reduce complexity.
		
		

\subsection{Problem Formulation}
During the training phase, transmitter and receiver use a training precoder ${\mathbf {F}}_{{{\text{tr}}}}^{(m)}\in \mathbb {C}^{ N_{\mathrm {t}}\times L_{\mathrm {t}}}$ and a training combiner ${\mathbf {W}}_{{{\text{tr}}}}^{(m)}\in \mathbb {C}^{ N_{\mathrm {r}}\times L_{\mathrm {r}}}$ for the $\nth{m}$ pilot training frame, respectively. The precoders and combiners considered in this phase are frequency-flat  to keep the complexity of the sparse recovery algorithms low. The transmitted symbols are assumed to satisfy $\mathbb {E}\{ {\boldsymbol {\mathsf {s}}}^{(m)}[k] {\boldsymbol {\mathsf {s}}}^{(m)*}[k]\} = \frac {P}{N_{{{\text{s}}}}} {\mathbf {I}}_{N_{{{\text{s}}}}}$, where $P$ is the total transmitted power and $N_{\mathrm {s}}= L_{\mathrm {t}}$. The transmitted symbol ${\boldsymbol {\mathsf {s}}}^{(m)}[k]$ is decomposed as ${\boldsymbol {\mathsf {s}}}^{(m)}[k] = {\mathbf {q}}^{(m)} {\mathsf {t}}^{(m)}[k]$, with ${\mathbf {q}}^{(m)} \in \mathbb {C}^{ L_{\mathrm {t}}\times 1}$ is a frequency-flat vector and ${\mathsf {t}}^{(m)}[k]$ is a pilot symbol known at the receiver. This decomposition is used to reduce computational complexity since it allows simultaneous use of the $L_{\mathrm {t}}$ spatial degrees of freedom coming from $L_{\mathrm {t}}$ RF chains and enables channel estimation using a single subcarrier-independent measurement matrix. 
Moreover, each entry in ${\mathbf {F}}_{{{\text{tr}}}}^{(m)}$ and in ${\mathbf {W}}_{{{\text{tr}}}}^{(m)}$ are normalized such that their squared-modulus would be $\tfrac{1}{N_{\mathrm {t}}}$ and $\tfrac{1}{N_{\mathrm {r}}}$ , respectively. Then, the received samples in the frequency domain for the $\nth{m}$ training frame can be expressed as
\begin{equation} 
	{\boldsymbol {\mathsf {y}}}^{(m)}[k] = { {\mathbf {W}}_{{{\text{tr}}}}^{(m)}}^{*} \boldsymbol {\mathsf {H}}[k] {\mathbf {F}}_{{{\text{tr}}}}^{(m)} {\mathbf {q}} ^{(m)} {\mathsf {t}}^{(m)}[k] + {\boldsymbol {\mathsf {n}}}_{{{\text{c}}}}^{(m)}[k], 
\end{equation}
where $\boldsymbol {\mathsf {H}}[k] \in \mathbb {C}^{N_{r}\times N_{t}}$ denotes the frequency-domain MIMO channel response at the $\nth{k}$ subcarrier and ${\boldsymbol {\mathsf {n}}}_{{{\text{c}}}}^{(m)}[k] \in \mathbb {C}^{L_{r}\times 1}$, ${\boldsymbol {\mathsf {n}}}_{{{\text{c}}}}^{(m)}[k] = { {\mathbf {W}}_{{{\text{tr}}}}^{(m)}}^{*} {\boldsymbol {\mathsf {n}}}^{(m)}[k]$, represents the frequency-domain combined noise vector received at the $\nth{k}$ subcarrier. The average received SNR is given by ${{\text{SNR}}} = \frac {P}{\rho _{{{\text{L}}}} \sigma ^{2}}$. Furthermore, the channel coherence time is assumed to be larger than the frame duration and that the same channel can be considered for several consecutive frames.

\subsubsection{Measurement Matrix}\label{sec:MeasMat}
In order to apply sparse reconstruction with a single subcarrier-independent measurement matrix, we first remove the effect of the scalar $ {\mathsf {t}}^{(m)}[k]$ by multiplying the received signal by $ {\mathsf {t}}^{(m)}[k]^{-1}$. Using the following property $\mathop {\mathrm {vec}}\{ {\mathbf {A}} {\mathbf {X}} {\mathbf {C}}\} = ({\mathbf {C}}^{T} \otimes {\mathbf {A}}) \mathop {\mathrm {vec}}\{ {\mathbf {X}}\}$ , the vectorized received signal is given by 
\begin{align} 
	\mathop {\mathrm {vec}}\{ {\boldsymbol {\mathsf {y}}}^{(m)}[k]\} = ({\mathbf {q}}^{(m)T}{ {\mathbf {F}}}_{{{\text{tr}}}}^{(m)T} \otimes { {\mathbf {W}}}_{{{\text{tr}}}}^{(m)^{*}}) \mathop {\mathrm {vec}}\{ \boldsymbol {\mathsf {H}}[k]\} + {\boldsymbol {\mathsf {n}}}_{{{\text{c}}}}^{(m)}[k]. \notag \\ {}
\end{align}

The vectorized channel matrix can be expressed as 
\begin{equation}\label{Ch_vec}
  \mathop {\mathrm {vec}}\{ \boldsymbol {\mathsf {H}}[k]\} = (\bar {\tilde { {\mathbf {A}}}}_{\mathrm{T}} \otimes \tilde { {\mathbf {A}}}_{\mathrm{R}}) \mathop {\mathrm {vec}}\{\boldsymbol{\Delta }^{\mathrm{v}}[k]\}.  
\end{equation}
Furthermore, we define the measurement matrix $\boldsymbol{\Phi }^{(m)} \in \mathbb {C}^{L_{{{\text{r}}}} \times N_{{{\text{t}}}} N_{{{\text{r}}}}}$:
\begin{equation} 
	\boldsymbol{\Phi }^{(m)} = ({\mathbf {q}}^{(m)T}{ {\mathbf {F}}_{{{\text{tr}}}}^{(m)T}}\otimes { {\mathbf {W}}_{{{\text{tr}}}}^{(m)*}}), 
\end{equation}
and the dictionary $\boldsymbol{\Psi } \in \mathbb {C}^{N_{{{\text{t}}}} N_{{{\text{r}}}} \times G_{\mathrm{t}} G_{\mathrm{r}}}$ as
\begin{equation}
	\boldsymbol{\Psi } = (\bar {\tilde { {\mathbf {A}}}}_{\mathrm{T}} \otimes \tilde { {\mathbf {A}}}_{\mathrm{R}}), 
\end{equation}

Then, the vectorized received pilot signal $L_{\mathrm{r}} \times 1$ at the $\nth{m}$ training symbol can be written as
\begin{equation} \label{equ:rec_vec}
	\mathop {\mathrm {vec}}\{ {\boldsymbol {\mathsf {y}}}^{(m)}[k]\} = \boldsymbol{\Phi }^{(m)} \boldsymbol{\Psi } {\boldsymbol {\mathsf {h}}}^{\mathrm{v}}[k] + {\boldsymbol {\mathsf {n}}}_{{{\text{c}}}}^{(m)}[k], 
\end{equation}
where ${\boldsymbol {\mathsf {h}}}^{\mathrm{v}}[k] = \mathop {\mathrm {vec}}\{\boldsymbol{\Delta }^{\mathrm{v}}[k]\} \in \mathbb {C}^{G_{\mathrm{r}}G_{\mathrm{t}}\times 1}$ is the sparse vector containing the complex channel gains. Moreover, we use several training frames to get  enough measurements and accurately reconstruct the sparse vector ${\boldsymbol {\mathsf {h}}}^{\mathrm{v}}[k]$, especially in the very-low SNR regime. Therefore, when the transmitter and receiver communicate during $M$ training steps using different pseudorandomly built precoders and combiners, ~(\ref{equ:rec_vec}) can be extended to $M$ received signals given by 
\begin{equation} 
	\underbrace {\left [{\begin{array}{c} {\boldsymbol {\mathsf {y}}}^{(1) }[k] \\ \vdots \\ {\boldsymbol {\mathsf {y}}}^{(M)}[k] \end{array}}\right]}_{ {\boldsymbol {\mathsf {y}}}[k]} = \underbrace {\left [{\begin{array}{c} \boldsymbol{\Phi }^{(1) } \\ \vdots \\ \boldsymbol{\Phi }^{(M)} \end{array}}\right]^{T}}_{\boldsymbol \Phi } \boldsymbol{\Psi } {\boldsymbol {\mathsf {h}}}^{\mathrm{v}}[k] + \underbrace {\left [{\begin{array}{c} {\boldsymbol {\mathsf {n}}}_{{{\text{c}}}}^{(1) }[k] \\ \vdots \\ {\boldsymbol {\mathsf {n}}}_{{{\text{c}}}}^{(M)}[k] \end{array}}\right]}_{ {\boldsymbol {\mathsf {n}}}_{{{\text{c}}}}[k]}. 
\end{equation}

\begin{figure*}[t]
\centering
	\includegraphics[width=\textwidth]{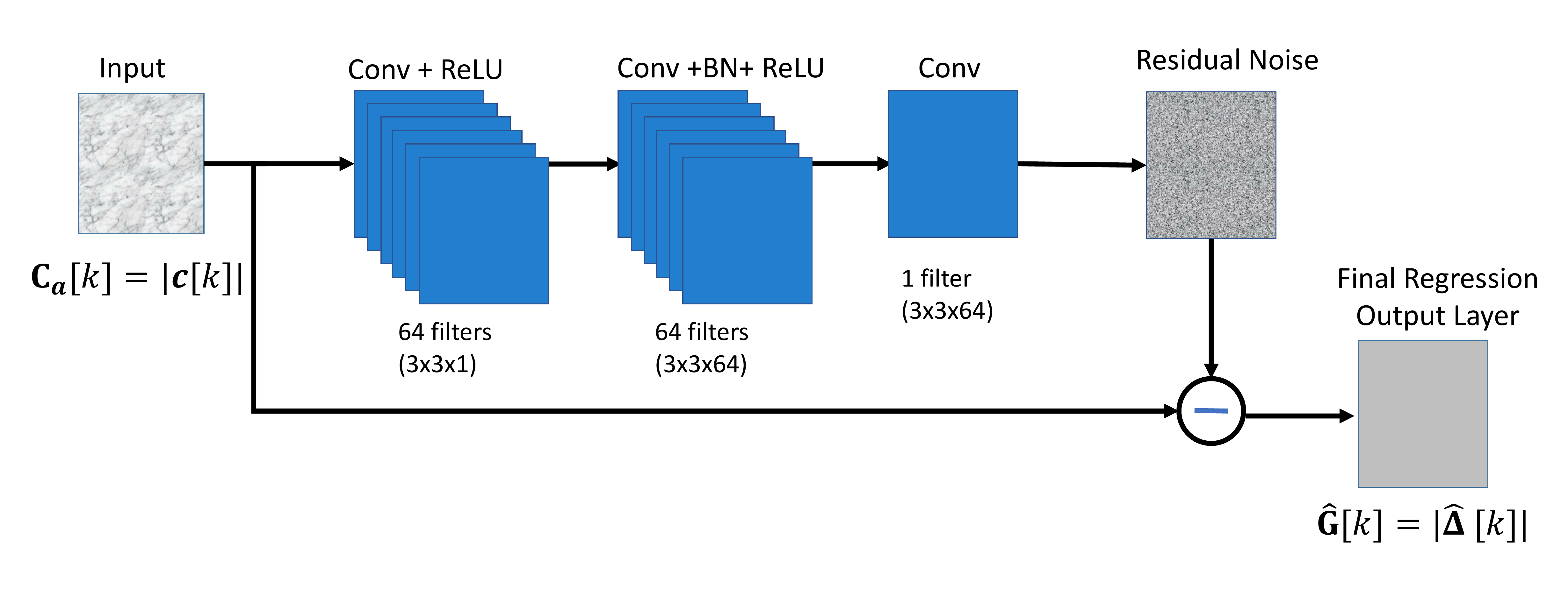}
	\centering
	\caption{Proposed denoising convolutional neural network (DnCNN) for multicarrier channel amplitude estimation.}
	\label{fig:Proposed_NN}\vspace{-0.15in}
\end{figure*}

Hence, the vector ${\boldsymbol {\mathsf {h}}}^{\mathrm{v}}[k]$ can be estimated by solving the sparse reconstruction problem as done in \cite{Heath2018CEMain},
	\begin{equation} \label{equ:prob}
		\min \| {\boldsymbol {\mathsf {h}}}^{\mathrm{v}}[k]\|_{1} \quad {{\text{subject to}}} ~\| {\boldsymbol {\mathsf {y}}}[k] - \boldsymbol{\Phi }\boldsymbol{\Psi } {\boldsymbol {\mathsf {h}}}^{\mathrm{v}}[k]\|_{2}^{2} < \epsilon, 
	\end{equation}
where $\epsilon$ represents a tunable parameter defining the maximum error between the reconstructed channel and the received signal. In realistic scenarios, the sparsity (number of channel paths) is usually unknown, therefore the choice of $\epsilon$ is critical to solve ~(\ref{equ:prob}) and estimate the sparsity level. The choice of this parameter is explained in Section \ref{sec:estimation_paths}.
			
Interestingly, the matrices in ~(\ref{equ:channel_model}) exhibit the same sparse structure for all $k$, since the AoA and AoD do not change with frequency in the transmission bandwidth. This is an interesting property that can be leveraged when solving the compressed channel estimation problem defined in ~(\ref{equ:prob}). 
Moreover, we denote the supports of the virtual channel matrices $\boldsymbol{\Delta }_{d}^{v}$ as ${\mathcal{ T}}_{0}, {\mathcal{ T}}_{1}, \ldots, {\mathcal{ T}}_{N_{{{\text{c}}}}-1}$, $d = 0, \ldots, N_{{{\text{c}}}}-1$. Then, knowing ${\boldsymbol {\mathsf {h}}}^{\mathrm{v}}[k] = \mathop {\mathrm {vec}}\{\boldsymbol{\Delta }^{\mathrm{v}}[k]\}$, with $\boldsymbol{\Delta }^{\mathrm{v}}[k] = \sum _{d=0}^{ N_{\mathrm {c}}-1} \boldsymbol{\Delta }_{d}^{v} e^{-j\frac {2\pi k}{N}d}$, $k = 0,\ldots,K-1$, the supports of ${\boldsymbol {\mathsf {h}}}^{\mathrm{v}}[k]$ are defined as
\begin{equation} \label{equ:supp2}
		\mathop {\mathrm {supp}}\{ {\boldsymbol {\mathsf {h}}}^{\mathrm{v}}[k]\} = \bigcup _{d=0}^{N_{{{\text{c}}}}-1}{ \mathop {\mathrm {supp}}\{  \mathop {\mathrm {vec}} \{\boldsymbol{\Delta }_{d}^{v}\}\}} \quad k = 0,\ldots,K-1, 
\end{equation}
where the union of the supports of the time-domain virtual channel matrices is due to the additive nature of the Fourier transform. Therefore, as shown in ~(\ref{equ:supp2}), where the union is independent of the subcarrier $k$, $\boldsymbol{\Delta }[k]$ has the same supports for all $k$. 
		

\subsubsection{Correlation Matrix}
To estimate multi-path components of the channel, i.e., AoAs/AoDs and channel gains, we first need to compute the atom, which is defined as the vector that produces the largest sum-correlation with the received signals in the measurement matrix. The sum-correlation is especially considered as the support of the different sparse vectors is the same over the $K$ subcarriers.   The correlation vector ${\boldsymbol {\mathsf {c}}}[k] \in \mathbb {C}^{G_{\mathrm{r}} G_{\mathrm{t}}}$ is given by
\begin{equation}\label{equ:Corr}
	{\boldsymbol {\mathsf {c}}}[k] = \boldsymbol \Upsilon ^{*} {\boldsymbol {\mathsf {y}}}[k],
\end{equation}
where $\boldsymbol \Upsilon \in \mathbb {C}^{M L_{{{\text{r}}}} \times G_{\mathrm{t}} G_{\mathrm{r}}}$, $\boldsymbol \Upsilon = \boldsymbol \Phi \boldsymbol \Psi$ represents the equivalent measurement matrix which is the same $\forall k$ and ${\boldsymbol {\mathsf {y}}}[k] \in \mathbb {C}^{M L_{{{\text{r}}}} \times 1}$ is the received signal for a given $k$ , $k = 0,\ldots,K-1$ . 
		
One can note that  if there exists a {correlation between noise components}, the atom estimated from the projection in ~(\ref{equ:Corr}) might not be the correct one. In order to compensate for this error in estimation, we consider the noise covariance matrix when performing the correlation step. In particular, we consider two arbitrary (hybrid) combiners ${ {\mathbf {W}}_{{{\text{tr}}}}^{(m)}}^{(i)}$, ${ {\mathbf {W}}_{{{\text{tr}}}}^{(m)}}^{(j)} \in \mathbb {C}^{N_{{{\text{r}}}} \times L_{{{\text{r}}}}}$ for two arbitrary training steps $i ,j$ and a given subcarrier $k$. Hence, the combined noise at a given training step $i$ and subcarrier $k$ is represented as ${\boldsymbol {\mathsf {n}}}_{{{\text{c}}}}^{(i)}[k] = { {\mathbf {W}}_{{{\text{tr}}}}^{(i)*}} {\boldsymbol {\mathsf {n}}}^{(i)}[k]$ , with ${\boldsymbol {\mathsf {n}}}^{(i)}[k] \sim {\mathcal{ N}}(\boldsymbol 0,\sigma ^{2} {\mathbf {I}}_{L_{{{\text{r}}}}})$, which results in  noise cross-covariance matrix given by $\mathbb {E}\{ {\boldsymbol {\mathsf {n}}}_{{{\text{c}}}}^{(i)}[k] {\boldsymbol {\mathsf {n}}}_{{{\text{c}}}}^{(j)*}[k]\} = { {\mathbf {W}}_{{{\text{tr}}}}^{(i)*}}\sigma ^{2}\delta [i-j] { {\mathbf {W}}_{{{\text{tr}}}}^{(j)}}$. We can further write the noise covariance matrix of ${\boldsymbol {\mathsf {y}}}[k]$ as a block diagonal matrix ${\mathbf {C}_{\text{w}}}\in \mathbb {C}^{M L_{{{\text{r}}}} \times M L_{{{\text{r}}}}}$, 
	\begin{align}
		&{\mathbf {C}}_{{{\text{w}}}}= \mathop {\mathrm {blkdiag}}\{{ {\mathbf {W}}_{{{\text{tr}}}}}^{(1) *} { {\mathbf {W}}_{{{\text{tr}}}}}^{(1) },\ldots, { {\mathbf {W}}_{{{\text{tr}}}}}^{(M)*} { {\mathbf {W}}_{{{\text{tr}}}}}^{(M)}\}.
	\end{align}
Moreover, {Cholesky factorization} can be used to factorize ${\mathbf {C}}_{{{\text{w}}}}$  into ${\mathbf {C}}_{{{\text{w}}}}= {\mathbf {D}}_{{{\text{w}}}}^{*} {\mathbf {D}}_{{{\text{w}}}}$, where ${\mathbf {D}}_{{{\text{w}}}} \in \mathbb {C}^{M L_{{{\text{r}}}} \times M L_{{{\text{r}}}}}$is an upper triangular matrix. Then, by taking into consideration the noise covariance matrix, the correlation step is given by
\begin{equation} \label{equ:corrw}
	{\boldsymbol {\mathsf {c}}}[k] = \boldsymbol \Upsilon _{{{\text{w}}}}^{*} {\boldsymbol {\mathsf {y}}}_{{{\text{w}}}}[k], 
\end{equation}
where $\boldsymbol \Upsilon _{{{\text{w}}}} \in \mathbb {C}^{M L_{\mathrm {r}}\times G_{\mathrm {t}} G_{\mathrm {r}} }$ represents the whitened measurement matrix given by $\boldsymbol \Upsilon _{{{\text{w}}}} = {\mathbf {D}}_{{{\text{w}}}}^{-*}\boldsymbol \Upsilon$. And, the $M L_{\mathrm {r}}\times 1$ whitened received signal ${\boldsymbol {\mathsf {y}}}_{{{\text{w}}}}[k]$ is given by ${\boldsymbol {\mathsf {y}}}_{{{\text{w}}}}[k] = {\mathbf {D}}_{{{\text{w}}}}^{-*} {\boldsymbol {\mathsf {y}}}[k]$. The matrix ${\mathbf {D}}_{{{\text{w}}}}^{-1} \in \mathbb {C}^{M L_{\mathrm {r}}\times M L_{\mathrm {r}}}$ is given by ${\mathbf {D}}_{{{\text{w}}}}^{-1} = \mathop {\mathrm {blkdiag}}\left \{{\left ({{\mathbf {D}}_{{{\text{w}}}}^{(1) }}\right)^{-1},\ldots,\left ({{\mathbf {D}}_{{{\text{w}}}}^{(M)}}\right)^{-1}}\right \}$, where $\left ({{\mathbf {D}}_{{{\text{w}}}}^{(m)}}\right)^{-1}$ can be considered as a frequency-flat baseband combiner ${\mathbf {W}}_{{{\text{BB}}},tr}^{(m)}$ used in the $m$-th training step. Therefore, by applying the whitened measurement matrix, the resulting correlation would simultaneously whiten the spatial noise components and estimate a more accurate support index in the sparse vectors ${\mathbf {h}}^{\mathrm{v}}[k]$.

\begin{figure*}[t]
	\includegraphics[width=0.8\textwidth]{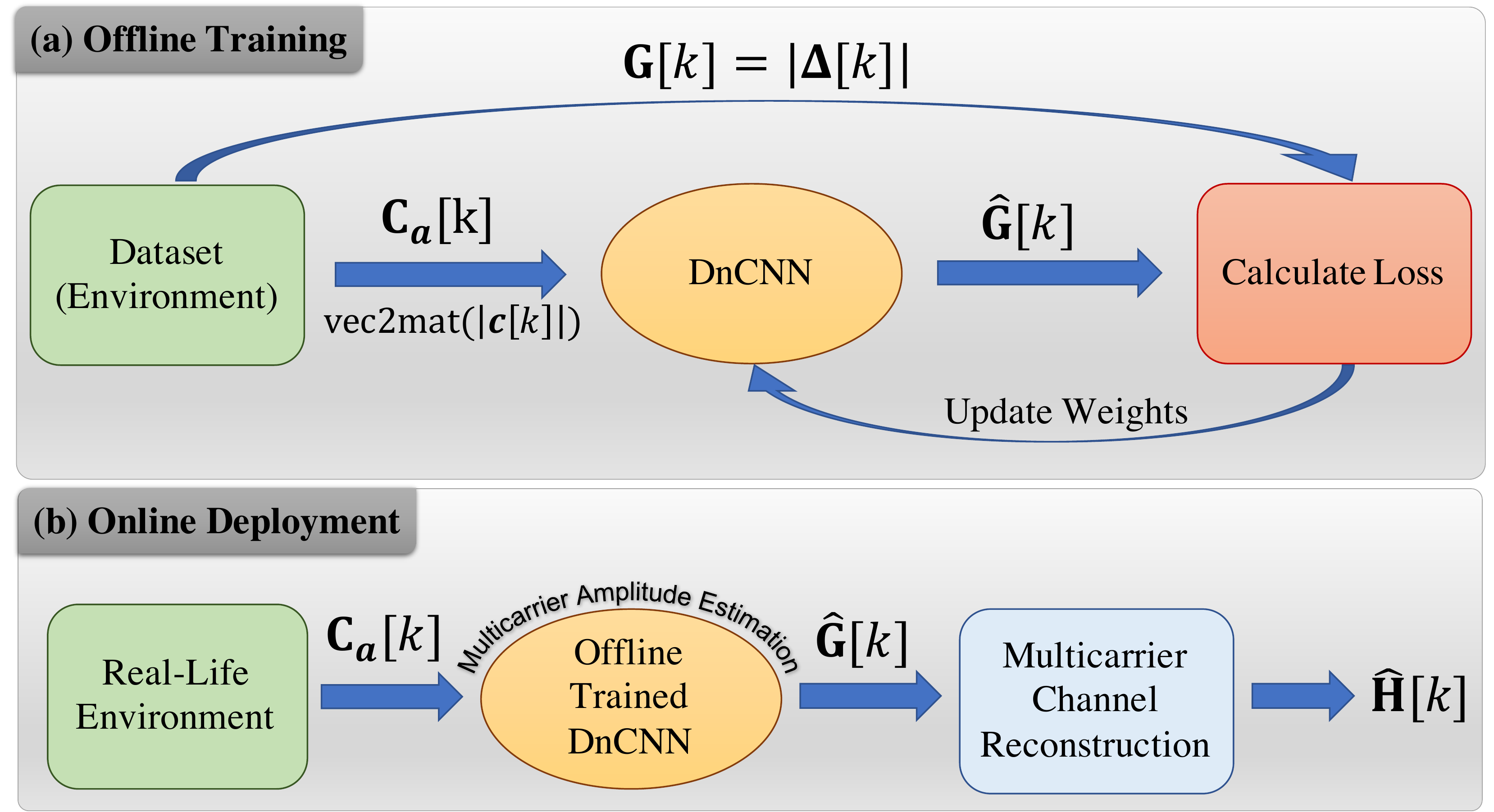}
	\centering
	\caption{Block diagram of the DL-CS-CE Scheme: offline training and online deployment.}\label{fig:Proposed_scheme}
\end{figure*}

\section{Deep Learning and Compressive-Sensing Based Channel Estimation (DL-CS-CE)}
\label{sec:proposed_schem}

To solve the CS channel estimation problem formulated above, this section proposes two DL-based algorithms. Both leverage the common support between the channel matrices for every subcarrier and provide different complexity-performance trade-offs. The former simultaneously estimate the support using an offline-trained DnCNN and then reconstruct the channel. On the other hand, the latter applies further fine-tuning to accurately estimate the AoAs and AoDs with higher resolution dictionary matrices while keeping computational complexity low.

\subsection{Offline Training and Online Deployment of DnCNN}
Before delving into the proposed solutions' details, let us first provide insights into the considered DnCNN architecture as well as its offline training and online deployment.

\subsubsection{DnCNN Architecture}
Fig. \ref{fig:Proposed_NN} illustrates the network architecture of the DnCNN denoiser that consists of $L_{\mathrm{C}}$ convolutional (Conv) layers. Each layer uses $c_{\mathrm{CL}}^{(l)}$ different $D_{x}^{(l)} \times D_{y}^{(l)} \times D_{z}^{(l)}$ filters.  The first convolutional layer is followed by a rectified linear unit (ReLU). The succeeding $L_{\mathrm{C}}-2$ convolutional layers are followed by batch-normalization (BN) and a ReLU. The final $\nth{L_{\mathrm{C}}}$ convolutional layer uses one separate $D_{x}^{(L_{\mathrm{C}})} \times D_{y}^{(L_{\mathrm{C}})} \times D_{z}^{(L_{\mathrm{C}})} $ filter to reconstruct the signal. Here, $D_x^{(l)}$, $ D_y^{(l)}$ and $D_z^{(l)}$ are the convolutional kernel dimensions,  and $c_{\mathrm{CL}}^{(l)}$ is the number of filters in the $\nth{l}$ layer.  


We present three pseudo-color images of the noisy channel, residual noise, and estimated output channel in Fig. \ref{fig:Proposed_NN}. The DnCNN considers the amplitude of the correlation $G_{\mathrm{r}} \times G_{\mathrm{t}}$ matrix, i.e., 
\begin{equation}\label{eq:Calpha}
 {\boldsymbol {\mathsf {C}}}_{\bm{\alpha}}[k]=\mathrm{vec2mat}(|{\boldsymbol {\mathsf {c}}}[k]|, [G_{\mathrm{r}}, G_{\mathrm{t}}]), \forall k,
\end{equation}
as input and produces residual noise as an output, rather than estimated channel amplitudes, where we define a $G_{\mathrm{r}} \times G_{\mathrm{t}}$ matrix of channel amplitudes as 
\begin{equation}\label{equ:amps}
{\bf  G}[k]=|\boldsymbol{\Delta }^{\mathrm{v}}[k]| \in \mathbb {R}^{G_{\mathrm{r}}\times G_{\mathrm{t}} },  \forall k.
\end{equation}
The DnCNN aims to learn a mapping function $\mathcal {F}( {\boldsymbol {\mathsf {C}}}_{\bm{\alpha}}[k]) = {\bf  G}[k]$ to predict the latent clean image from noisy observation ${\boldsymbol {\mathsf {C}}}_{\bm{\alpha}}[k]$. We adopt the residual learning formulation to train a residual mapping $\mathcal {R}({\boldsymbol {\mathsf {C}}}_{\bm{\alpha}}[k]) \approx \mathbf {V}$ where $\bf{V}$ is the residual noise, and then we have ${\bf  G}[k]={\boldsymbol {\mathsf {C}}}_{\bm{\alpha}}[k]- \mathcal {R}({\boldsymbol {\mathsf {C}}}_{\bm{\alpha}}[k])$. Instead of learning a mapping directly from a noisy image to a denoised image, learning the residual noise is beneficial \cite{Zhang2017Dncnn,he2016deep}. Furthermore, the averaged mean squared error between the desired residual images and estimated ones from noisy input is adopted as the loss function to learn the trainable parameters $\boldsymbol{\Theta }$ of the DnCNN. This loss function is given by
\begin{equation}
\ell (\boldsymbol{\Theta }) = \frac {1}{2N}\sum _{i=1}^{N}\|\mathcal {R}( {{\boldsymbol {\mathsf {C}}}_{\bm{\alpha}}[k]}^{i}; \boldsymbol{\Theta }) - ( {\boldsymbol {\mathsf {C}}}_{\bm{\alpha}}[k]^{i} - {\bf  G}[k]^{i}) \|_{F}^{2}
\end{equation}
where ${( {\boldsymbol {\mathsf {C}}}_{\bm{\alpha}}[k]^{i} , {\bf  G}[k]^{i})}_{i=1}^{N}$ represents $N$ noisy-clean training patch pairs.
This method is also known as residual learning \cite{he2016deep} and renders the DnCNN to remove the highly structured natural image rather than the unstructured noise. Consequently, residual learning improves both the training times and accuracy of a network. In this way, combining batch normalization and residual learning techniques can accelerate the training speed and improve the denoising performance. Besides, batch normalization has been shown to offer some merits for residual learning, such as alleviating internal covariate shift problem in \cite{Zhang2017Dncnn,Jin2019CellFreeDLCE}. 

\subsubsection{Offline Training of the DnCNN} 

During offline training of the DnCNN, the dataset of $ {\boldsymbol {\mathsf {C}}}_{\bm{\alpha}}[k], \forall k$ and ${\bf  G}[k], \forall k$ is generated based on the realistic Raymobtime dataset for mmWave frequency selective channel environment\footnote{Raymobtime is developed based on collecting realistic datasets collected by ray-tracing and realistic 3D scenarios that considers mobility, time, frequency, and space. Available at https://www.lasse.ufpa.br/raymobtime/}. With the mmWave channel amplitude in ~(\ref{equ:amps}) and the correlation of the received signals and the
measurement matrix in ~(\ref{eq:Calpha}), the training data of ${\boldsymbol {\mathsf {C}}}_{\bm{\alpha}}[k]$ and ${\bf  {G}}[k]$ can
be obtained. In particular, the process to obtain ${\boldsymbol {\mathsf {C}}}_{\bm{\alpha}}[k]$ and ${\bf  {G}}[k]$ involves the following four steps: i) generation of channel matrices based on the mmWave channel model from the Raymobtime dataset ii) 
obtaining ${\bf  {G}}[k]$ based on ~(\ref{equ:amps}); iii) computing the whitened received signal
vector ${\boldsymbol {\mathsf {y}}}_{{{\text{w}}}}[k] \:\: \forall k$; and iv) acquiring the amplitudes of the correlation vector ${\boldsymbol {\mathsf {c}}}[k]$ and transforming it into a matrix form 
${\boldsymbol {\mathsf {C}}}_{\bm{\alpha}}[k]$ as per ~(\ref{eq:Calpha}). 

\subsubsection{Online Deployment of the DnCNN}\label{subsec:Onlinesubcarr}
During the online deployment of the DL-CS-CE, we obtain the measured received signal ${\boldsymbol {\mathsf {y}}}_{{{\text{w}}}}[k]$  from the realistic mmWave channel environments. We compute ${\boldsymbol {\mathsf {C}}}_{\bm{\alpha}}[k]$ based on (\ref{eq:Calpha}), which is then fed to the offline-trained DnCNN. 
Then, the trained DnCNN would predict  ${\bf  \hat{G}}[k]$, from which we can estimate the supports of $\boldsymbol{\Delta }^{\mathrm{v}}[k]$. An interesting and noteworthy issue is that we can feed the trained DnCNN a subset ${K}_p$ of $K$ subcarriers of the amplitudes of the correlation matrices ${\boldsymbol {\mathsf {C}}}_{\bm{\alpha}}[k]$, to eventually estimate the support of $\boldsymbol{\Delta }^{\mathrm{v}}[k]$,  since as shown in Section \ref{sec:MeasMat} $\boldsymbol{\Delta }^{\mathrm{v}}[k]$ have the same support for all $k$. In particular, the support can be estimated if a small number of subcarriers $K_p \ll K$ is used instead. This will eliminate the need for computing  ${\boldsymbol {\mathsf {C}}}_{\bm{\alpha}}[k]$ for all subcarriers and eventually reduce the overall computational complexity at the cost of a negligible performance degradation. By leveraging from triangle inequality, $|| {\boldsymbol {\mathsf {y}}}[k]||_{2}^{2} \leq ||\boldsymbol \Phi {\boldsymbol {\mathsf {h}}}^{\mathrm{v}}[k]||_{2}^{2} + || {\boldsymbol {\mathsf {n}}}_{{{\text{c}}}}[k]||_{2}^{2}$, such that the $K_p$ selected signals are expected to exhibit the strongest channel response. Therefore, the $K_p$ subcarriers having largest $\ell_2$-norm will be exploited to derive an estimate of the support of the already defined sparse channel matrix $\boldsymbol{\Delta }^{\mathrm{v}}[k]$,  $k = 0,.\dots,K -1$.
	
\subsection{Algorithm 1: DL-CS-CE}

The state-of-the-art sparse channel estimation schemes \cite[and references therein]{Heath2018CEMain} depend on greedy algorithms to detect the supports sequentially, which naturally yield suboptimal solutions. This motivated us to exploit the neural networks to estimate all supports simultaneously rather than sequentially. The algorithmic implementation of the proposed DL-CS-CE solution is presented in Algorithm \ref{Alg:Ch_Est}. After initialization steps between lines 1-3 and the computation of the whitened equivalent observation matrix in line 4, DL-CS-CE is structured based on three main procedures: 
\begin{itemize}
\item Estimation of the channel amplitudes by using an offline-trained DnCNN,

\item Sorting the estimated channel amplitudes in descending order to select the supports of dominant entries, 
 
\item Reconstruction of the channel according to the selected indices, 
\end{itemize}
which are explained in the sequel.

\begin{algorithm}[t!]
 \caption{DL-CS-CE}
  \label{Alg:Ch_Est}
\begin{algorithmic}[1]
 \renewcommand{\algorithmicrequire}{\textbf{Input:}}
 \renewcommand{\algorithmicensure}{\textbf{Output:}}
\Require ${\boldsymbol {\mathsf {y}}}[k]$, $\boldsymbol {\Phi}$, $\boldsymbol {\Psi}$, $ {\tilde { \mathbf {A}}}_{\mathrm{T}}$, $ {\tilde { \mathbf {A}}}_{\mathrm{R}}$, $K_p$, $\epsilon$

\State ${\boldsymbol {\mathsf {y}}}_{{{\text{w}}}}[k] \gets {\mathbf {D}}_{{{\text{w}}}}^{-*} {\boldsymbol {\mathsf {y}}}[k]\:\:\:\forall k$

 \State ${\boldsymbol {\mathsf {r}}}[k] \gets {\boldsymbol {\mathsf {y}}}_{{{\text{w}}}}[k]\:\:\:\forall k$

\State $\hat{\mathcal{T}}, {\mathcal{K}} \gets \{\emptyset \}$

\State	$\boldsymbol{\Upsilon }_{{{\text{w}}}} \gets {\mathbf {D}}_{{{\text{w}}}}^{-*}\boldsymbol{\Phi } \boldsymbol{\Psi}$

\State $\mathcal{K} \gets \textsc{Find Strongest Subcarriers (${\boldsymbol {\mathsf {y}}}[k]$)}$

\State $\hat{\boldsymbol {g}}[k] \gets \textsc{Estimate Amplitudes ($\boldsymbol{\Upsilon }_{{{\text{w}}}}^{*}, {\boldsymbol {\mathsf {r}}}[k], \mathcal{K}$)}$

\State $\hat{\boldsymbol {\mathsf {H}}}[k] \gets \textsc{Reconstruct Channel ($\hat{\boldsymbol {g}}[k]  $)}$

\hspace{-40pt} \Return $\hat{\boldsymbol {\mathsf {H}}}[k]$
\vspace{2pt}
\hrule 
\vspace{2pt}
\Procedure{Find Strongest Subcarriers}{${\boldsymbol {\mathsf {y}}}[k]$}

\For  {$i=1: K_p$} 
	$${\mathcal{ K}} = {\mathcal{ K}} \cup \mathop{\arg \,\max }\limits_{k\not \in {\mathcal{ K}}}\,\| {\boldsymbol {\mathsf {y}}}[k]\|_{2}^{2}$$
\EndFor

\hspace{-9pt} \Return $\mathcal{K}$ 
\EndProcedure

\vspace{2pt}
\hrule 
\vspace{2pt}
\Procedure{Estimate Amplitudes}{$\boldsymbol{\Upsilon }_{{{\text{w}}}}^{*}, {\boldsymbol {\mathsf {r}}}[k], \mathcal{K}$}

\State ${\boldsymbol {\mathsf {c}}}[k] \gets \boldsymbol{\Upsilon }_{{{\text{w}}}}^{*} {\boldsymbol {\mathsf {r}}}[k]$, $k \in \mathcal{K} \:\:\: // \text{ as per \eqref{equ:corrw}}$
\State ${\boldsymbol {\mathsf {C}}}_{\bm{\alpha}}[k] \gets  \mathrm{vec2mat}(|{\boldsymbol {\mathsf {c}}}[k]|, [G_{\mathrm{r}}, G_{\mathrm{t}}])\:\:\: // \text{ as per \eqref{eq:Calpha}}$ 

\State $\hat{\mbf {G}}[k] \xleftarrow[\text{DnCNN}]{\text{Online}} {\boldsymbol {\mathsf {C}}}_{\bm{\alpha}}[k] \:\:\:// \: $ [c.f. Fig. \ref{fig:Proposed_scheme}.b] 

\State $\hat{\boldsymbol {g}}[k] \gets \text{vec}({\hat{\mbf {G}}}[k]) \:\:\: // \text{ as per \eqref{eq:g_vector}}$

\hspace{-9pt} \Return $\hat{\boldsymbol {g}}[k], \: \forall k \in \mathcal{K}$ 
\EndProcedure

\vspace{2pt}
\hrule 
\vspace{2pt}

\Procedure{Reconstruct Channel}{$\hat{\boldsymbol {g}}[k],\: \forall k$}
\State MSE $\gets \infty$
\State $i \gets 1$
\State $\mathcal{I} \gets \textsc{IndexSortDescend}\left(\sum_{k\in {\mathcal{K}}} |\hat{\boldsymbol {g}}[k]|\right)$

\While {MSE $> \epsilon \:\: \& \:\: i\leq G_{\mathrm{t}}G_{\mathrm{r}}$}

	\State 
	$\hat {\mathcal{T}} \gets \hat {\mathcal{T}} \cup \mathcal{I}(i)$

	\State 
	$\hat {\tilde {\boldsymbol \xi }}[k] \gets \left ({\left [{\boldsymbol \Upsilon _{{{\text{w}}}}}\right]_{:,\hat {\mathcal{ T}}}}\right)^\dagger {\boldsymbol {\mathsf {y}}} _{{{\text{w}}}}[k]$,  $\forall k $
				
	\State 
	${\boldsymbol {\mathsf {r}}}[k] \gets {\boldsymbol {\mathsf {y}}}_{{{\text{w}}}}[k]-\left [{\boldsymbol \Upsilon _{{{\text{w}}}}}\right]_{:,\hat {\mathcal{ T}}} \hat {\tilde {\boldsymbol \xi }}[k]$, $\forall k $ 
				
	\State 
	${{\text{MSE}}} \gets \frac {1}{KML_{{{\mathrm{r}}}}}\sum _{k=0}^{K-1}{ {\boldsymbol {\mathsf {r}}}^{*}[k] {\boldsymbol {\mathsf {r}}}[k]}$
	
	\State $i\gets i+1$
\EndWhile

\State $\hat L \gets i \:\:\: //$ \: Estimate \# paths [c.f. Section \ref{sec:estimation_paths}] 

\State $\hat{\boldsymbol {\mathsf {h}}}^{\mathrm{v}}[k] \gets $  as per \eqref{eq:hv[k]}.

\State $\mathop {\mathrm {vec}}\{\hat{\boldsymbol{\Delta }}^{\mathrm{v}}[k]\} \gets \hat{\boldsymbol {\mathsf {h}}}^{\mathrm{v}}[k]$
				
\State 
$\mathop {\mathrm {vec}}\{ \hat{\boldsymbol {\mathsf {H}}}[k]\} \gets (\bar {\tilde { {\mathbf {A}}}}_{\mathrm{T}} \otimes \tilde { {\mathbf {A}}}_{\mathrm{R}}) \mathop {\mathrm {vec}}\{\hat{\boldsymbol{\Delta }}^{\mathrm{v}}[k]\}$.

\hspace{-9pt} \Return $ \hat{\boldsymbol {\mathsf {H}}} [k]$ 
\EndProcedure
  \end{algorithmic}
 \end{algorithm}

\subsubsection{Strongest Subcarriers Selection}
This procedure is represented in lines 8-11 of Algorithm \ref{Alg:Ch_Est}, where the algorithm iteratively finds a subset $\mathcal{K} \in K$ containing the $K_p$ strongest subcarriers which are expected to exhibit the strongest channel response as explained in Section \ref{subsec:Onlinesubcarr}.

\subsubsection{Amplitude Estimation}
As depicted in Fig. \ref{fig:Proposed_scheme}, the lines 13 and 14 of Algorithm \ref{Alg:Ch_Est} first compute the correlation vector as per (\ref{equ:corrw}) and then create the DnCNN input ${\boldsymbol {\mathsf {C}}}_{\bm{\alpha}}[k]$ by putting correlation vectors into a matrix form as per \eqref{eq:Calpha}, respectively. In line 15, the offline trained DnCNN is used as the kernel of the channel amplitude estimation to obtain the DnCNN output ${\bf  \hat{G}}[k]$ of size $G_{\mathrm{r}} \times G_{\mathrm{t}}$, which is the estimate of ${\bf G}[k]$ given in \eqref{equ:amps}. It is worth noting that we only use a subset $\mathcal{K}$ of the correlation matrices ${\boldsymbol {\mathsf {C}}}_{\bm{\alpha}}[k] \: \forall k \in \mathcal{K}$ as an input to the DnCNN. In line 16, the output channel amplitude estimation matrix ${\bf  \hat{G}}[k]$ is then vectorized into the following $G_{\mathrm{t}}G_{\mathrm{r}} \times 1$ vector form
\begin{equation}
\label{eq:g_vector}
    \hat{\boldsymbol {g}}[k]=\text{vec}({\hat{\mbf{G}}}[k]), \: \forall k \in \mathcal{K}
\end{equation}
where the indices of the maximum amplitudes of $\hat{\boldsymbol {g}}[k]$ will be exploited for support detection. 

\subsubsection{Multicarrier Channel Reconstruction}

This procedure corresponds to the last block depicted in the last stage of the block diagram in Fig.\ref{fig:Proposed_scheme}.b. It detects supports by iteratively updating residual until the MSE falls below a predetermined threshold, $\epsilon$. After initialization steps in lines 19 and 20, line 19 first sums the amplitudes of predicted $\hat {\boldsymbol {g}}[k]$ over the subcarriers $k \in \mathcal{K}$ as the supports are the same for all $k$ [c.f. Section ~\ref{sec:MeasMat}]. Then, \textsc{IndexSortDescend} function sorts the sum vector in descending order and return corresponding index set $\mathcal{I}$, $|\mathcal{I}|=G_{\mathrm{r}}G_{\mathrm{t}}$. Thereafter, the while loop between lines 22 and 28 follows the below steps until the termination condition is satisfied:

Line 23 updates the detected support set $\hat{\mathcal{T}}$ by adding the $\nth{i}$ element of ordered index set $\mathcal{I}$. Then, line 24 projects the input signal ${\boldsymbol {\mathsf {y}}} _{{{\text{w}}}}[k] \:\:\: \forall k$ onto the subspace given by the detected support $\mathcal{T}$ using Weighted Least-Squares (WLS) $\left ({\left [{\boldsymbol \Upsilon _{{{\text{w}}}}}\right]_{:,\hat {\mathcal{ T}}}}\right)^\dagger$, which is followed by residual update and MSE computation in lines 25 and 26, respectively. It is also worth noting that $\left ({\left [{\boldsymbol \Upsilon _{{{\text{w}}}}}\right]_{:,\hat {\mathcal{ T}}}}\right)^\dagger$ corresponds to a WLS estimator, with the corresponding weights given by the inverse noise covariance matrix.
Lastly, line 26 increments the loop index $i$ for the next iteration. The final value of $i=|\hat{\mathcal{T}}|$ provides us with one of the key parameters: $\hat{L}$, the estimate of the sufficient number of paths that guarantees MSE$>\epsilon$, i.e., $L$. Thereby, it is closely tied with the choice of $\epsilon$, which will be explained in details in Section \ref{sec:estimation_paths}. We should also note that the while loop is terminated by the MSE$>\epsilon$  condition almost all the time since $G_{\mathrm{t}}G_{\mathrm{r}} \gg \hat{L}$ as shown in Table \ref{table:supp}\footnote{This assumption holds since mmWave channels are known to have limited number of paths.}. 

Since the support of sparse channel vectors is already estimated by $\hat {\mathcal{T}}$, the measurement matrix can now be defined as $\left [{\boldsymbol{\Upsilon }}\right]_{:,\hat {\mathcal{ T}}} \in \mathbb {C}^{M L_{{{\text{r}}}} \times \hat {L}}$  such that $\left [{\boldsymbol{\Upsilon }}\right]_{:,\hat {\mathcal{ T}}} = \left [{\boldsymbol{\Phi } \boldsymbol{\Psi }}\right]_{:,\hat {\mathcal{ T}}}$. Hence, the received signal model for the $\nth{k}$ subcarrier can be rewritten as
\begin{equation} \label{equ:recvec}
{\boldsymbol {\mathsf {y}}}[k] = \left [{\boldsymbol{\Upsilon }}\right]_{:,\hat {\mathcal{ T}}} \tilde {\boldsymbol {\xi }}[k] + \tilde { {\boldsymbol {\mathsf {n}}}}_{{{\text{c}}}}[k], \end{equation}
where $\tilde { {\boldsymbol {\mathsf {n}}}}_{{{\text{c}}}}[k] \in \mathbb {C}^{M L_{\mathrm {r}}\times 1}$ represents the residual noise after estimating the channel support and $\tilde {\boldsymbol \xi }[k] \in \mathbb{C}^{\hat {L} \times 1}$ is the vector containing the channel gains to be estimated after sparse recovery. If the support estimation is accurate enough, $\tilde { {\boldsymbol {\mathsf {n}}}}_{{{\text{c}}}}[k]$ will be approximately similar to the post-combining noise vector ${\boldsymbol {\mathsf {n}}}_{{{\text{c}}}}[k]$ \cite{Heath2018CEMain}. 
It is important to remark that the indices obtained by the trained DnCNN may be different from the actual channel support. In this case, the support detected $\hat {\mathcal{T}}$ may also be different from the actual support. Likewise, the channel gains to be estimated $\tilde {\boldsymbol \xi }[k]$, can also be different from actual vector, $\boldsymbol \xi [k] = \mathop {\mathrm {vec}}\{ \mathop {\mathrm {diag}}\{\boldsymbol \Delta [k]\}\}$.

The mathematical model in \eqref{equ:recvec} is usually considered as the General Linear Model (GLM), where the solution of $\tilde {\boldsymbol \xi }[k]$ for real parameters is provided in \cite{kay1993fundamentals}. For the case with complex-valued parameters, the solution is straightforward and given by
\begin{equation}
\hat {\tilde {\boldsymbol \xi }}[k] = \left ({\left [{\boldsymbol{\Upsilon }}\right]_{:,\hat {\mathcal{ T}}}^{*} {\mathbf {C}}_{{{\text{w}}}}^{-1}\left [{\boldsymbol{\Upsilon }}\right]_{:,\hat {\mathcal{ T}}}}\right)^{-1}\left [{\boldsymbol{\Upsilon }}\right]_{:,\hat {\mathcal{ T}}}^{*} {\mathbf {C}}_{{{\text{w}}}}^{-1} {\boldsymbol {\mathsf {y}}}[k], \end{equation}
which can be further reduced to
\begin{equation}
\hat {\tilde {\boldsymbol \xi }}[k] = \left ({\left [{\boldsymbol \Upsilon _{{{\text{w}}}}}\right]_{:,\hat {\mathcal{ T}}}}\right)^\dagger {\boldsymbol {\mathsf {y}}} _{{{\text{w}}}}[k]. 
\end{equation}
Therefore, $\hat {\tilde {\boldsymbol \xi }}[k]$ is considered as the Minimum Variance Unbiased (MVU) estimator for the complex parameter vector $\tilde {\boldsymbol \xi }[k]$, $k = 0,\ldots,K-1$. Hence, it is unbiased and attains the Cramér-Rao Lower Bound (CRLB) if the support is correctly estimated \cite{Heath2018CEMain} \footnote{ This is considered as Cramér-Rao Lower Bound of a Genie-aided estimation problem, in which the estimator knows the location of the nonzero taps i.e., $\mathcal{T}$, as if a Genie has aided the estimator with the location of the taps \cite{Jutt2011CRLB}}. 

Once all the supports are detected, line 29 computes the sparse channel vector ${\hat{\boldsymbol {\mathsf {h}}}}^{\mathrm{v}}[k]$ where its non-zero elements are obtained according to 
\begin{equation}
\label{eq:hv[k]}
    [{\hat{\boldsymbol {\mathsf {h}}}}^{\mathrm{v}}[k]]_{\hat {\mathcal{ T}}}= \left ({\left [{\boldsymbol \Upsilon _{{{\text{w}}}}}\right]_{:,\hat {\mathcal{ T}}}}\right)^\dagger {\boldsymbol {\mathsf {y}}} _{{{\text{w}}}}[k]. 
\end{equation}
Finally, line 32 reconstructs the channel based on ~(\ref{Ch_vec}) as follows
\begin{equation}
    \mathop {\mathrm {vec}}\{ \hat{\boldsymbol {\mathsf {H}}}[k]\}= (\bar {\tilde { {\mathbf {A}}}}_{\mathrm{T}} \otimes \tilde { {\mathbf {A}}}_{\mathrm{R}}) \mathop {\mathrm {vec}}\{\hat{\boldsymbol{\Delta }}^\mathrm{v}[k]\},
\end{equation}
such that $\mathop {\mathrm {vec}}\{\hat{\boldsymbol{\Delta }}^{\mathrm{v}}\}[k]={\hat{\boldsymbol {\mathsf {h}}}}^{\mathrm{v}}[k]$.

\subsection{Algorithm 2: Refined DL-CS-CE} 

The sparsity of $\boldsymbol{\mathsf {h}}^{\mathrm{v}}[k]$ can be impaired by channel power leakage caused by the limited resolution of the chosen dictionary matrices \cite{qin2018sparse}. Although the DL-CS-CE provides reasonable AoD/AoA estimates, the adopted virtual quantized dictionary matrices may not obtain the exact AoDs/AoAs that really lies in the off-grid regions of the dictionary. In this section, we combat this issue by developing a method to obtain more accurate AoDs/AoAs. This new procedure is called refined DL-CS-CE and improves NMSE performance of Algorithm \ref{Alg:Ch_Est} while reducing the incurring computational complexity at the same time. 

Using the superscript $\mathrm{r}$ for referring to the refining phase, we consider higher resolution refining dictionary matrices ${\tilde { \mathbf {A}}}_{\mathrm{R}}^{\mathrm{r}}$ and ${\tilde { \mathbf {A}}}_{\mathrm{T}}^{\mathrm{r}}$ with grid sizes $G_{\mathrm{r}}^{\mathrm{r}}$ and $G_{\mathrm{t}}^{\mathrm{r}}$, respectively. Based on this notation, the refined DL-CS-CE summarized in Algorithm \ref{Alg:Ch_Est_ref} follows the same implementation as that of Algorithm \ref{Alg:Ch_Est} except some technical differences during the channel reconstruction stage, on which we focus our attention in the sequel.

\begin{algorithm}[t]
\caption{Refined DL-CS-CE }\label{Alg:Ch_Est_ref}
\begin{algorithmic}[1]
\renewcommand{\algorithmicrequire}{\textbf{Input:}}
 \renewcommand{\algorithmicensure}{\textbf{Output:}}
\Require ${\boldsymbol {\mathsf {y}}}[k]$, $\boldsymbol {\Phi}$, $\boldsymbol {\Psi}$, $ {\tilde { \mathbf {A}}}_{\mathrm{T}}$, $ {\tilde { \mathbf {A}}}_{\mathrm{R}}$, $ {\tilde { \mathbf {A}}}_{\mathrm{T}}^{\mathrm{r}}$, $ {\tilde { \mathbf {A}}}_{\mathrm{R}}^{\mathrm{r}}$, $K_p$, $\epsilon$

\State ${\boldsymbol {\mathsf {y}}}_{{{\text{w}}}}[k] \gets {\mathbf {D}}_{{{\text{w}}}}^{-*} {\boldsymbol {\mathsf {y}}}[k]\:\:\:\forall k$

 \State ${\boldsymbol {\mathsf {r}}}[k] \gets {\boldsymbol {\mathsf {y}}}_{{{\text{w}}}}[k]\:\:\:\forall k$

\State $ \hat {\mathcal{ T}}, {\mathcal{K}} \gets \{\emptyset \}$

\State $ \boldsymbol{\Phi}_{\text{w}}={\mathbf {D}}_{{{\text{w}}}}^{-*}\boldsymbol{\Phi }$

\State $\boldsymbol{\Psi }\gets (\bar {\tilde { \mathbf {A}}}_{\mathrm{T}}\otimes  {\tilde { \mathbf {A}}}_{\mathrm{R}})  \:\:\:// \:\:\:For \:\:\: Detection$

\State	$\boldsymbol{\Upsilon }_{{{\text{w}}}} \gets {\mathbf {D}}_{{{\text{w}}}}^{-*}\boldsymbol{\Phi } \boldsymbol{\Psi}$

\State $\boldsymbol{\Psi }^{\mathrm{r}} = (\bar {\tilde { \mathbf {A}}}_{\mathrm{T}}^{\mathrm{r}} \otimes  {\tilde { \mathbf {A}}}_{\mathrm{R}}^{\mathrm{r}})  \:\:\:// \:\:\:For \:\:\: Refining $

\State
$\boldsymbol{\Upsilon }_{{{\text{w}}}}^{\mathrm{r}} \gets {\mathbf {D}}_{{{\text{w}}}}^{-*}\boldsymbol{\Phi } \boldsymbol{\Psi }^{\mathrm{r}}$

\State $\mathcal{K} \gets \textsc{Find Strongest Subcarriers (${\boldsymbol {\mathsf {y}}}[k]$)}$ 

\State $\hat{\boldsymbol {g}}[k] \gets \textsc{Estimate Amplitudes ($\boldsymbol{\Upsilon }_{{{\text{w}}}}^{*}, {\boldsymbol {\mathsf {r}}}[k], \mathcal{K}$)}$

\State $\hat{\boldsymbol {\mathsf {H}}}[k] \gets \textsc{Reconstruct Channel \& Refine($\hat{\boldsymbol {g}}[k]  $)}$

\hspace{-40pt} \Return $\hat{\boldsymbol {\mathsf {H}}}[k]$
\vspace{2pt}
\hrule 
\vspace{2pt}
\Procedure{Find Strongest Subcarriers}{${\boldsymbol {\mathsf {y}}}[k]$}
\State Lines 9-10 in Algorithm \ref{Alg:Ch_Est}

\EndProcedure

\vspace{2pt}
\hrule 
\vspace{2pt}
\Procedure{Estimate Amplitudes}{$\boldsymbol{\Upsilon }_{{{\text{w}}}}^{*}, {\boldsymbol {\mathsf {r}}}[k], \mathcal{K}$}

\State Lines 13-16 in Algorithm \ref{Alg:Ch_Est}

\EndProcedure

\vspace{2pt}
\hrule 
\vspace{2pt}

\Procedure{Reconstruct Channel \& Refine}{$\hat{\boldsymbol {g}}[k]$}
\State $\mathcal{I} \gets \textsc{IndexSortDescend}\left(\sum_{k\in {\mathcal{K}}} |\hat{\boldsymbol {g}}[k]|\right)$
\State MSE $\gets \infty$
\State $i \gets 1$
\While { MSE $> \epsilon \:\: \& \:\: i\leq G_{\mathrm{t}}G_{\mathrm{r}}$}
				
	
	\State 
	$[i_{\text{AoA}}^\mathrm{d}, i_{\text{AoD}}^\mathrm{d}]\gets \mathrm{ind2sub}([G_{\mathrm{r}}, G_{\mathrm{t}}],  \mathcal{I}(i))$
	
	\State $\hat {\mathcal{ T}}  \gets \textsc{Refine ($i_{\text{AoA}}^\mathrm{d}, i_{\text{AoD}}^\mathrm{d} $)}$

	\State 
	$\hat {\tilde {\boldsymbol \xi }}[k] \gets \left ({\left [{\boldsymbol \Upsilon _{{{\text{w}}}}^{\mathrm{r}} }\right]_{:,\hat {\mathcal{ T}}}}\right)^\dagger {\boldsymbol {\mathsf {y}}} _{{{\text{w}}}}[k], \forall k $

	\State 
	${\boldsymbol {\mathsf {r}}}[k] \gets {\boldsymbol {\mathsf {y}}}_{{{\text{w}}}}[k]-\left [{\boldsymbol \Upsilon _{{{\text{w}}}}^{\mathrm{r}} }\right]_{:,\hat {\mathcal{ T}}} \hat {\tilde {\boldsymbol \xi }}[k], \forall k $ 
				
	\State 
	${{\text{MSE}}} \gets \frac {1}{KML_{{{\mathrm{r}}}}}\sum _{k=0}^{K-1}{ {\boldsymbol {\mathsf {r}}}^{*}[k] {\boldsymbol {\mathsf {r}}}[k]}$
	
	\State $i\gets i+1$
\EndWhile

\State $\hat L \gets i \:\:\: //$ \: Estimate \# paths [c.f. Section \ref{sec:estimation_paths}] 

\State $\hat{\boldsymbol {\mathsf {h}}}^{\mathrm{v}}[k] \gets$ as per \eqref{eq:hv[k]} but using $\left [{\boldsymbol \Upsilon _{\text{w}}^{\text{r}}}\right]_{:,\hat {\mathcal{ T}}}$ instead
				
\State $\mathop {\mathrm {vec}}\{\hat{\boldsymbol{\Delta }}^{\mathrm{v}}[k]\} \gets \hat{\boldsymbol {\mathsf {h}}}^{\mathrm{v}}[k]$
				
\State 
$\mathop {\mathrm {vec}}\{ \hat{\boldsymbol {\mathsf {H}}}[k]\} \gets \boldsymbol{\Psi }^{\mathrm{r}}  \mathop {\mathrm {vec}}\{\hat{\boldsymbol{\Delta }}^{\mathrm{v}}[k]\}$.

\hspace{-9pt} \Return $ \hat{\boldsymbol {\mathsf {H}}} [k]$ 
\EndProcedure
\vspace{2pt}
\hrule 
\vspace{2pt}
\Procedure{Refine}{$i_{\text{AoA}}^\mathrm{d}, i_{\text{AoD}}^\mathrm{d}$ }

\State 
${i_{\text{AoA}}^\mathrm{r}} \gets$ as per in \eqref{eq:refine_1}

\State
${i_{\text{AoD}}^\mathrm{r}} \gets$  as per \eqref{eq:refine_2}

\State 
${i_{\text{AoA}}^\mathrm{r}}^\star \gets$ as per \eqref{eq:refine_3}

\State
${i_{\text{AoD}}^\mathrm{r}}^\star \gets$  as per \eqref{eq:refine_2} by using ${i_{\text{AoA}}^\mathrm{r}}^\star$ instead of ${i_{\text{AoA}}^\mathrm{r}}$

\State
$j^\star \gets \mathrm{sub2ind}([G_{\mathrm{r}}^\mathrm{r}, G_{\mathrm{t}}^\mathrm{r}], [{i_{\text{AoA}}^\mathrm{r}}^\star, {i_{\text{AoD}}^\mathrm{r}}^\star])$ 

\State 	$\hat {\mathcal{ T}} \gets \hat {\mathcal{ T}}\cup j^\star$

\hspace{-9pt} \Return $\hat {\mathcal{ T}}$ 
\EndProcedure
\end{algorithmic}
\end{algorithm}

\subsubsection*{Multicarrier Channel Reconstruction and Refinement}

The while loop between lines 22 and 29 refines the path components by iterative projections. In line 23, the detected support ${\mathcal{I}}(i)$ is first transformed into column and row indices of a $G_{\mathrm{r}}\times G_{\mathrm{t}}$ matrix representing the indices ($i_{\text{AoA}}^\mathrm{d}, i_{\text{AoD}}^\mathrm{d}$) of the detected AoAs and AoDs in the original lower resolution dictionary matrices ${\tilde { \mathbf {A}}}_{\mathrm{R}}$ and ${\tilde { \mathbf {A}}}_{\mathrm{T}}$, respectively. In line 24, a multi-resolution fine-tuning method is applied to enhance the resolution of the detected AoAs and AoDs. The refining procedure consists of two steps as shown between lines 36 and 39 of Algorithm \ref{Alg:Ch_Est_ref}. In what follows, these steps are explained based on the column index set notation $\Omega_{\rm{K,q}}$
, where $\rm K\in\{A,D\}$ represents arrival or departure, and $\rm q \in\{d,r\}$ refer to detection or refinement, respectively.
 

\begin{enumerate}
\item The first step starts with line 36 which refine the angle components with the highest number of antennas. For instance, let’s assume that $N_{\mathrm{r}} > N_{\mathrm{t}}$. By increasing the resolution of $\hat \phi_l$ to $G_{\mathrm{r}}^\mathrm{r}\gg G_{\mathrm{r}}$, the maximum projection along the refined receiving array steering matrix ${\tilde { \mathbf {A}}}_{\mathrm{R}}^{\mathrm{r}}$, while the corresponding AoD $\hat{\theta}_l$ is fixed, can be expressed as
\begin{equation}
\label{eq:refine_1}
{i_{\text{AoA}}^\mathrm{r}}= \mathop{\arg \,\max }\limits_{i} \left[\sum_{k\in \mathcal{K}} \left \vert \left(\left[\boldsymbol{\Upsilon}_{\text{w}}^{\text{d}}\right]_{:,\Omega_{\text{D,d}}}\right)^* \boldsymbol {\mathsf {y}} _{{{\text{w}}}}[k] \right \vert \right]_{i}, 
\end{equation} 
 where $ \boldsymbol{\Upsilon}_{\text{w}}^{\text{d}}$ is an ${M L_{{{\text{r}}}} \times  G_{\mathrm{r}}^\mathrm{r}G_{\mathrm{t}}}$ matrix such that $\boldsymbol{\Upsilon}_{\text{w}}^{\text{d}}=\boldsymbol{\Phi}_{\text{w}}(\bar {\tilde { \mathbf {A}}}_{\mathrm{T}} \otimes  {\tilde { \mathbf {A}}}_{\mathrm{R}}^{\mathrm{r}})$, and $\left[\boldsymbol{\Gamma}_{\text{w}}^{\text{d}}\right]_{:,\Omega_{\text{D,d}}}$ is an ${M L_{{{\text{r}}}} \times  G_{\mathrm{r}}^\mathrm{r}}$ sub-matrix with the column index set defined as  $\Omega_{\text{D,d}}=\{i_{\text{AoD}}^{\mathrm{d}}:i_{\text{AoD}}^\mathrm{d}.G_{\mathrm{r}}^{\mathrm{r}}\}$;  $i_{\text{AoD}}^\mathrm{d}$ corresponds to the index of the previously detected AoD before refining. Then, line 37 continues with the remaining angle by increasing the resolution of $\hat \theta_l$ to $G_{\mathrm{t}}^\mathrm{r}\gg G_{\mathrm{t}}$. Similar to \eqref{eq:refine_1}, the maximum projection along the refined transmit array steering matrix ${\tilde { \mathbf {A}}}_{\mathrm{T}}^{\mathrm{r}}$, while the corresponding obtained refined AoA $\hat\phi_l$ is fixed,  can be expressed as
 \begin{equation}
 \label{eq:refine_2}
{i_{\text{AoD}}^\mathrm{r}}= \mathop{\arg \,\max }\limits_{i} \left[\sum_{k\in \mathcal{K}} \left \vert\left(\left[\boldsymbol{\Upsilon}_{\text{w}}^{\text{r}}\right]_{:,\Omega_{\text{A,r}}}\right)^* \boldsymbol {\mathsf {y}} _{{{\text{w}}}}[k]\right \vert\right]_{i}
\end{equation}
where $ \boldsymbol{\Upsilon}_{\text{w}}^{\text{r}}$ is an  ${M L_{{{\text{r}}}} \times  G_{\mathrm{r}}^\mathrm{r}G_{\mathrm{t}}^\mathrm{r}}$ matrix such that $\boldsymbol{\Upsilon}_{\text{w}}^{\text{r}}=\boldsymbol{\Phi}_{\text{w}}(\bar {\tilde { \mathbf {A}}}_{\mathrm{T}}^{\mathrm{r}} \otimes  {\tilde { \mathbf {A}}}_{\mathrm{R}}^{\mathrm{r}})$, and $\left[\boldsymbol{\Upsilon}_{\text{w}}^{\text{r}}\right]_{:,\Omega_{\text{A,r}}}$ is an ${M L_{{{\text{r}}}} \times  G_{\mathrm{t}}^\mathrm{r}}$ sub-matrix with the column index set defined as
\begin{equation}\label{eq:OmegaR}
    \Omega_{\text{A,r}}=\{0: G_{\mathrm{r}}^{\mathrm{r}}G_{\mathrm{t}}^{\mathrm{r}}-1\}\bmod G_{\mathrm{r}}^{\mathrm{r}}+i_{\text{AoA}}^{\mathrm{r}}.
\end{equation}
Here, ${i_{\text{AoA}}^\mathrm{r}}$ is the index of the refined AoA obtained from ~(\ref{eq:refine_1}). 
 \item The second step:
 In line 38, after removing the angle uncertainty caused by the detection phase, we can proceed to repeat the same step by substituting all angles with their corresponding refined angles. The maximum projection along the refined received array is given by
\begin{equation}
\label{eq:refine_3}
{i_{\text{AoA}}^\mathrm{r}}^\star= \mathop{\arg \,\max }\limits_{i} \left[\sum_{k\in \mathcal{K}} \left \vert \left(\left[\boldsymbol{\Upsilon}_{\text{w}}^{\text{r}}\right]_{:,\Omega_{\text{D,r}}}\right)^* \boldsymbol {\mathsf {y}} _{{{\text{w}}}}[k] \right \vert \right]_{i}, 
\end{equation} 
where $\Omega_{\text{D,r}}=\{i_{\text{AoD}}^{\mathrm{r}}:i_{\text{AoD}}^\mathrm{r}.G_{\mathrm{r}}^{\mathrm{r}}\}$, and  $i_{\text{AoD}}^\mathrm{r}$ corresponds to the index obtained from the previous step in \eqref{eq:refine_2}.
\end{enumerate}

Similarly in line 39, ${i_{\text{AoD}}^\mathrm{r}}^\star$ is now obtained using equation ~(\ref{eq:refine_2}) but by substituting ${i_{\text{AoA}}^\mathrm{r}}$ in ~(\ref{eq:OmegaR}) with the obtained ${i_{\text{AoA}}^\mathrm{r}}^\star$ (the result from equation ~(\ref{eq:refine_3})).

Next, line 40 transforms the row and column indices $[{i_{\text{AoA}}^\mathrm{r}}^\star, {i_{\text{AoD}}^\mathrm{r}}^\star]$ into a linear index $j^\star$. The refining procedure lastly updates the refined support estimation set $\hat {\mathcal{ T}} $ by admitting index $j^\star$ into $\hat {\mathcal{ T}} $. 


\begin{table}[t!]
	\centering
	\caption{\textsc{Average size of estimated support $\hat{L}=|\hat{\mathcal{T}}|$ }}
	\footnotesize
	\label{table:supp}
	\begin{tabular}{|l|c|c|c|c|c|}
		\hline
		\textbf{SNR} & $\unit[-15]{dB}$& $\unit[-10]{dB}$& $\unit[-5]{dB}$ & $\unit[-0]{dB}$& $\unit[5]{dB}$    \\\hline
		$\hat{L}$  & $4$& $5$& $9$ & $12$& $15$    \\\hline
	\end{tabular}
\end{table}

\subsection{Estimation of The Sufficient Number of Paths}\label{sec:estimation_paths}

After estimating the channel amplitudes using the trained DnCNN, it is necessary to determine the sufficient support indices representing the number of paths needed to reconstruct the channel. To solve this detection problem, some prior information is needed to compare the received signals ${\boldsymbol {\mathsf {y}}}[k]$ with the reconstructed signals $\hat { {\boldsymbol {\mathsf {x}}}}_{{{\text{rec}}}}[k] = \left [{\boldsymbol \Upsilon }\right]_{:,\hat {\mathcal{ T}}}\hat {\boldsymbol \xi }[k]$. For instance, the noise variance is assumed to be known at the receiver in which the receiver can accurately estimate the noise variance before the training stage takes place. Hence, the received signal ${\boldsymbol {\mathsf {y}}}[k]$ can be approximately modeled as ${\boldsymbol {\mathsf {y}}}[k] \approx \hat { {\boldsymbol {\mathsf {x}}}}_{{{\text{rec}}}}[k] + \tilde { {\boldsymbol {\mathsf {n}}}}_{{{\text{c}}}}[k]$, since $\hat { {\boldsymbol {\mathsf {x}}}}_{{{\text{rec}}}}[k]$ is an estimate of the mean of ${\boldsymbol {\mathsf {y}}}[k] $.

The estimation of the noise variance can be formulated as a Maximum-Likelihood estimation problem \cite{kay1993fundamentals,Heath2018CEMain}:
		\begin{equation} \hat {\sigma ^{2}}_{{{\text{ML}}}} = \underset {\sigma ^{2}}{\arg \,\max }\,\mathcal{L}({\boldsymbol {\mathsf {y}}},\hat { {\boldsymbol {\mathsf {x}}}}_{{{\text{rec}}}},\sigma ^{2}), \end{equation}
where ${\boldsymbol {\mathsf {y}}}\triangleq \mathop {\mathrm {vec}}\{ {\boldsymbol {\mathsf {y}}}[{0}],\ldots, {\boldsymbol {\mathsf {y}}}[K-1]\}$ represents the complete received signal, $\hat { {\boldsymbol {\mathsf {x}}}}_{{{\text{rec}}}} \triangleq \mathop {\mathrm {vec}} \{\hat { {\boldsymbol {\mathsf {x}}}}_{{{\text{rec}}}}[{0}],\ldots,\hat { {\boldsymbol {\mathsf {x}}}}_{{{\text{rec}}}}[K-1]\}$ is the complete reconstructed signal, and $\mathcal{L}({\boldsymbol {\mathsf {y}}},\hat { {\boldsymbol {\mathsf {x}}}}_{{{\text{rec}}}},\sigma ^{2})$ denotes the log likelihood function of ${\boldsymbol {\mathsf {y}}}$. This log likelihood function is given by
\begin{align}
&\hspace {-2pc}\mathcal{L}({\boldsymbol {\mathsf {y}}},\hat { {\boldsymbol {\mathsf {x}}}}_{{{\text{rec}}}},\sigma ^{2}) \notag \\=&-K M L_{\mathrm {r}}\ln {\pi \sigma ^{2}} -\ln {\det \{ {\mathbf {C}}_{{{\text{w}}}}\}} \notag \\&-\frac {1}{\sigma ^{2}}\sum _{k=0}^{K-1}{\left ({{\boldsymbol {\mathsf {y}}}[k] - \hat { {\boldsymbol {\mathsf {x}}}}_{{{\text{rec}}}}[k])^{*} {\mathbf {C}}_{{{\text{w}}}}^{-1}({\boldsymbol {\mathsf {y}}}[k] - \hat { {\boldsymbol {\mathsf {x}}}}_{{{\text{rec}}}}[k]}\right)}.\quad ~ 
\end{align}
The ML estimator of the noise variance is then obtained by taking partial derivative with respect to $\sigma ^{2}$ where $\partial \mathcal{L}({\boldsymbol {\mathsf {y}}},\hat { {\boldsymbol {\mathsf {x}}}}_{{{\text{rec}}}},\sigma ^{2})/\partial \sigma ^{2} = 0$. Hence, $\widehat {\sigma ^{2}}_{{{\text{ML}}}}$ is given by
\begin{align} \widehat {\sigma ^{2}}_{{{\text{ML}}}} = \frac {1}{KM L_{\mathrm {r}}}\sum _{k=0}^{K-1}{\underbrace {\left ({{\boldsymbol {\mathsf {y}}}[k] - \hat { {\boldsymbol {\mathsf {x}}}}_{{{\text{rec}}}}[k]}\right)^{*} {\mathbf {C}}_{{{\text{w}}}}^{-1}\left ({{\boldsymbol {\mathsf {y}}}[k] - \hat { {\boldsymbol {\mathsf {x}}}}_{{{\text{rec}}}}[k]}\right)}_{{ {\boldsymbol {\mathsf {r}}}}^{*}[k]{ {\boldsymbol {\mathsf {r}}}}[k]}}\!\!\notag \\ {}\end{align}
where the $M L_{\mathrm {r}}\times 1$ vector ${ {\boldsymbol {\mathsf {r}}}}[k] \triangleq {\boldsymbol {\mathsf {y}}} _{{{\text{w}}}}[k] - {\mathbf {D}}_{{{\text{w}}}}^{-*}\hat { {\boldsymbol {\mathsf {x}}}}_{{{\text{rec}}}}$ is the residual. One can note that ${ {\boldsymbol {\mathsf {r}}}}[k]$ can also be expressed as ${ {\boldsymbol {\mathsf {r}}}}[k] = \left ({{\mathbf {I}}_{M L_{\mathrm {r}}} - {\mathbf {P}}}\right) {\boldsymbol {\mathsf {y}}}_{{{\text{w}}}}[k]$ , where ${\mathbf {P}}\in \mathbb {C}^{M L_{\mathrm {r}}\times M L_{\mathrm {r}}}$ represents the projection matrix given by ${\mathbf {P}}= \left [{\boldsymbol \Upsilon _{{{\text{w}}}}}\right]_{:,\hat {\mathcal{ T}}}^\dagger \left [{\boldsymbol \Upsilon _{{{\text{w}}}}}\right]_{:,\hat {\mathcal{ T}}}$ .

Therefore, for a sufficient number of iterations, $\hat {L}$ sufficient paths are expected to be detected as those $\hat {L}$ paths correspond to the dominant $\hat {L}$ entries of $\sum_{k\in {\mathcal{K}}} |{\boldsymbol {\mathsf {h}}}^{\mathrm{v}}[k]|$. Moreover, the detection process is achieved when the estimated noise variance becomes equal to the true noise variance of the received signal by setting $\epsilon$  to $\sigma^2$ in ~(\ref{equ:prob}).


\section{Convergence and Complexity Analysis}\label{sec:Comp_analysis}
In this section, we analyze the convergence of the proposed algorithms to a local optimum, which is then followed by their step-by-step computational complexity analysis. 
\subsection{Convergence Analysis}
We assume that the dictionary sizes $G_{\text {t}}$ and $G_{\text {r}}$ are large enough\footnote{This assumption holds for large enough values of $M$  and $K$ \cite{Heath2016shiftOrSwitches}.} to have coarsely quantized AoAs/AoDs are accurately estimated. For the sake of simplicity, we build the convergence analysis based on the notation for Algorithm \ref{Alg:Ch_Est} to analyze the convergence, which is also applicable for Algorithm \ref{Alg:Ch_Est_ref}. In order to insure convergence to a local optimum, the energy of the residual computed at the $\nth{n+1}$ iteration should be strictly smaller than that of the previous $\nth{n}$ iteration, i.e.,
\begin{equation}\label{equ:qu:residual}
 || {\boldsymbol {\mathsf {r}}}^{(n+1)}[k]||_{2}^{2} < || {\boldsymbol {\mathsf {r}}}^{(n)}[k]||_{2}^{2},\quad k = 0,\ldots,K-1. 
 \end{equation}
Noting that the residual computation for SW-OMP in \cite{Heath2018CEMain} and proposed algorithms are identical as they follow the same analysis, the residual for a given iteration $n$ is expressed as
 \begin{equation} 
 {\boldsymbol {\mathsf {r}}}^{(n)}[k] = \left ({{\mathbf {I}}_{M L_{\mathrm {r}}} - {\mathbf {P}}^{(n)}}\right) {\boldsymbol {\mathsf {y}}}_{{{\text{w}}}}[k], 
 \end{equation}
where ${\mathbf {P}}^{(n)} \in \mathbb {C}^{M L_{\mathrm {r}}\times M L_{\mathrm {r}}}$ corresponds to a projection matrix given by ${\mathbf {P}}^{(n)} \triangleq \left [{\boldsymbol \Upsilon _{{{\text{w}}}}}\right]_{:,{\hat {\mathcal{ T}}^{(n)}}} \left [{\boldsymbol \Upsilon _{{{\text{w}}}}}\right]_{:,{\hat {\mathcal{ T}}^{(n)}}}^\dagger$. It is worth mentioning that the residual ${\boldsymbol {\mathsf {r}}}^{(n)}[k]$ is the vector resulting from projecting ${\boldsymbol {\mathsf {y}}}_{{{\text{w}}}}[k]$ onto the subspace orthogonal to the column space of $\left [{\boldsymbol \Upsilon _{{{\text{w}}}}}\right]_{:,{\hat {\mathcal{ T}}^{(n)}}}$. Moreover, we can use the projection  onto the column space of $\left [{\boldsymbol \Upsilon _{{{\text{w}}}}}\right]_{:,{\hat {\mathcal{ T}}^{(n)}}}$ to rewrite the condition in ~(\ref{equ:qu:residual}) as follows
 \begin{equation} \label{equ:qu:proj}
 || {\mathbf {P}}^{(n+1)} {\boldsymbol {\mathsf {y}}}_{{{\text{w}}}}[k]||_{2}^{2} > || {\mathbf {P}}^{(n)} {\boldsymbol {\mathsf {y}}}_{{{\text{w}}}}[k]||_{2}^{2}. 
 \end{equation}
Following the notation used in Algorithm \ref{Alg:Ch_Est}, the term inside the $\ell _{2}$-norm on the left side of (\ref{equ:qu:proj}) can be expressed as
 \begin{align}
  {\mathbf {P}}^{(n+1)} {\boldsymbol {\mathsf {y}}}_{{{\text{w}}}}[k]=&\left [{\begin{array}{cc} \left [{\boldsymbol \Upsilon _{{{\text{w}}}}}\right]_{:,{\hat {\mathcal{ T}}^{(n)}}} &\quad \left [{\boldsymbol \Upsilon _{{{\text{w}}}}}\right]_{:,{\hat {p}^{(n+1)*}}} \end{array}}\right] \notag \\&\times \left [{\begin{array}{cc} \left [{\boldsymbol \Upsilon _{{{\text{w}}}}}\right]_{:,{\hat {\mathcal{ T}}^{(n)}}} &\quad \left [{\boldsymbol \Upsilon _{{{\text{w}}}}}\right]_{:,{\hat {p}^{(n+1)*}}} \end{array}}\right]^\dagger {\boldsymbol {\mathsf {y}}} _{{{\text{w}}}}[k],\quad \qquad 
 \end{align}
 where $\hat {p}^{(n+1)*}$ is the estimate for the support index found during the $\nth{n+1}$ iteration, such that $\hat {p}^{(n+1)*} \not \in \hat {\mathcal{ T}}^{(n)}$.

\begin{table}[t!]
	\centering
	\captionsetup{justification=centering}
	\caption{ \textsc{Online Computational Complexity \\ of Algorithm \ref{Alg:Ch_Est}.}}
	\footnotesize
	\label{table:c1}
	\begin{tabular}{|l|c|}
		\hline
		\textbf{Operation} & \textbf{Complexity}    \\\hline
		$K_p \times {\boldsymbol {\mathsf {c}}}[k] = \boldsymbol{\Upsilon }_{{{\text{w}}}}^{*} {\boldsymbol {\mathsf {r}}}[k]$& $\mathcal{O} (K_p (G_{\mathrm{r}} G_{\mathrm{t}} )M L_{\mathrm{r}})$ \\
			Estimation using DnCNN  & \eqref{eq:comp_dnCNN}  \\
		$\mathop{ \max }\limits_{p} \sum_{k\in {\mathcal{K}}} |{\boldsymbol {\mathsf {h}}}^{\mathrm{v}}[k]|$ & {$\mathcal{O}(K_p(G_ t G_{\mathrm{r}})\hat{L})$}   \\
		$(K) \times {\boldsymbol {\mathsf {x}}}_{\hat {\mathcal{ T}}}[k]\!\!\! = \!\!\! \left ({\left [{\boldsymbol \Upsilon _{{{\text{w}}}}}\right]_{:,\hat {\mathcal{ T}}}}\right)^\dagger {\boldsymbol {\mathsf {y}}} _{{{\text{w}}}}[k]$&  $\mathcal{O}(2 \hat{L}^2 {L}_r M +\hat{L}^3)$    \\
	$(K)\times {\boldsymbol {\mathsf {r}}}[k] \!\!=\!\! {\boldsymbol {\mathsf {y}}}_{{{\text{w}}}}[k]\!-\!\left [{\boldsymbol \Upsilon _{{{\text{w}}}}}\right]_{:,\hat {\mathcal{ T}}} \hat {\tilde {\boldsymbol \xi }}[k]$ & $\mathcal{O}(K L_{\mathrm{r}}  M\hat{L})$    \\
		$\text{MSE}= \frac {1}{KML_{{{\text{r}}}}}\sum _{k=0}^{K-1}{ {\boldsymbol {\mathsf {r}}}^{*}[k] {\boldsymbol {\mathsf {r}}}[k]}$&  $\mathcal{O}(K L_{\mathrm{r}}  M\hat{L})$    \\\hline
		\textbf{Overall}&  $\mathcal{O}( K_p (G_{\mathrm{r}} G_{\mathrm{t}} )M L_{\mathrm{r}})$    \\\hline
	\end{tabular}
\end{table}
\normalsize

\begin{table}[t!]
	\centering
    \captionsetup{justification=centering}
	\caption{ \textsc{Online Computational Complexity \\ of Algorithm \ref{Alg:Ch_Est_ref}.}}
	\footnotesize
	\label{table:c2}
	\begin{tabular}{|l|c|}
			\hline
		\textbf{Operation} & \textbf{Complexity}   \\\hline
			$K_p \times {\boldsymbol {\mathsf {c}}}[k] = \boldsymbol{\Upsilon }_{{{\text{w}}}}^{*} {\boldsymbol {\mathsf {r}}}[k]$& $\mathcal{O}(K_p (G_{\mathrm{r}} G_{\mathrm{t}} )M L_{\mathrm{r}})$ \\
		Estimation using DnCNN  & \eqref{eq:comp_dnCNN}  \\
		$\mathop{ \max }\limits_{p} \sum_{k\in {\mathcal{K}}} |{\boldsymbol {\mathsf {h}}}^{\mathrm{v}}[k]|$ & {$\mathcal{O}(K_p(G_ t G_{\mathrm{r}}) \hat{L})$}   \\
		$\mathop{\arg \,\max }\limits_{i} \left[\sum_{k\in \mathcal{K}} \left \vert\left(\left[\boldsymbol{\Upsilon}_{\text{w}}^{\text{d}}\right]_{:,\Omega}\right)^* \boldsymbol {\mathsf {y}} _{{{\text{w}}}}[k]\right \vert\right]_{i}$ & {$\mathcal{O}({K_p} M L_{\mathrm{r}} G_{\mathrm{r}}^\mathrm{r}\hat{L})$}   \\
		$ \mathop{\arg \,\max }\limits_{i} \left[\sum_{k\in \mathcal{K}} \left \vert\left(\left[\boldsymbol{\Upsilon}_{\text{w}}^{\text{r}}\right]_{:,\Omega}\right)^* \boldsymbol {\mathsf {y}} _{{{\text{w}}}}[k]\right \vert\right]_{i}$&  {$\mathcal{O}({K_p} M L_{\mathrm{r}} G_{\mathrm{t}}^\mathrm{r} \hat{L})$}      \\
		$(K) \times {\boldsymbol {\mathsf {x}}}_{\hat {\mathcal{ T}}}[k] = \left ({\left [{\boldsymbol \Upsilon _{{{\text{w}}}}^{\mathrm{r}}}\right]_{:,\hat {\mathcal{ T}}}}\right)^\dagger {\boldsymbol {\mathsf {y}}} _{{{\text{w}}}}[k]$&  $\mathcal{O}(2 \hat{L}^2 {L}_r M +\hat{L}^3)$    \\
		$(K)\times {\boldsymbol {\mathsf {r}}}[k] = {\boldsymbol {\mathsf {y}}}_{{{\text{w}}}}[k]-\left [{\boldsymbol \Upsilon _{{{\text{w}}}}^{\mathrm{r}}}\right]_{:,\hat {\mathcal{ T}}} \hat {\tilde {\boldsymbol \xi }}[k]$ & $\mathcal{O}(K L_{\mathrm{r}}  M\hat{L})$    \\
		$\text{MSE}= \frac {1}{KML_{{{\text{r}}}}}\sum _{k=0}^{K-1}{ {\boldsymbol {\mathsf {r}}}^{*}[k] {\boldsymbol {\mathsf {r}}}[k]}$&  $\mathcal{O}(K L_{\mathrm{r}}  M\hat{L})$    \\\hline
		\textbf{Overall}&  $\mathcal{O}({K_p} M L_{\mathrm{r}} G_{\mathrm{r}}^\mathrm{r}\hat{L})$    \\\hline
	\end{tabular}
\end{table}
\normalsize

\begin{table}[t!]
	\centering
    \captionsetup{justification=centering}
	\caption{\textsc{Online Computational Complexity \\ of SW-OMP \cite{Heath2018CEMain}.}}
	\footnotesize
	\label{table:SWOMP}
	\begin{tabular}{|l|c|}
		\hline
		\textbf{Operation} & \textbf{Overall Complexity}   \\\hline
		For grid size dictionary matrices $G_{\mathrm{r}} G_{\mathrm{t}} $  &  $\mathcal{O} (K (G_{\mathrm{r}} G_{\mathrm{t}} )M L_{\mathrm{r}}  L)$    \\\hline
		For grid size dictionary matrices $G_{\mathrm{r}}^{\mathrm{r}} G_{\mathrm{t}} ^{\mathrm{r}} $  &  $ \mathcal{O}(K (G_{\mathrm{r}}^{\mathrm{r}}  G_{\mathrm{t}}^{\mathrm{r}}  )M L_{\mathrm{r}}  L)$    \\\hline
	\end{tabular}
\end{table}
\normalsize

 By using the formula for the inverse of a $2\times 2$ block matrix (from Appendix 8B in \cite{kay1993fundamentals}), the projection matrix ${\mathbf {P}}^{(n+1)}$ can be recursively written as a function of ${\mathbf {P}}^{(n)}$ as
 \begin{align} \label{equ:proj2}
&{ {\mathbf {P}}^{(n+1)}=\mathbf {P}}^{(n)}\notag \\&+\underbrace {\frac {\left ({{\mathbf {I}}_{M L_{\mathrm {r}}} \!-\! {\mathbf {P}}^{(n)}}\right)\!\left [{\boldsymbol \Upsilon _{{{\text{w}}}}}\right]_{:,{\hat {p}}^{(n+1)*}}\!\left [{\boldsymbol \Upsilon _{{{\text{w}}}}}\right]_{:,{\hat {p}}^{(n+1)*}}^{*}\!\left ({{\mathbf {I}}_{M L_{\mathrm {r}}} \!- \! {\mathbf {P}}^{(n)}}\right)}{\left [{\boldsymbol \Upsilon _{{{\text{w}}}}}\right]_{:,{\hat {p}}^{(n+1)*}}^{*}\!\left ({{\mathbf {I}}_{M L_{\mathrm {r}}}\!-\! {\mathbf {P}}^{(n)}}\right)\!\left [{\boldsymbol \Upsilon _{{{\text{w}}}}}\right]_{:,{\hat {p}}^{(n+1)*}}}}_{\boldsymbol \Delta {\mathbf {P}}^{(n+1)}}, \!\!\!\!\notag \\ {}
 \end{align}
 with $\boldsymbol \Delta {\mathbf {P}}^{(n+1)} \in \mathbb {C}^{M L_{\mathrm {r}}\times M L_{\mathrm {r}}}$ is another projection matrix that considers the relation between the projections at the $\nth{n}$ and $\nth{n+1}$ iterations. The equation in ~(\ref{equ:proj2}) can be easily shown to fulfill the orthogonality principle, i.e., ${\mathbf {P}}^{(n+1)}\boldsymbol \Delta {\mathbf {P}}^{(n+1)} = \boldsymbol 0$. The left-handed term in ~(\ref{equ:qu:proj}) then can be expressed as
 \begin{align} \label{equ:proj3}
 || {\mathbf {P}}^{(n+1)} {\boldsymbol {\mathsf {y}}}_{{{\text{w}}}}[k]||_{2}^{2}=&|| {\mathbf {P}}^{(n)} {\boldsymbol {\mathsf {y}}}_{{{\text{w}}}}[k] + \boldsymbol \Delta {\mathbf {P}}^{(n+1)} {\boldsymbol {\mathsf {y}}}_{{{\text{w}}}}[k]||_{2}^{2} \notag \\=&|| {\mathbf {P}}^{(n)} {\boldsymbol {\mathsf {y}}}_{{{\text{w}}}}[k]||_{2}^{2} + ||\boldsymbol \Delta {\mathbf {P}}^{(n+1)} {\boldsymbol {\mathsf {y}}}_{{{\text{w}}}}[k]||_{2}^{2}, \qquad \quad
  \end{align}
which satisfies the triangle equality. Moreover, $\boldsymbol \Delta {\mathbf {P}}^{(n+1)}$ is \textit{idempotent} \cite{kay1993fundamentals} in which, using straight-forward linear algebraic manipulations, it is easy to show that $\boldsymbol \Delta {\mathbf {P}}^{(n+1)}= \left(\boldsymbol{\Delta} {\mathbf {P}}^{(n+1)}\right)^{2}$. Hence, the eigen values of $\boldsymbol \Delta {\mathbf {P}}^{(n+1)}$ are either $0$ or $1$, thereby,  $|| {\mathbf {P}}^{(n+1)} {\boldsymbol {\mathsf {y}}}_{{{\text{w}}}}[k]||_{2}^{2} > || {\mathbf {P}}^{(n)} {\boldsymbol {\mathsf {y}}}_{{{\text{w}}}}[k]||_{2}^{2}$. Since the condition in (\ref{equ:qu:proj}) is satisfied, the proposed algorithms are therefore guaranteed to converge to a local optimum. Moreover, Table \ref{table:supp} shows the average number of sufficient iterations $|\hat{\mathcal{T}}|=\hat{L}$ for a range of SNR values. The results in the table confirms that the proposed support detection method using the trained DnCNN needs few iterations to converge.

\subsection{Computational Analysis}
The computational complexity for Algorithm \ref{Alg:Ch_Est} and  Algorithm \ref{Alg:Ch_Est_ref} are provided in Table \ref{table:c1} and Table \ref{table:c2}, respectively. For comparison purposes, the overall computational complexity of SW-OMP~\cite{Heath2018CEMain} benchmark is also provided in Table \ref{table:SWOMP}. Since some steps can be performed before running the channel estimation algorithms, we will distinguish between online and offline operations. For instance, the matrices $\boldsymbol \Upsilon _{{{\text{w}}}} = {\mathbf {D}}_{{{\text{w}}}}^{-*}\boldsymbol \Upsilon$, ${\mathbf {C}}_{{{\text{w}}}}$, ${\mathbf {D}}_{{{\text{w}}}}$, $\boldsymbol \Upsilon_{\text{w}}^{\text{d}} $, and $\boldsymbol \Upsilon_{\text{w}}^{\text{r}} $  can be computed offline before explicit channel estimation. 

Besides, the computational complexity of the proposed DnCNN arises from both online deployment and offline training. Although the online complexity is easier to compute, the offline training complexity is still an open issue due to a more involved implementation of the backpropagation process during training \cite{debbah2020EEPC}. Therefore, we only consider the complexity of the online deployment which is based on simple matrix-vector multiplications. 

		\begin{table}[t!]
			\centering
			\caption{\textsc{Simulation Parameters}}
			\label{table:1}
			\small
			\begin{tabular}{l|c}
				\hline
				\textbf{Parameter} & \textbf{Value}   \\\hline
				Total size of dataset  & $10,000$   \\\hline
		     	Total number of subcarriers ($K$) & {$16$}   \\\hline
		     	Subset number of subcarriers ($K_p$) & {$K/4$}   \\\hline
				Operating frequency &  $\unit[60]{GHz}$    \\\hline
				Number of TX (RX) antennas $N_{\mathrm{t}} \: (N_{\mathrm{r}})$ & $16( 64)$    \\\hline
				Number of TX (RX) RF chains $L_{\mathrm{t}} \:(L_{\mathrm{r}})$&  $2(4)$    \\\hline
				Grid size of  TX (RX)  detecting \\dictionary steering vectors  $G_ t\:(G_{\mathrm{r}})$&  $2N_{\mathrm{t}} (2N_{\mathrm{r}})$    \\\hline
				Grid size of  TX (RX) refining \\dictionary steering vectors  $G_ t^{\mathrm{r}} \:(G_{\mathrm{r}}^{\mathrm{r}})$&  $8N_{\mathrm{t}} (8N_{\mathrm{r}})$    \\\hline
				Channel paths $L$ &  $16$    \\\hline
			    Number of delay taps of the channel $N_{\mathrm{c}}$ &  $16$    \\\hline
			    Distribution of AoAs/AoDs &  $\mathcal{U}(0, \pi)$  \\\hline
			\end{tabular}
		\end{table}
		\normalsize

For a deep neural network with $L_{\mathrm{C}}$ convolutional layers \cite{simonyan2014very}, the total time complexity of is given by 
\begin{equation}
\label{eq:comp_dnCNN}
    \mathcal{O}\left(\sum_{l=1}^{L_{\mathrm{C}}} D_x^{(l)} D_y^{(l)}  D_z^{(l)} b_x^{(l)} b_y^{(l)}  c_{\mathrm{CL}}^{(l-1)} c_{\mathrm{CL}}^{(l)}\right)
\end{equation}
where $D_x^{(l)}$, $ D_y^{(l)}$ and $D_z^{(l)}$ are the convolutional kernel dimensions, $b_x^{(l)}$ and $ b_y^{(l)}$ are the dimensions of the $\nth{l}$ convolutional layer output; and $c_{\mathrm{CL}}^{(l)}$ is the number of filters in the $\nth{l}$ layer. We should also note that DL enjoys the advantages of graphics processing units (GPUs) and parallel processing, and hence, the overall time complexity is dominated by the analytical operations performed in the proposed algorithms. 

Moreover, we observe that the overall computational complexity of DL-CS-CE is lower than SW-OMP specially for small grid sizes (for instance, when $G_{\mathrm{t}}$ and $G_{\mathrm{r}}$ are twice the size of the transmit and receive antennas). Moreover, when the refined algorithm is applied with the new refining higher resolution $G_{\mathrm{t}}^{\mathrm{r}}$ and $G_{\mathrm{r}}^{\mathrm{r}}$, the computational complexity is still less than that of SW-OMP applied with the same higher resolution grid sizes applied ($G_{\mathrm{t}}^{\mathrm{r}}$ and $G_{\mathrm{r}}^{\mathrm{r}}$).  In Section \ref{sec:time_comp}, we compare the computation times of the proposed methods with that of SW-OMP.

\section{Simulation Results}\label{sec:simulation}
This section evaluates the performance of the proposed algorithms and compares empirical results with benchmark frequency-domain channel estimation algorithms, including SW-OMP \cite{Heath2018CEMain}. The results are obtained through extensive Monte Carlo simulations to evaluate the average normalized mean squared error (NMSE), and the ergodic rate as a function of SNR and the number of training frames $M$. The simulations are performed based on realistic channel realizations from Raymobtime channel datasets\footnote{Available at https://www.lasse.ufpa.br/raymobtime/}.

The main parameters used for system configuration are as follows.  The phase-shifters used in both the transmitter and the receiver are assumed to have $N_{{{\text{Q}}}}$ quantization bits, so that the entries of the training vectors ${\mathbf {f}}_{{{\text{tr}}}}^{(m)}$, ${\mathbf {w}}_{{{\text{tr}}}}^{(m)}$, $m = 1,2,\ldots,M$ are drawn from the set ${\mathcal{ A}} = \left \{{0,\frac {2\pi }{2^{N_{{{\text{Q}}}}}},\ldots,\frac {2\pi (2^{N_{{{\text{Q}}}}}-1)}{2^{N_{{{\text{Q}}}}}} }\right \}$. The number of quantization bits is set to $N_{\text{Q}}=2$. The band-limiting filter $p_{\mathrm {rc}}(t)$ is assumed to be a raised-cosine filter with roll-off factor of $0.8$. 
		 
		 The DnCNN adopted in this work has $L_{\mathrm{C}}=3$ convolutional layers. The first convolutional layer uses $c_{\mathrm{CL}}^{1}=64$ different $3  \times 3 \times 1$ filters. The succeeding convolutional layer uses $64$ different $3 \times 3 \times 64$ filters. The final convolutional layer uses one separate $3 \times 3 \times 64$ filter.
		 Moreover, we divide the dataset into the training set
and the validation set randomly, where the size of the training
set is $\unit[70]{\%}$ of the total set and the validation set is the other $\unit[30]{\%}$. 
We adopt the adaptive moment estimation (Adam) optimizer to train the DnCNN. The DnCNN is trained for $10$ epochs, where $256$ mini-batches are utilized in each epoch. The learning rate is set to $0.01$. The training process terminates when the validation accuracy does not improve in ten consecutive iterations. 
		 
		 Unless stated explicitly otherwise, the default system parameters used throughout the experimental simulations are summarized in Table \ref{table:1}, where $\mathcal{U}(\cdot,\cdot)$ represents the uniform distribution. 
		 
\begin{figure}[t!]
    \begin{subfigure}[b]{0.49\textwidth}
        \includegraphics[width=\columnwidth]{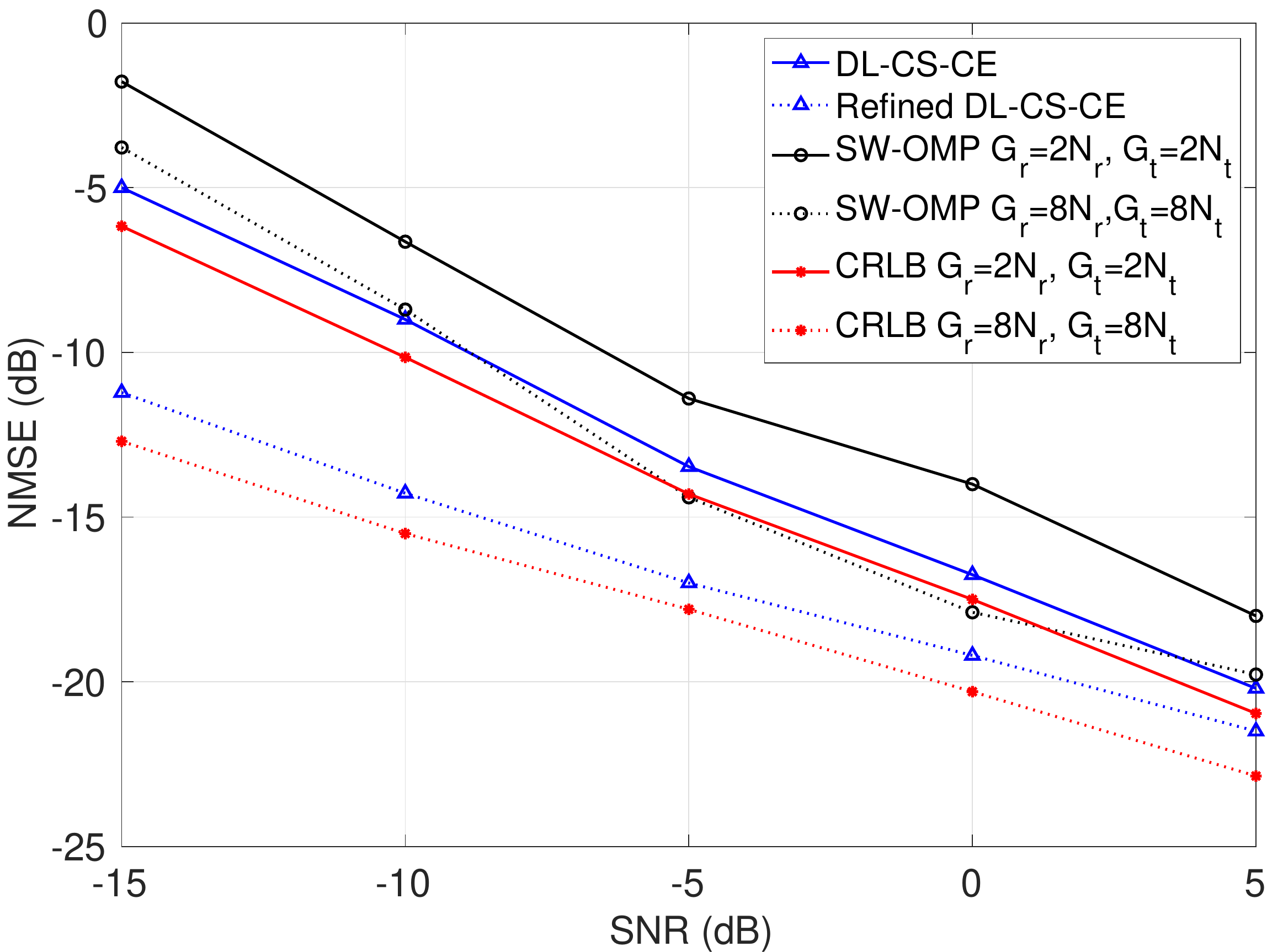}
        \caption{$M=100$}
			\label{fig:NMSESWOMP100}
    \end{subfigure}
    \begin{subfigure}[b]{0.49\textwidth}
        \includegraphics[width=\columnwidth]{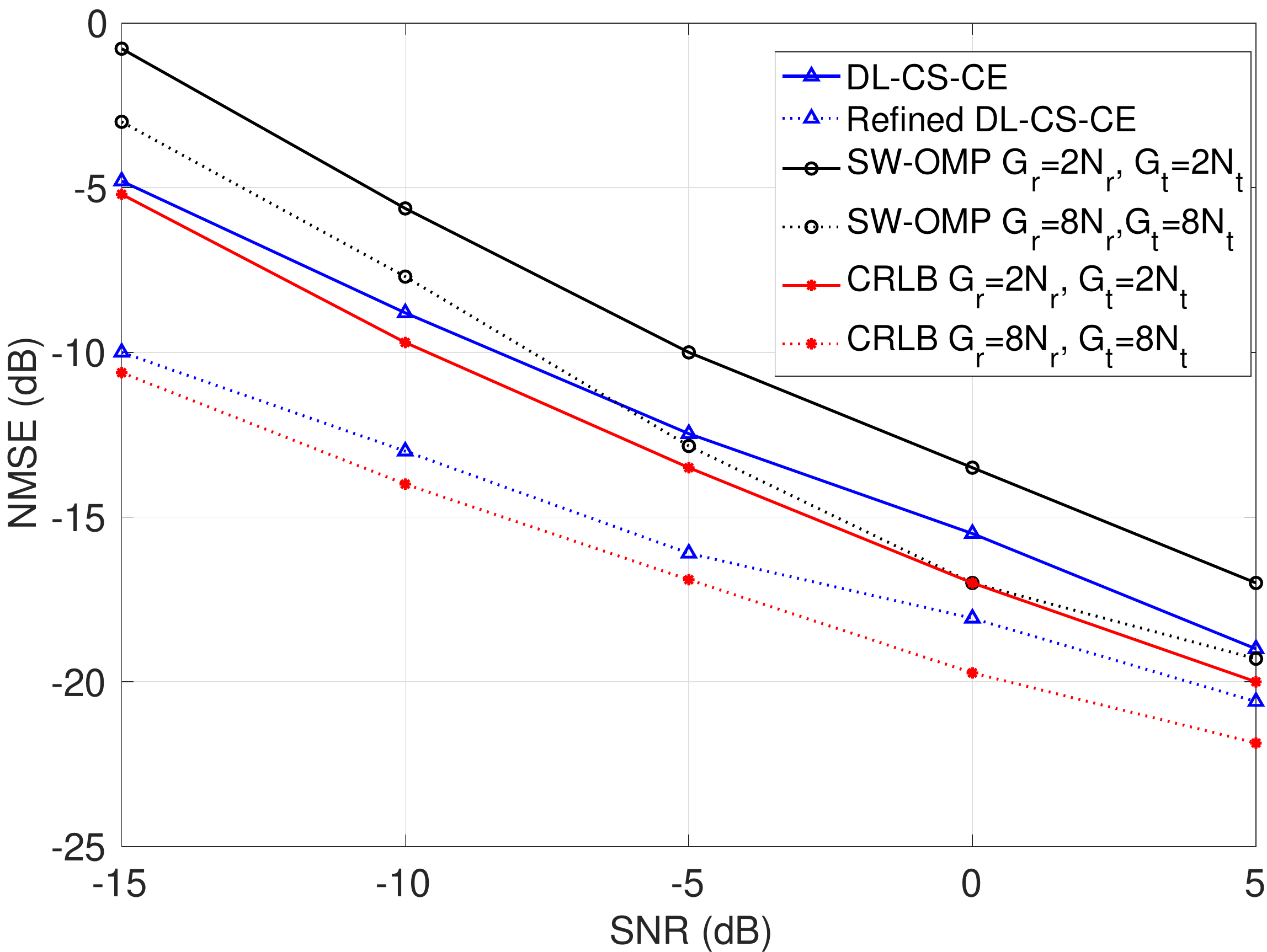}
        \caption{$M=80$}
		\label{fig:NMSESWOMP80}
    \end{subfigure}
    \begin{subfigure}[b]{0.49\textwidth}
        \includegraphics[width=\columnwidth]{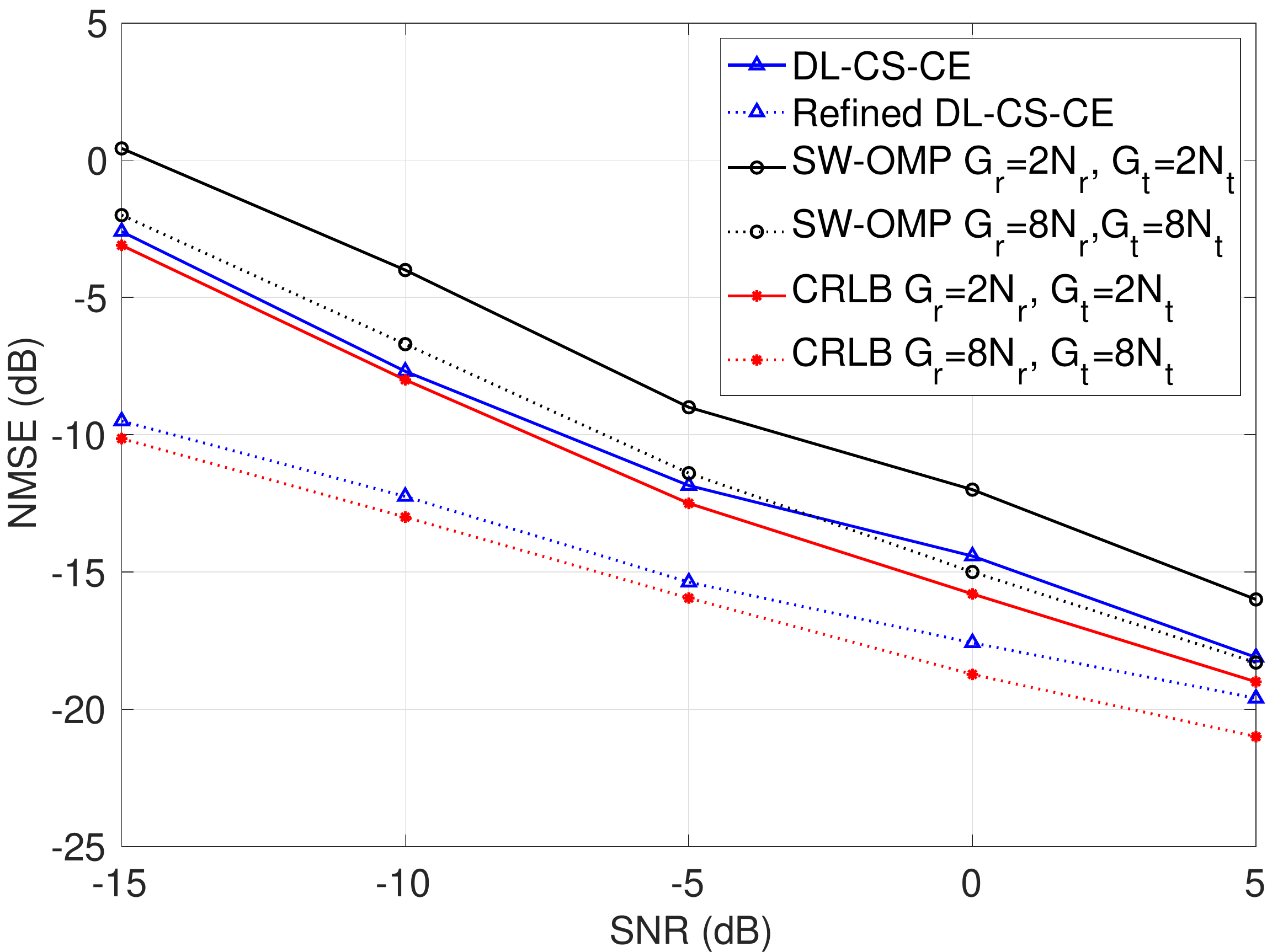}
        \caption{$M=60$}
	\label{fig:NMSESWOMP60}
    \end{subfigure}
    \caption{The NMSE vs. SNR for the DL-CS-CE, the refined DL-CS-CE, and the SW-OMP ($N_{\mathrm{t}}=16 , N_{\mathrm{r}}= 64, K=16$).}\label{fig:NMSE}
\end{figure}
	
	


\subsection{Comparison of the Normalized Mean Squared Errors}
One of the key performance metrics for the channel estimate $\hat { \boldsymbol {\mathsf {H}}}[k]$ is the NMSE, which is expressed for a given realization as
\begin{equation}
 {{\text{NMSE}}} = \frac {\sum _{k=0}^{K-1}{\|\hat { \boldsymbol {\mathsf {H}}}[k]- \boldsymbol {\mathsf {H}}[k]\|_{F}^{2}}}{\sum _{k=0}^{K-1}{\| \boldsymbol {\mathsf {H}}[k]\|_{F}^{2}}}. 
 \end{equation}
 The NMSE is considered our baseline metric to compute the proposed algorithms' performance and will be averaged over many channel realizations. The normalized CRLB (NCRLB), from which the supports are perfectly estimated \cite{Heath2018CEMain}, is also provided to compare each algorithm's average performance with the lowest achievable NMSE.

We compare the average NMSE versus SNR obtained for the different channel estimation algorithms in Figs. \ref{fig:NMSE} for a practical SNR range of $\unit[-15]{dB}$ to $\unit[5]{dB}$ and three different lengths of training frames $M=\{100,80,60\}$. It is worth noting that the choice of the SNR range is based on the fact that the SNR expected in mmWave communication systems is in the order of  $\unit[-20]{dB}$  up to  $\unit[0]{dB}$.  Using a large number of training frames $M$ increases performance at the cost of both higher overhead and computational complexity since the complexity of estimating the support, channel gains, and noise variance grows linearly with $L_{\mathrm{r}} M$.  

In Fig. \ref{fig:NMSE}, DL-CS-CE with refining performs the best, achieving NMSE values very close to the NCRLB (around $\unit[1]{dB}$ gap). 
The performance difference between SW-OMP and proposed algorithms is noticeable, which comes from the fact that SW-OMP estimates the mmWave channel dominant entries sequentially rather than at a single shot. The DL-CS-CE obviously deliver an NMSE lower than that of SW-OMP by $\unit[-3]{dB}$. The refined DL-CS-CE achieves even lower NMSE values below $\unit[-10]{dB}$ especially for low SNR values such as $\text{SNR}=\unit[-15]{dB}$ whereas SW-OMP with higher resolution grid sizes achieves NMSE around $\unit[-3]{dB}$ and $\unit[-4]{dB}$ for $\text{SNR}=\unit[-15]{dB}$.

\begin{figure}[t]
	\includegraphics[width=\columnwidth]{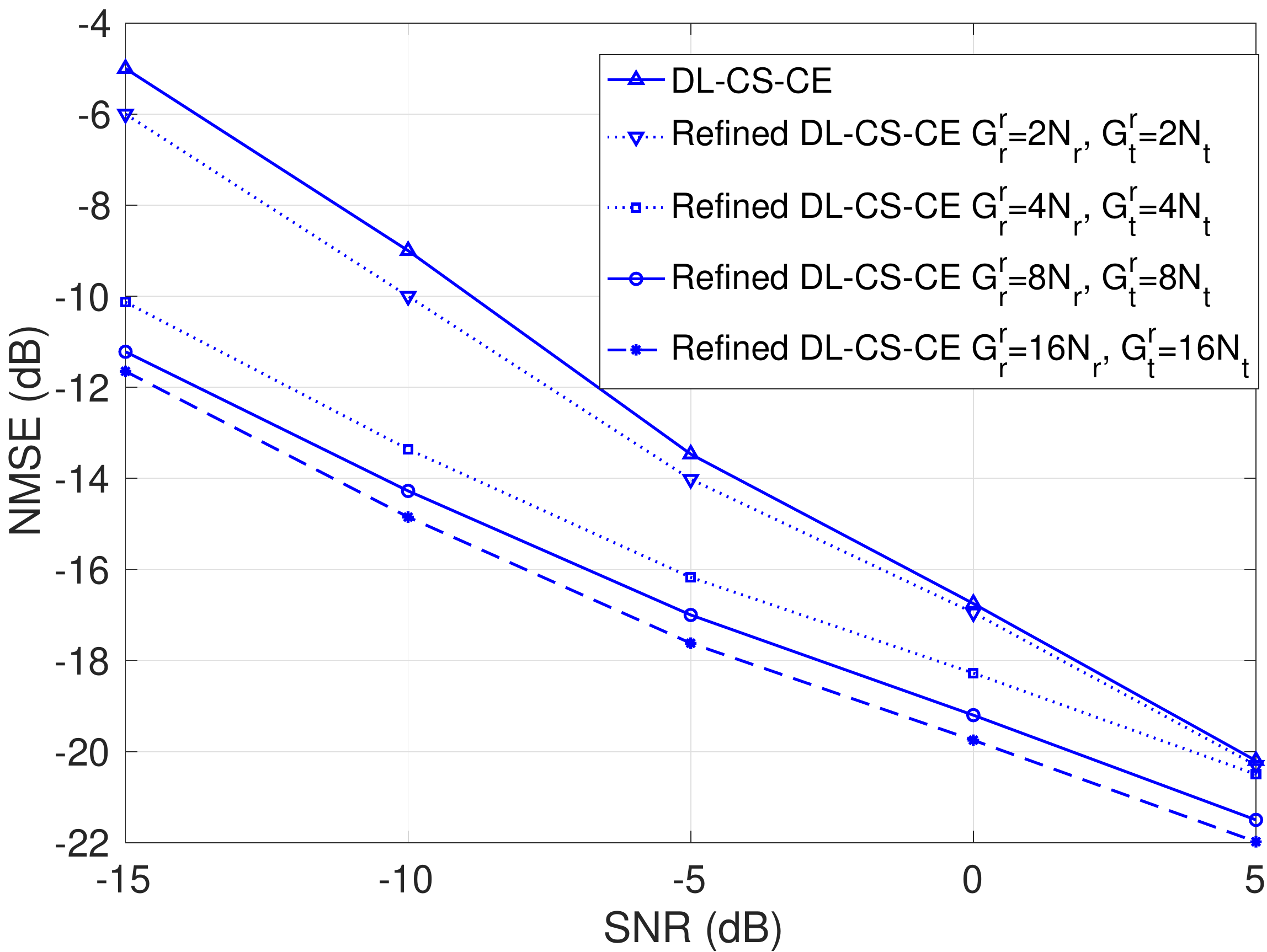}
	\caption{The NMSE vs. SNR for the DL-CS-CE and the refined DL-CS-CE under different refining grid sizes of $G_{\mathrm{r}}^{\mathrm{r}}=\{2 N_{\mathrm{r}}, 4 N_{\mathrm{r}}, 8 N_{\mathrm{r}}, 16 N_{\mathrm{r}}\}$ and $G_{\mathrm{t}}^{\mathrm{r}}=\{2 N_{\mathrm{t}}, 4 N_{\mathrm{t}}, 8 N_{\mathrm{t}}, 16 N_{\mathrm{t}}\}$ ($N_{\mathrm{t}}=16 , N_{\mathrm{r}}= 64, K=16, M=100$).}
	\label{fig:NMSE100_r}
\end{figure}

In Fig. \ref{fig:NMSE100_r}, we compare the NMSE of the DL-CS-CE with $G_{\mathrm{r}}=2 N_{\mathrm{r}}$ and $G_{\mathrm{t}}=2 N_{\mathrm{t}}$ and the refined DL-CS-CE with refining grid sizes $G_{\mathrm{r}}^{\mathrm{r}}=\{2 N_{\mathrm{r}}, 4 N_{\mathrm{r}}, 8N_{\mathrm{r}}, 16N_{\mathrm{r}}\}$ and $G_{\mathrm{t}}^{\mathrm{r}}=\{2 N_{\mathrm{t}}, 4 N_{\mathrm{t}}, 8N_{\mathrm{t}}, 16N_{\mathrm{t}}\}$. It is obvious from Fig. \ref{fig:NMSE100_r} that setting the dictionary sizes to twice the number of antennas at transmitter and receiver is not enough to estimate the exact AoDs/AoAs that lie in the off grid regions of the dictionary. 
At this very point, the refining method introduced in Algorithm \ref{Alg:Ch_Est_ref} is shown to greatly enhance the NMSE performance especially for the low SNR regime, at the cost of increased computational complexity as the refining resolution increases as shown in Table \ref{table:c2}. 
Hence, a trade-off exists between attaining good NMSE performance and keeping the computational complexity order low. However, even with the proposed refining approach, the complexity remains lower than that of SW-OMP for the same high resolution dictionary matrices by at least two orders of magnitude. For instance, by taking $M=100, K_p= K/4, G_{\mathrm{t}}^{\mathrm{r}}=8N_{\mathrm{t}}$, and $G_{\mathrm{r}}^{\mathrm{r}}=8N_{\mathrm{r}}$, the complexity order of SW-OMP is $\mathcal{O}(K (G_{\mathrm{r}}^{\mathrm{r}}  G_{\mathrm{t}}^{\mathrm{r}}  )M L_{\mathrm{r}}  L)= \mathcal{O}( 6.7 \times 10^9)$, while the complexity order of the refined DL-CS-CE  is $\mathcal{O}(K_p G_{\mathrm{r}}^{\mathrm{r}} M L_{\mathrm{r}}  L)= \mathcal{O}( 1.3 \times 10^7)$.
Moreover, Fig. \ref{fig:NMSE100_r} shows that as the refining resolution increases (i.e., $G_{\mathrm{r}}^{\mathrm{r}}> 8 N_{\mathrm{r}}$, $G_{\mathrm{t}}^{\mathrm{r}}> 8N_{\mathrm{t}}$), the NMSE enhancement becomes gradual as no further gains are attained from further refinement.

\subsection{Comparisons for the Probability of Successful Support Estimation for $L$ Paths}

\begin{figure}[t]
	\includegraphics[width=\columnwidth]{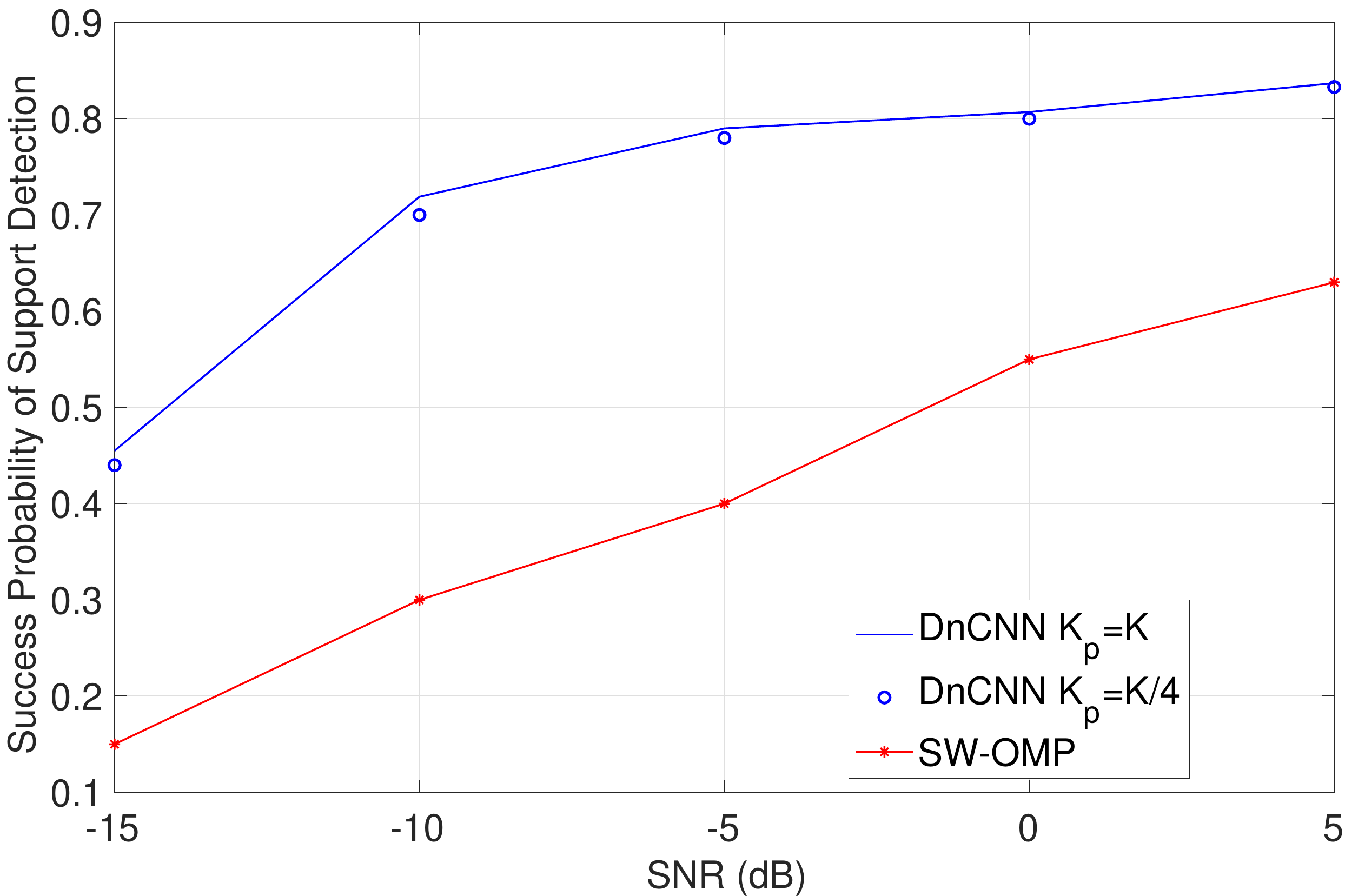}
	\caption {Probability of successfully detecting the supports vs. SNR for the DL-CS-CE, the refined DL-CS-CE, and the SW-OMP ($N_{\mathrm{t}}=16 , N_{\mathrm{r}}= 64, K=16, M=100$).}
	\label{fig:sucProb}
\end{figure}

In Fig.~\ref{fig:sucProb}, we compare the successful support detection probability versus SNR for the proposed DnCNN-based amplitude estimation and that of SW-OMP. It can be seen that the proposed DnCNN outperforms SW-OMP over the whole SNR range as the trained DnCNN can efficiently denoise the correlated input image and obtain a sparse matrix of the channel amplitudes. From this denoised sparse matrix, the indices of the supports (i.e., dominant entries of th obtained sparse matrix) are detected. Moreover, we show that when we set $K_p \ll K$, the support detection is not affected, since as shown in Section \ref{sec:MeasMat} $\boldsymbol{\Delta }[k]$ have the same support for all $k$. Therefore, we can reduce computational complexity since there is no need to compute the correlation step (given in ~(\ref{equ:corrw}) for all subcarriers. Thus, a smaller subset of subcarriers can also provide a high probability of correct support detection.

\subsection{Spectral Efficiency Comparison}

Another key performance metric is the spectral efficiency, which is computed by assuming fully-digital precoding and combining.  In this way, using estimates for the $N_{\mathrm{s}}$ dominant left and right singular vectors of the channel estimate gives $K$ parallel effective channels $\boldsymbol {\mathsf {H}}_{{{\text{eff}}}}[k] = \left [{\hat { \boldsymbol {\mathsf {U}}}[k]}\right]_{:,1: N_{\mathrm {s}}}^{*} \boldsymbol {\mathsf {H}}[k]\left [{\hat { \boldsymbol {\mathsf {V}}}[k]}\right]_{:,1: N_{\mathrm {s}}}$. Accordingly, the average spectral efficiency can be expressed as 
\begin{equation} R = \frac {1}{K}\sum _{k=0}^{K-1} \sum _{n=1}^{N_{{{\text{s}}}}} \log_2 \left ({1 + \frac {{{\text{SNR}}}}{N_{{{\text{s}}}}} \lambda _{n}(\boldsymbol {\mathsf {H}}_{{{\text{eff}}}}[k])^{2} }\right), \end{equation}
with $\lambda _{n}(\boldsymbol {\mathsf {H}}_{{{\text{eff}}}}[k])$, $n = 1,\ldots, N_{\mathrm {s}} $ the eigenvalues of each effective channel $\boldsymbol {\mathsf {H}}_{{{\text{eff}}}}[k]$.
		
\begin{figure}[t]
	\includegraphics[width=\columnwidth]{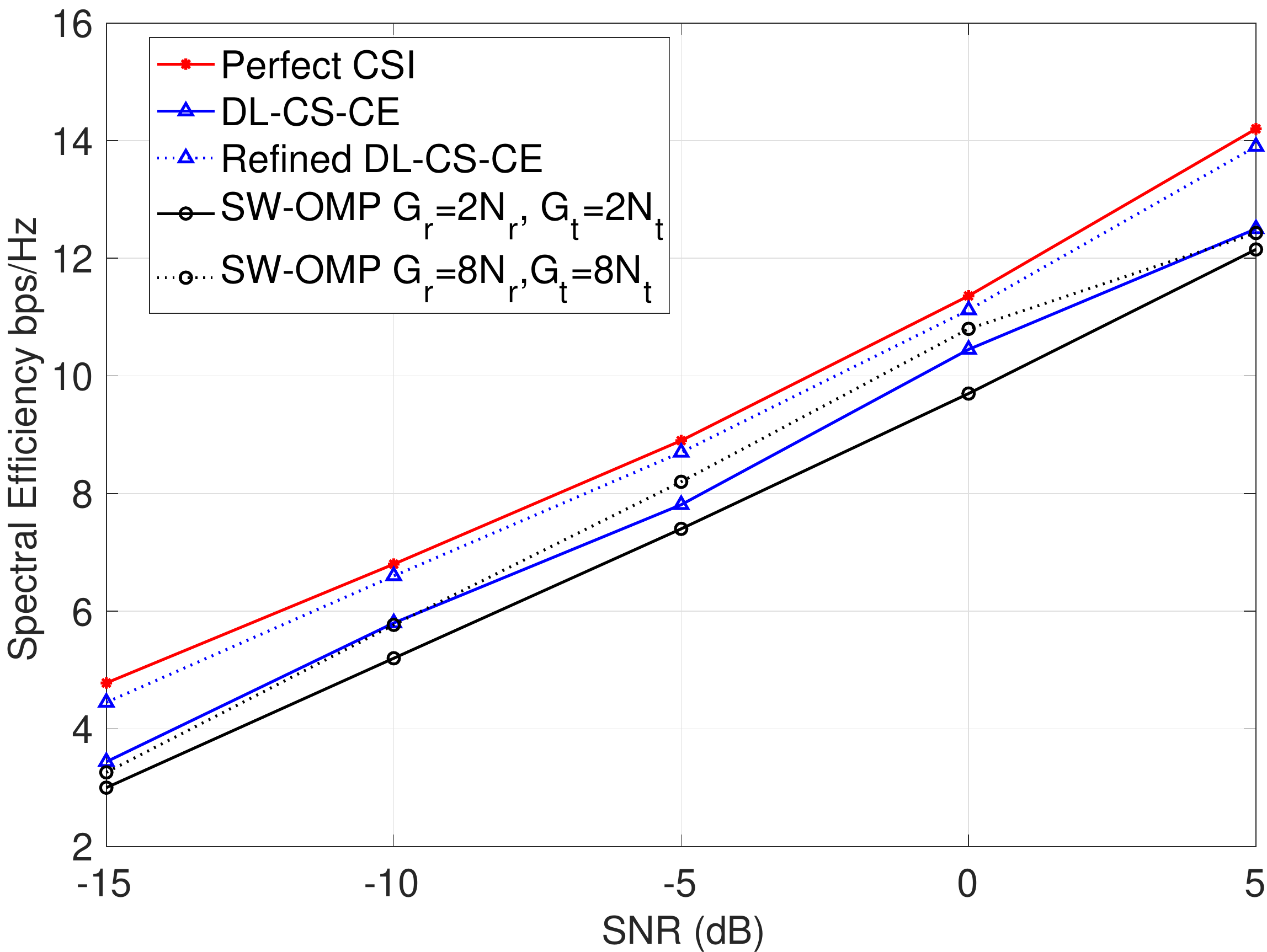}
	\caption{Spectral efficiency vs. SNR ($N_{\mathrm{t}}=16 , N_{\mathrm{r}}= 64, K=16, M=100$).}
	\label{fig:SE100}
\end{figure}

\begin{figure}[t]
	\includegraphics[width=\columnwidth]{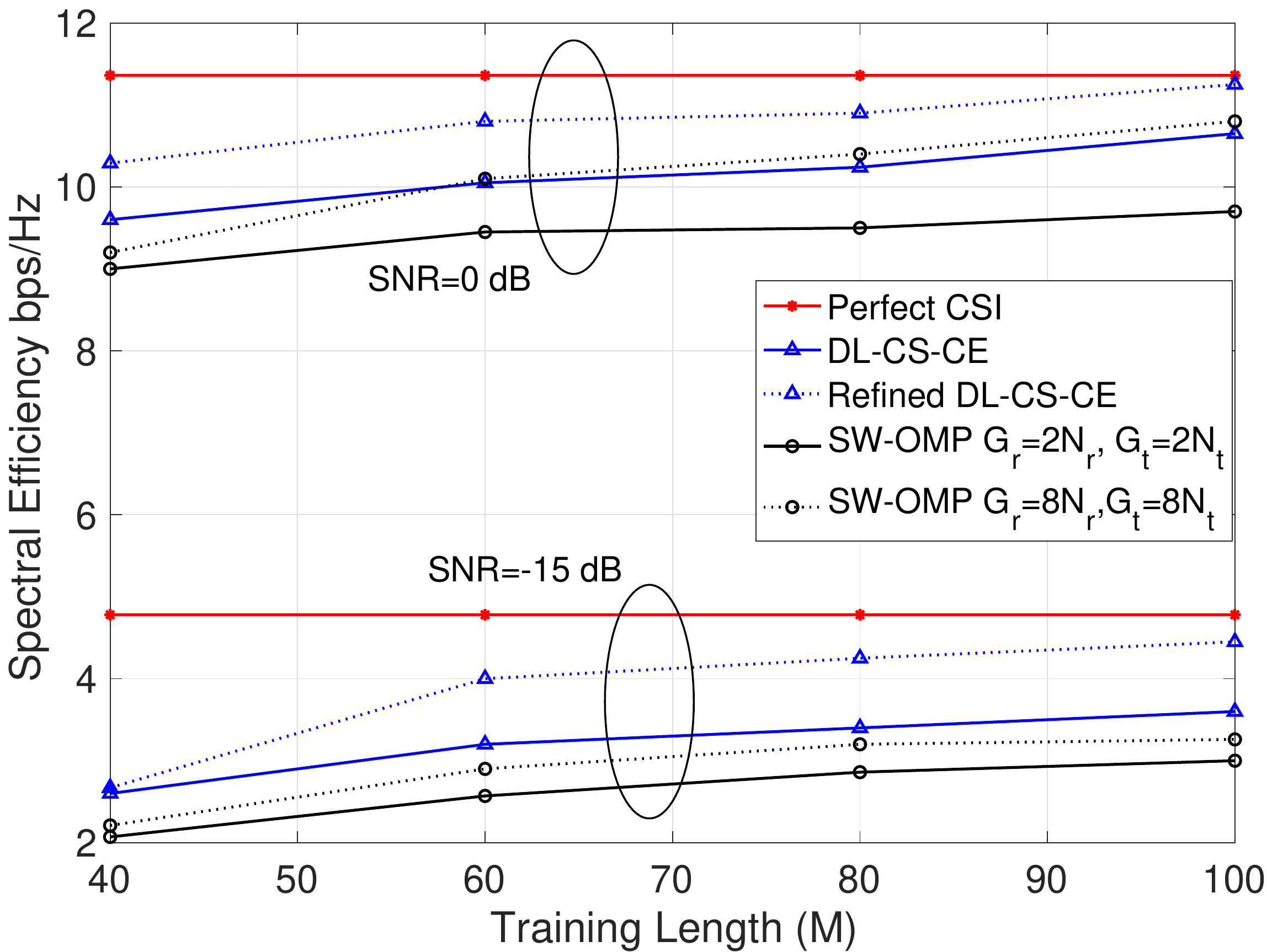}
	\caption{Spectral efficiency vs. $M$ training lengths ($N_{\mathrm{t}}=16 , N_{\mathrm{r}}= 64, K=16, \text{SNR}=\{\unit[-15, 0]{dB}\} $).}
	\label{fig:SEM}
\end{figure}		

In Fig.\ref{fig:SE100}, we show the achievable spectral efficiency as a function of the SNR for the different channel estimation algorithms. The proposed DL-CS-CE approach provides at least $3.6\%$  performance improvement over the SW-OMP. The refined DL-CS-CE provides near-optimal achievable rates with at least $12.6\%$ performance improvement over the other schemes. The spectral efficiency gap of the different schemes is smaller than that of the NMSE gap, since the NMSE performance is much more sensitive to the success rate of the sparse recovery. However, the spectral efficiency performance is determined by the beamforming gain and is less sensitive to the success rate of the sparse recovery.
		
In Fig. \ref{fig:SEM}, we show the achievable spectral efficiency as a function of different training lengths for the proposed schemes under different SNRs.  We observe that using $M > 40$ frames does not significantly improve performance, which leverages the robustness of the two proposed approaches. Simulations also show that near-optimal achievable rates can be achieved by using a reasonable number of frames, i.e., $60 \leq M \leq 100$. Therefore, with the proposed schemes, we can save in training overhead.

\subsection{Time Complexity Analysis}\label{sec:time_comp}
Table \ref{table:time} shows online estimation stage computational times of the proposed frameworks and SW-OMP \cite{Heath2018CEMain}. SW-OMP is the slowest to solve the inherent optimization problem, especially for high-resolution dictionary matrices.  The running time of the DL-CS-CE without refining exhibits shorter computational times than the SW-OMP algorithm. However, for fair comparison when refining is applied, we compare the running time of the refined DL-CS-CE with higher resolution SW-OMP where $G_{\mathrm{r}}=G_{\mathrm{r}}^{\mathrm{r}}=8N_{\mathrm{r}}$ and $G_{\mathrm{t}}=G_{\mathrm{t}}^{\mathrm{r}}=8N_{\mathrm{t}}$, and it is shown that the refined DL-CS-CE takes less time to perform the channel estimation. Hence, we conclude that the proposed DL-CS-CE frameworks are computationally efficient and tolerant, especially for higher resolution dictionary matrices.

\begin{table}[t]
	\centering
\captionsetup{justification=centering}
	\caption{\textsc{Average Running Time for $M=100$ and SNR$=\unit[-5]{dB}$}}
		\footnotesize
	\begin{tabular}{|l|c|}
		\hline
		\textbf{Algorithm} & \textbf{ Run time [seconds]} \\\hline
		DL-CS-CE \newline $G_{\mathrm{r}}=2N_{\mathrm{r}}$ and $G_{\mathrm{t}}=2N_{\mathrm{t}}$ &  $0.144$\\\hline
		Refined DL-CS-CE $G_{\mathrm{r}}^{\mathrm{r}}=2N_{\mathrm{r}}$ and $G_{\mathrm{t}}^{\mathrm{r}}=2N_{\mathrm{t}}$ & $0.201$ \\\hline
		Refined DL-CS-CE $G_{\mathrm{r}}^{\mathrm{r}}=8N_{\mathrm{r}}$ and $G_{\mathrm{t}}^{\mathrm{r}}=8N_{\mathrm{t}}$ & $0.464$ \\\hline
		SW-OMP for grids $G_{\mathrm{r}}=2N_{\mathrm{r}}$ and $G_{\mathrm{t}}=2N_{\mathrm{t}}$  &  $0.25$\\\hline
		SW-OMP for grids $G_{\mathrm{r}}=8N_{\mathrm{r}}$ and $G_{\mathrm{t}}=8N_{\mathrm{t}}$ & $0.97$ \\\hline
	\end{tabular}
	\label{table:time}
\end{table}

\section{Conclusion}\label{sec:conclusion}
In this work, we have proposed two DL-CS-based frequency-selective channel estimation approaches for mmWave wideband communication systems under hybrid architectures. The developed algorithms are based on joint-sparse recovery to exploit information on the common basis shared for every subcarrier. Compared to the state-of-the-art channel estimation techniques that estimate supports iteratively, the proposed solutions reduce computational complexity and estimation error by detecting all supports simultaneously.  Simulation results have shown that the DL-CS-CE and the refined DL-CS-CE schemes have better channel estimation performance than existing schemes using a reasonably small training length and low complexity order. It has also been shown that a small number of subcarriers are sufficient for successful support detection during the deep learning prediction phase. Thus, the proposed schemes are able to attain good NMSE performance with low computational complexity.


\bibliographystyle{IEEEtran}
\bibliography{IEEEabrv,bio1}

\begin{thebibliography}{10}
\providecommand{\url}[1]{#1}
\csname url@samestyle\endcsname
\providecommand{\newblock}{\relax}
\providecommand{\bibinfo}[2]{#2}
\providecommand{\BIBentrySTDinterwordspacing}{\spaceskip=0pt\relax}
\providecommand{\BIBentryALTinterwordstretchfactor}{4}
\providecommand{\BIBentryALTinterwordspacing}{\spaceskip=\fontdimen2\font plus
\BIBentryALTinterwordstretchfactor\fontdimen3\font minus
  \fontdimen4\font\relax}
\providecommand{\BIBforeignlanguage}[2]{{%
\expandafter\ifx\csname l@#1\endcsname\relax
\typeout{** WARNING: IEEEtran.bst: No hyphenation pattern has been}%
\typeout{** loaded for the language `#1'. Using the pattern for}%
\typeout{** the default language instead.}%
\else
\language=\csname l@#1\endcsname
\fi
#2}}
\providecommand{\BIBdecl}{\relax}
\BIBdecl

\bibitem{Pi2011InroMmwave}
Z.~{Pi} and F.~{Khan}, ``An introduction to millimeter-wave mobile broadband
  systems,'' \emph{{IEEE} Commun. Mag.}, vol.~49, no.~6, pp. 101--107, Jun.
  2011.

\bibitem{Heath2016OverviewMmwave}
R.~W. {Heath}, N.~{González-Prelcic}, S.~{Rangan}, W.~{Roh}, and A.~M.
  {Sayeed}, ``An overview of signal processing techniques for {M}illimeter
  {W}ave {MIMO} systems,'' \emph{{IEEE} J. Sel. Topics Signal Process.},
  vol.~10, no.~3, pp. 436--453, Apr. 2016.

\bibitem{Khateeb2014covcap}
T.~{Bai}, A.~{Alkhateeb}, and R.~W. {Heath}, ``Coverage and capacity of
  millimeter-wave cellular networks,'' \emph{{IEEE} Commun. Mag.}, vol.~52,
  no.~9, pp. 70--77, Sep. 2014.

\bibitem{Andrews20145g}
J.~G. {Andrews}, S.~{Buzzi}, W.~{Choi}, S.~V. {Hanly}, A.~{Lozano}, A.~C.~K.
  {Soong}, and J.~C. {Zhang}, ``What will 5g be?'' \emph{{IEEE} J. Sel. Areas
  Commun.}, vol.~32, no.~6, pp. 1065--1082, Jun. 2014.

\bibitem{Khateeb2014mmwave}
A.~{Alkhateeb}, J.~{Mo}, N.~{Gonzalez-Prelcic}, and R.~W. {Heath}, ``Mimo
  precoding and combining solutions for millimeter-wave systems,'' \emph{{IEEE}
  Commun. Mag.}, vol.~52, no.~12, pp. 122--131, Dec. 2014.

\bibitem{Heath2016shiftOrSwitches}
R.~{Méndez-Rial}, C.~{Rusu}, N.~{González-Prelcic}, A.~{Alkhateeb}, and R.~W.
  {Heath}, ``Hybrid {MIMO} architectures for millimeter wave communications:
  Phase shifters or switches?'' \emph{IEEE Access}, vol.~4, pp. 247--267, Jan.
  2016.

\bibitem{Khateeb2014Chanest}
A.~{Alkhateeb}, O.~{El Ayach}, G.~{Leus}, and R.~W. {Heath}, ``Channel
  estimation and hybrid precoding for millimeter wave cellular systems,''
  \emph{{IEEE} J. Sel. Topics Signal Process.}, vol.~8, no.~5, pp. 831--846,
  Oct. 2014.

\bibitem{eldar2011CS}
M.~F. {Duarte} and Y.~C. {Eldar}, ``Structured compressed sensing: From theory
  to applications,'' \emph{{IEEE} Trans. Signal Process.}, vol.~59, no.~9, pp.
  4053--4085, Sep. 2011.

\bibitem{Gao2016CEfreqSelec}
Z.~{Gao}, C.~{Hu}, L.~{Dai}, and Z.~{Wang}, ``Channel estimation for
  millimeter-wave massive {MIMO} with hybrid precoding over frequency-selective
  fading channels,'' \emph{{IEEE} Commun. Lett.}, vol.~20, no.~6, pp.
  1259--1262, Jun. 2016.

\bibitem{Heath2017TDCE}
K.~{Venugopal}, A.~{Alkhateeb}, R.~W. {Heath}, and N.~G. {Prelcic},
  ``Time-domain channel estimation for wideband millimeter wave systems with
  hybrid architecture,'' in \emph{Proc. IEEE Int. Conf. Acoustics, Speech, and
  Signal Process. (ICASSP)}, 2017, pp. 6493--6497.

\bibitem{Heath2018CEMain}
J.~{Rodríguez-Fernández}, N.~{González-Prelcic}, K.~{Venugopal}, and R.~W.
  {Heath}, ``Frequency-domain compressive channel estimation for
  frequency-selective hybrid millimeter wave{MIMO} systems,'' \emph{{IEEE}
  Trans. Wireless Commun.}, vol.~17, no.~5, pp. 2946--2960, May 2018.

\bibitem{Ma2018CE}
W.~{Ma} and C.~{Qi}, ``Beamspace channel estimation for millimeter wave massive
  {MIMO} system with hybrid precoding and combining,'' \emph{{IEEE} Trans.
  Signal Process.}, vol.~66, no.~18, pp. 4839--4853, Sep. 2018.

\bibitem{Ye2018PwrDLCE}
H.~{Ye}, G.~Y. {Li}, and B.~{Juang}, ``Power of deep learning for channel
  estimation and signal detection in ofdm systems,'' \emph{{IEEE} Microw.
  Wireless Compon. Lett.}, vol.~7, no.~1, pp. 114--117, Feb. 2018.

\bibitem{DOng2019DLDNNCE}
P.~{Dong}, H.~{Zhang}, G.~Y. {Li}, I.~S. {Gaspar}, and N.~{Naderi Alizadeh},
  ``Deep {CNN}-based channel estimation for mmwave massive {MIMO} systems,''
  \emph{{IEEE} J. Sel. Topics Signal Process.}, vol.~13, no.~5, pp. 989--1000,
  Sep. 2019.

\bibitem{He2018DLCE}
H.~{He}, C.~{Wen}, S.~{Jin}, and G.~Y. {Li}, ``Deep learning-based channel
  estimation for beamspace mmwave massive {MIMO} systems,'' \emph{{IEEE}
  Commun. Lett.}, vol.~7, no.~5, pp. 852--855, Oct. 2018.

\bibitem{M2019DLCE}
M.~{Soltani}, V.~{Pourahmadi}, A.~{Mirzaei}, and H.~{Sheikhzadeh}, ``Deep
  learning-based channel estimation,'' \emph{{IEEE} Commun. Lett.}, vol.~23,
  no.~4, pp. 652--655, Apr. 2019.

\bibitem{Ma2020SparseDLCE}
W.~{Ma}, C.~{Qi}, Z.~{Zhang}, and J.~{Cheng}, ``Sparse channel estimation and
  hybrid precoding using deep learning for millimeter wave massive {MIMO},''
  \emph{{IEEE} Trans. Commun.}, vol.~68, no.~5, pp. 2838--2849, May 2020.

\bibitem{wei2019knowledge}
X.~Wei, C.~Hu, and L.~Dai, ``Knowledge-aided deep learning for beamspace
  channel estimation in millimeter-wave massive {MIMO} systems,'' \emph{arXiv
  preprint arXiv:1910.12455}, 2019.

\bibitem{Chun2019DLCEmassiveMIMO}
C.~{Chun}, J.~{Kang}, and I.~{Kim}, ``Deep learning-based channel estimation
  for massive {MIMO} systems,'' \emph{{IEEE} Microw. Wireless Compon. Lett.},
  vol.~8, no.~4, pp. 1228--1231, Aug. 2019.

\bibitem{Jin2019CellFreeDLCE}
Y.~{Jin}, J.~{Zhang}, S.~{Jin}, and B.~{Ai}, ``Channel estimation for cell-free
  mmwave massive {MIMO} through deep learning,'' \emph{{IEEE} Trans. Veh.
  Technol.}, vol.~68, no.~10, pp. 10\,325--10\,329, Oct. 2019.

\bibitem{andrew2020MIMOCEDL}
E.~{Balevi}, A.~{Doshi}, and J.~G. {Andrews}, ``Massive mimo channel estimation
  with an untrained deep neural network,'' \emph{{IEEE} Trans. Wireless
  Commun.}, vol.~19, no.~3, pp. 2079--2090, Mar. 2020.

\bibitem{Bjornson2020CEbayes}
{\"O}.~T. Demir and E.~Bj{\"o}rnson, ``Channel estimation in massive mimo under
  hardware non-linearities: Bayesian methods versus deep learning,'' \emph{IEEE
  Open Journal of the Commun. Soc.}, vol.~1, pp. 109--124, 2020.

\bibitem{long2018DD}
Y.~{Long}, Z.~{Chen}, J.~{Fang}, and C.~{Tellambura}, ``Data-driven-based
  analog beam selection for hybrid beamforming under mm-wave channels,''
  \emph{{IEEE} J. Sel. Topics Signal Process.}, vol.~12, no.~2, pp. 340--352,
  May 2018.

\bibitem{Hodge2019RFmeta}
J.~A. {Hodge}, K.~{Vijay Mishra}, and A.~I. {Zaghloul}, ``Multi-discriminator
  distributed generative model for multi-layer rf metasurface discovery,'' in
  \emph{Proc. IEEE Global Conf. on Signal and Inform. Process. (GlobalSIP)},
  Ottawa, ON, Canada, 2019, pp. 1--5.

\bibitem{Khateeb2018DLbeam}
A.~{Alkhateeb}, S.~{Alex}, P.~{Varkey}, Y.~{Li}, Q.~{Qu}, and D.~{Tujkovic},
  ``Deep learning coordinated beamforming for highly-mobile millimeter wave
  systems,'' \emph{IEEE Access}, vol.~6, pp. 37\,328--37\,348, 2018.

\bibitem{Huang2019DLHP}
H.~{Huang}, Y.~{Song}, J.~{Yang}, G.~{Gui}, and F.~{Adachi},
  ``Deep-learning-based millimeter-wave massive {MIMO} for hybrid precoding,''
  \emph{{IEEE} Trans. Veh. Technol.}, vol.~68, no.~3, pp. 3027--3032, Mar.
  2019.

\bibitem{Elbir2019CNNHP}
A.~M. {Elbir}, ``Cnn-based precoder and combiner design in mmwave {MIMO}
  systems,'' \emph{{IEEE} Commun. Lett.}, vol.~23, no.~7, pp. 1240--1243, Jul.
  2019.

\bibitem{Elbir2020JASHP}
A.~M. {Elbir} and K.~V. {Mishra}, ``Joint antenna selection and hybrid
  beamformer design using unquantized and quantized deep learning networks,''
  \emph{{IEEE} Trans. Wireless Commun.}, vol.~19, no.~3, pp. 1677--1688, Mar.
  2020.

\bibitem{elbir2019online}
A.~M. Elbir and K.~V. Mishra, ``Online and offline deep learning strategies for
  channel estimation and hybrid beamforming in multi-carrier mm-wave massive
  {MIMO} systems,'' \emph{arXiv preprint arXiv:1912.10036}, 2019.

\bibitem{dorner2018DLair}
S.~{Dörner}, S.~{Cammerer}, J.~{Hoydis}, and S.~t.~{Brink}, ``Deep learning
  based communication over the air,'' \emph{{IEEE} J. Sel. Topics Signal
  Process.}, vol.~12, no.~1, pp. 132--143, Feb. 2018.

\bibitem{Xu2019DLCEMultiuser}
J.~{Xu}, P.~{Zhu}, J.~{Li}, and X.~{You}, ``Deep learning-based pilot design
  for multi-user distributed massive {MIMO} systems,'' \emph{{IEEE} Commun.
  Lett.}, vol.~8, no.~4, pp. 1016--1019, Aug. 2019.

\bibitem{Kang2018DLCEEnergy}
J.~{Kang}, C.~{Chun}, and I.~{Kim}, ``Deep-learning-based channel estimation
  for wireless energy transfer,'' \emph{{IEEE} Commun. Lett.}, vol.~22, no.~11,
  pp. 2310--2313, Nov. 2018.

\bibitem{Asmaa2020CF}
A.~{Abdallah} and M.~M. {Mansour}, ``Efficient angle-domain processing for
  fdd-based cell-free massive mimo systems,'' \emph{IEEE Transactions on
  Communications}, vol.~68, no.~4, pp. 2188--2203, Apr. 2020.

\bibitem{Asmaa2019CF}
------, ``Angle-based multipath estimation and beamforming for fdd cell-free
  massive mimo,'' in \emph{Proc. IEEE Int. Sig. Process. Advances in Wireless
  Commun. Workshop (SPAWC)}, Cannes, France, 2019, pp. 1--5.

\bibitem{Heath2016FSF}
A.~{Alkhateeb} and R.~W. {Heath}, ``Frequency selective hybrid precoding for
  limited feedback millimeter wave systems,'' \emph{{IEEE} Trans. Commun.},
  vol.~64, no.~5, pp. 1801--1818, May 2016.

\bibitem{emil2019sub6mmWave}
E.~{Bjornson}, L.~{Van der Perre}, S.~{Buzzi}, and E.~G. {Larsson}, ``Massive
  {MIMO} in sub-6 ghz and mm{W}ave: Physical, practical, and use-case
  differences,'' \emph{{IEEE} Trans. Wireless Commun.}, vol.~26, no.~2, pp.
  100--108, Apr. 2019.

\bibitem{Zhang2017Dncnn}
K.~{Zhang}, W.~{Zuo}, Y.~{Chen}, D.~{Meng}, and L.~{Zhang}, ``Beyond a gaussian
  denoiser: Residual learning of deep cnn for image denoising,'' \emph{{IEEE}
  Trans. Image Process.}, vol.~26, no.~7, pp. 3142--3155, Jul. 2017.

\bibitem{he2016deep}
K.~He, X.~Zhang, S.~Ren, and J.~Sun, ``Deep residual learning for image
  recognition,'' in \emph{Proc. IEEE Conf.on Computer Vision and Pattern
  Recognition}, Las Vegas, NV, 2016, pp. 770--778.

\bibitem{kay1993fundamentals}
S.~M. Kay, \emph{Fundamentals of statistical signal processing}.\hskip 1em plus
  0.5em minus 0.4em\relax Prentice Hall PTR, 1993.

\bibitem{Jutt2011CRLB}
R.~{Niazadeh}, M.~{Babaie-Zadeh}, and C.~{Jutten}, ``On the achievability of
  {C}ramér–{R}ao bound in noisy compressed sensing,'' \emph{{IEEE} Trans.
  Signal Process.}, vol.~60, no.~1, pp. 518--526, Jan. 2012.

\bibitem{qin2018sparse}
Z.~{Qin}, J.~{Fan}, Y.~{Liu}, Y.~{Gao}, and G.~Y. {Li}, ``Sparse representation
  for wireless communications: A compressive sensing approach,'' \emph{{IEEE}
  Signal Process. Mag.}, vol.~35, no.~3, pp. 40--58, May 2018.

\bibitem{debbah2020EEPC}
B.~{Matthiesen}, A.~{Zappone}, K.~L. {Besser}, E.~A. {Jorswieck}, and
  M.~{Debbah}, ``A globally optimal energy-efficient power control framework
  and its efficient implementation in wireless interference networks,''
  \emph{{IEEE} Trans. Signal Process.}, vol.~68, pp. 3887--3902, 2020.

\bibitem{simonyan2014very}
K.~Simonyan and A.~Zisserman, ``Very deep convolutional networks for
  large-scale image recognition,'' \emph{arXiv preprint arXiv:1409.1556}, 2014.

\end{thebibliography}

\begin{IEEEbiography}[{\includegraphics[width=1in,height=1.25in,clip,keepaspectratio]{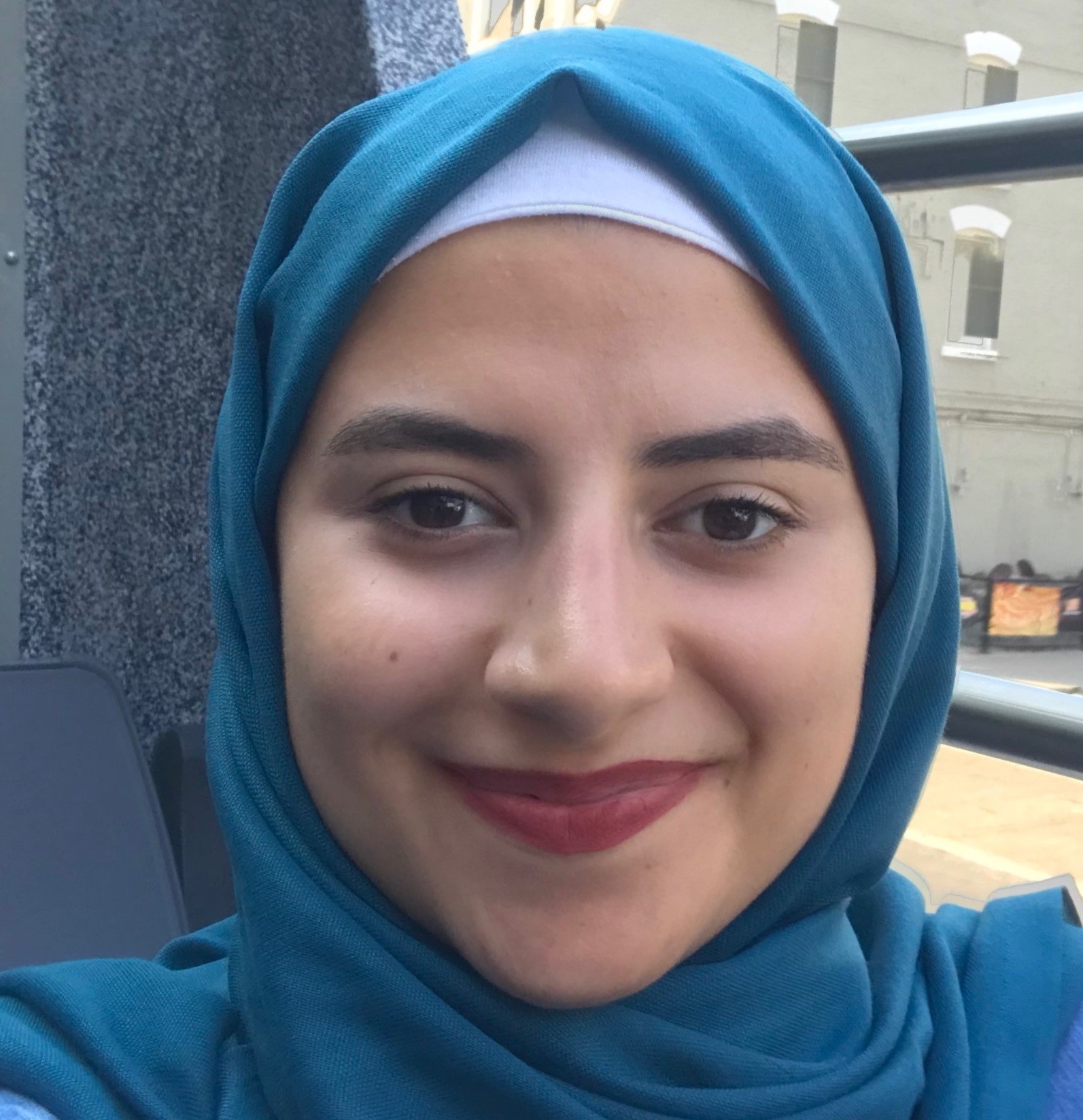}}]{Asmaa Abdallah}  received the B.S. (with High Distinction) and M.S degree in computer and communications engineering from Rafik Hariri University (RHU), Lebanon, in 2013 and 2015, respectively. In 2020, she received the Ph.D. degree in electrical and computer engineering at the American University of Beirut (AUB), Beirut, Lebanon. She is currently a post-doctoral fellow at King Abdullah University of Science and Technology (KAUST). She has been a research and teaching assistant at AUB since 2015. She was a research intern at Nokia Bell Labs in France from July 2019 till December 2019, where she worked on new hybrid automatic request (HARQ) mechanisms for long-delay channel in non-terrestrial networks (NTN).  Her research interests are in the area of communication theory, stochastic geometry for wireless communications, array signal processing, with emphasis on energy and spectral efficient algorithms for Device-to-Device (D2D) communications, massive multiple-input and multiple-output (MIMO) systems and cell free massive MIMO systems. Ms. Abdallah was the recipient of the Academic Excellence Award at RHU in 2013 for ranking first on the graduating class. She also received a scholarship from the Lebanese National Counsel for Scientific Research (CNRS-L/AUB) to support her doctoral studies. 
\end{IEEEbiography}
		
\begin{IEEEbiography}[{\includegraphics[width=1in,height=1.25in,clip,keepaspectratio]{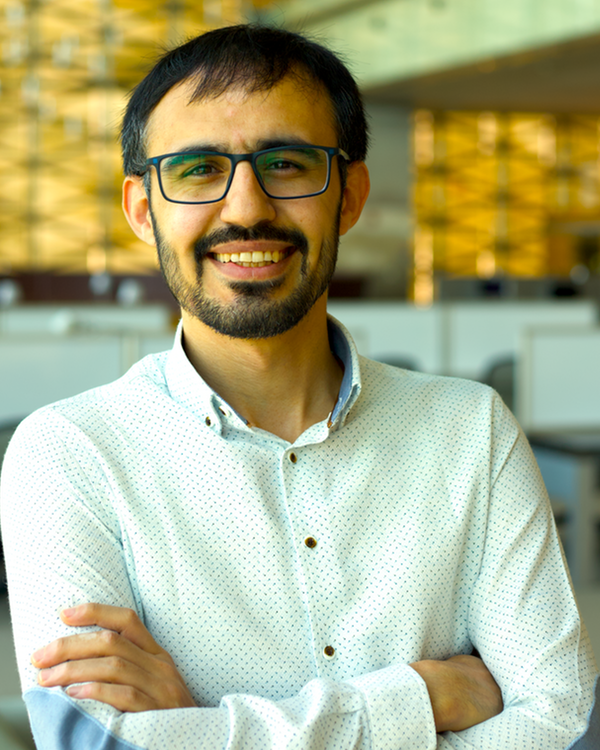}}]{Abdulkadir Celik} (S'14-M'16-SM'19) received the M.S. degree in electrical engineering in 2013, the M.S. degree in computer engineering in 2015, and the Ph.D. degree in co-majors of electrical engineering and computer engineering in 2016 from Iowa State University, Ames, IA, USA. He was a post-doctoral fellow at King Abdullah University of Science and Technology (KAUST) from 2016 to 2020. Since 2020, he has been a research scientist at the communications and computing systems lab at KAUST. His research interests are in the areas of wireless communication systems and networks. 
\end{IEEEbiography}

\begin{IEEEbiography}[{\includegraphics[width=1in,height=1.25in,clip,keepaspectratio]{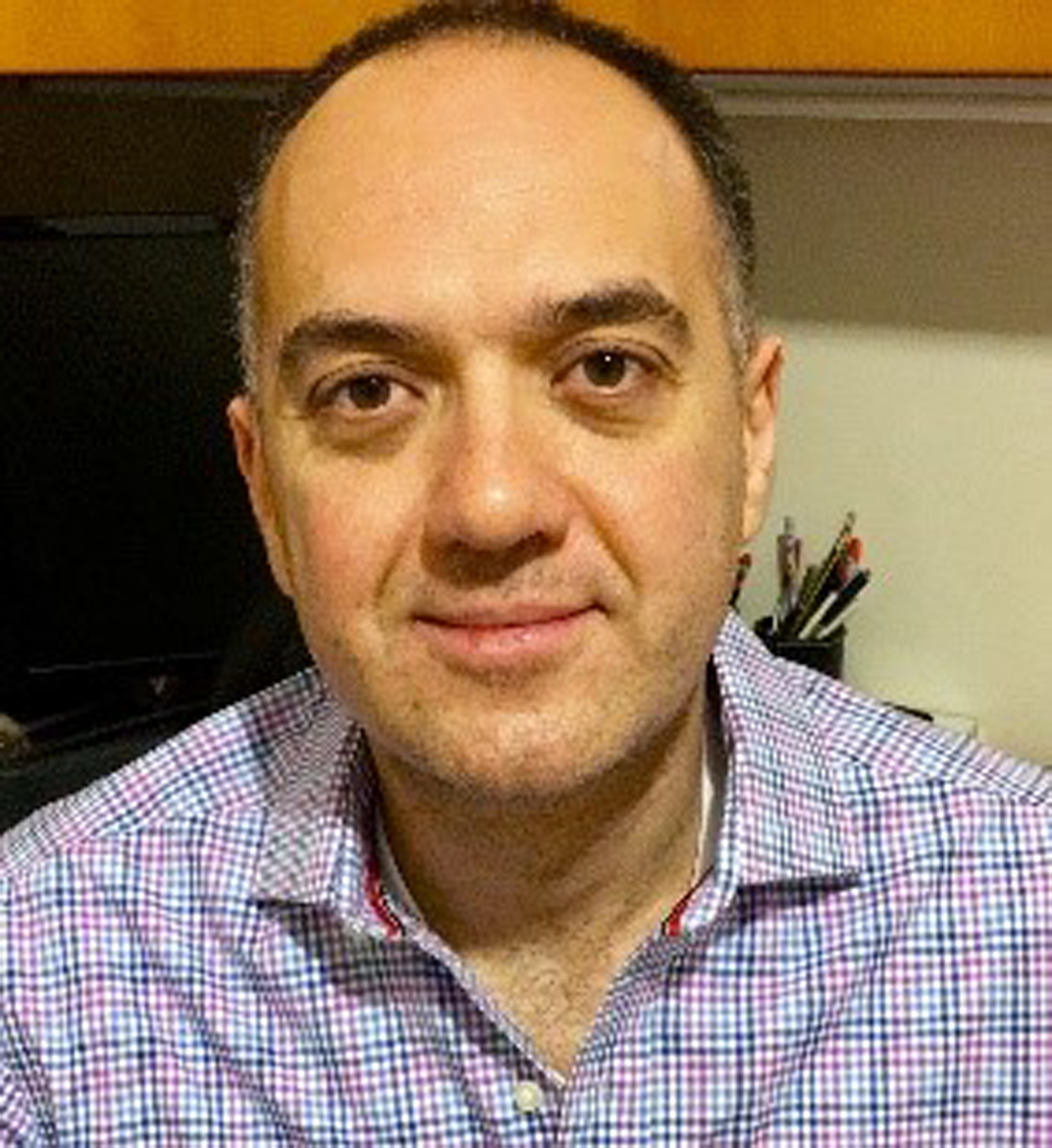}}]{Mohammad M. Mansour}(S'97-M'03-SM'08) received the B.E. (Hons.) and the M.E. degrees in computer and communications engineering from the American University of Beirut (AUB), Beirut, Lebanon, in 1996 and 1998, respectively, and the M.S. degree in mathematics and the Ph.D. degree in electrical engineering from the University of Illinois at Urbana–Champaign (UIUC), Champaign, IL, USA, in 2002 and 2003, respectively.
			
He was a Visiting Researcher at Qualcomm, San Jose, CA, USA, in summer of 2016, where he worked on baseband receiver architectures for the IEEE 802.11ax standard. He was a Visiting Researcher at Broadcom, Sunnyvale, CA, USA, from 2012 to 2014, where he worked on the physical layer SoC architecture and algorithm development for LTE-Advanced baseband receivers. He was on research leave with Qualcomm Flarion Technologies in Bridgewater, NJ, USA, from 2006 to 2008, where he worked on modem design and implementation for 3GPP-LTE, 3GPP2-UMB, and peer-to-peer wireless networking physical layer SoC architecture and algorithm development. He was a Research Assistant at the Coordinated Science Laboratory (CSL), UIUC, from 1998 to 2003. He worked at National Semiconductor Corporation, San Francisco, CA, with the Wireless Research group in 2000. He was a Research Assistant with the Department of Electrical and Computer Engineering, AUB, in 1997, and a Teaching Assistant in 1996. He joined as a faculty member with the Department of Electrical and Computer Engineering, AUB, in 2003, where he is currently a Professor. His research interests are in the area of energy-efficient and high-performance VLSI circuits, architectures, algorithms, and systems for computing, communications, and signal processing.
			
Prof. Mansour is a member of the Design and Implementation of Signal Processing Systems (DISPS) Technical Committee Advisory Board of the IEEE Signal Processing Society. He served as a member of the DISPS Technical Committee from 2006 to 2013. He served as an Associate Editor for IEEE TRANSACTIONS ON CIRCUITS AND SYSTEMS II (TCAS-II) from 2008 to 2013, as an Associate Editor for the IEEE SIGNAL PROCESSING LETTERS from 2012 to 2016, and as an Associate Editor of the IEEE TRANSACTIONS ON VLSI SYSTEMS from 2011 to 2016. He served as the Technical Co-Chair of the IEEE Workshop on Signal Processing Systems in 2011, and as a member of the Technical Program Committee of various international conferences and workshops. He was the recipient of the PHI Kappa PHI Honor Society Award twice in 2000 and 2001, and the recipient of the Hewlett Foundation Fellowship Award in 2006. He has seven issued U.S. patents.
\end{IEEEbiography}

\begin{IEEEbiography}[{\includegraphics[width=1in,height=1.25in,clip,keepaspectratio]{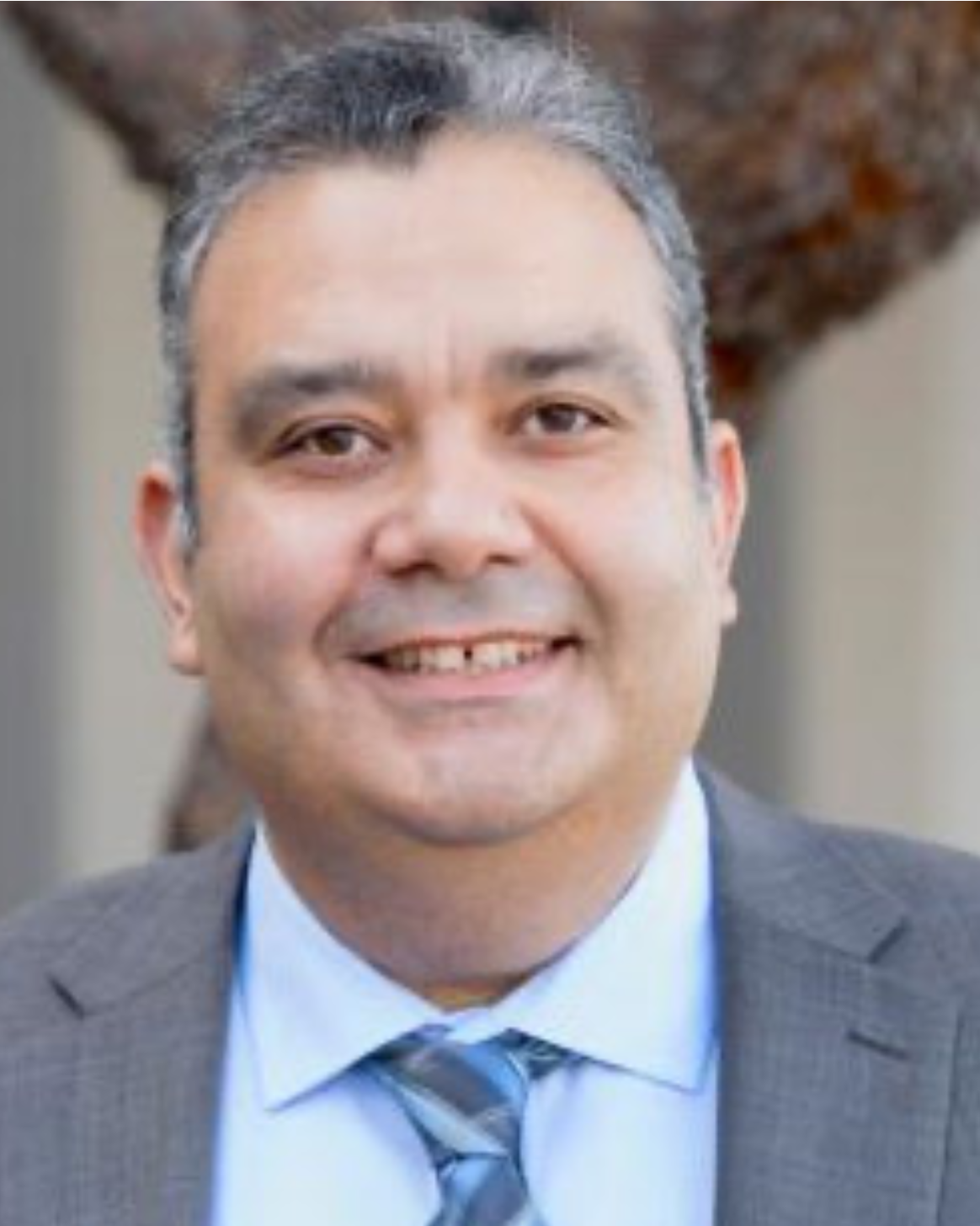}}]{Ahmed M. Eltawil}  	 (Senior Member, IEEE) received the M.Sc. and B.Sc. degrees (Hons.) from Cairo University, Giza, Egypt, in 1999 and 1997, respectively, and the Ph.D. degree from the University of California, Los Angeles, CA, USA, in 2003. Since 2019, he has been a Professor with the Computer, Electrical and Mathematical Science and Engineering Division (CEMSE), King Abdullah University of Science and Technology (KAUST), Thuwal, Saudi Arabia. Since 2005, he has been with the Department of Electrical Engineering and Computer Science,
University of California at Irvine, where he founded the Wireless Systems and Circuits Laboratory. His research interests are in the general area of low power digital circuit and signal processing architectures with an emphasis on mobile systems. He has been on the technical program committees and steering committees for numerous workshops, symposia, and conferences in the areas of low power computing and wireless communication system design. He received several awards, as well as distinguished grants, including the NSF CAREER Grant supporting his research in low power systems.
\end{IEEEbiography}

\end{document}